\documentclass[12pt]{iopart}
\usepackage{cite}
\usepackage{epsfig}
\usepackage{amssymb}
\usepackage{braket}
\usepackage{here}
\usepackage{color}
\usepackage{graphicx}
\usepackage{epstopdf}
\usepackage{float}
\usepackage{color}
\usepackage{siunitx}
\usepackage{soul,color}
\soulregister\cite7
\soulregister\ref7
\usepackage{ulem}
\usepackage{caption}
\usepackage{hyperref}
\usepackage{array,multirow,makecell} 
\usepackage{caption,tabularx,booktabs}
\usepackage{keyval}
\usepackage{graphicx}
\usepackage{comment}

\begin{document}

\title{Giant dipole resonance and shape evolution in Nd isotopes within TDHF method}

\author{A. Ait Ben Mennana$^1$ ,Y. EL Bassem$^1$, M. Oulne$^1$}

\address{ $^1$ High Energy Physics and Astrophysics Laboratory, Department of Physics,
	Faculty of Sciences SEMLALIA, Cadi Ayyad University,
	P.O.B. 2390, Marrakesh, Morocco.}
\ead{oulne@uca.ma}
\vspace{10pt}
\begin{indented}
\item[]December 2019
\end{indented}

\begin{abstract}
The isovector giant dipole resonance (IVGDR) in   even-even Nd isotopes from A=124 to A=160 is studied in the framework of  time-dependent Hartree-Fock (TDHF) with Skyrme forces SkI3, SVbas, SLy5 and SLy6. The dipole strength is calculated and compared with the experimental data on photon absorption cross section $\sigma_{\gamma}$.  An overall   agreement  between them is obtained. The dipole strengths in $ ^{124-140}\text{Nd}$ and $^{152-160}\text{Nd} $ are predicted. In addition, the correlation between the quadrupole deformation parameter $\beta_{2}$ and the splitting  $\Delta E/ \bar{E}_{m}$ of the giant dipole resonance (GDR) spectra is studied. The results confirm that $\Delta E/ \bar{E}_{m}$ is proportional to $\beta_{2}$. Shape phase transition in Nd isotopes is also investigated in the light of IVGDR.
\end{abstract}

%\keywords{Time-Dependent Hartree-Fock; giant diople resonance; Nd isotopes.}

\section{Introduction}
\qquad Giant resonances (GR) are an excellent example of collective modes in  nuclei. Among different types of GRs, the most known and oldest is the isovector giant diople resonance (IVGDR). It was interpreted by Goldhaber and Teller (GT) as collective vibrations  of the protons moving against the neutrons in the nucleus  with the centroid energy of the form $ E_{c}  \propto  A^{-1/6}$\cite{goldhaber1948}. On the other hand, Steinwedel and Jensen (SJ)  interpreted it as a vibration of proton fluid against neutron fluid with a fixed surface where the centroid energy has the form 
$ E_{c}  \propto  A^{-1/3}$~\cite{speth1981}. The experimental data are adjusted by a combination of these two \cite{berman1975}: in light nuclei, the data follow the law $ A^{-1/6} $, while the dependence $ A^{-1/3} $ becomes more and more dominant for increasing values of
A. Also, it has been much studied from the experimental point of view (see for example Refs.\cite{carlos1971,berman1975,donaldson2018}) as well as  from the theoritical one (Refs.~\cite{goeke1982,maruhn2005,reinhard2008,yoshida2011}).\\

Deformed nuclei provide an interesting testing ground since  there is a
strong interplay between the structure of the GDR and the ground-state deformation~\cite{danos1958}. The isotopic chain of neodymium (Nd) between A = 124 and A = 160 belongs to a known region with a sudden transition from spherical to deformed shape in nuclei and vice versa, and is therefore an excellent field to investigate the effects of deformation in GDRs in heavy deformed nuclei. GDRs in heavy deformed nuclei have been previously investigated by various microscopic methods such as time-dependent Skyrme-Hartree-Fock method ~\cite{maruhn2005,fracasso2012}, Separable Random-Phase-Approximation (SRPA) \cite{reinhard2008,reinhard2007c}, Relativistic Quasiparticle Random Phase Approximation (RQRPA) \cite{ring2009} and  Extended Quantum Molecular Dynamics (EQMD) \cite{wang2017}. The excitation of the GDRs in the experiment is induced by various ways, such as photoabsorption~\cite{carlos1971,Masur2006} inelastic scattering \cite{donaldson2018, ramakrishnan1996},$\gamma$-decay~\cite{gundlach1990}. \\

Many works based on the time-dependent Hartree-Fock (TDHF) approach have studied GRs in nuclei. The TDHF method provides a good approximation for GR. Early TDHF calculations concentrated on giant monopole resonance (GMR)\cite{blocki1979,chomaz1987} because they  require only a spherical one-dimensional code. In the last few years with the increase in computer power, large scale TDHF calculations become possible with no assumptions on the spatial symmetry of the system\cite{maruhn2005,maruhn2006,stevenson2003}. such calculations are performed by codes use a fully three dimensional (3D) Cartesian grid in coordinate space \cite{sky3d}.\\

The isotopic chain of neodymium (Nd) has been studied in many previous works \cite{ maruhn2005, reinhard2008,yoshida2011, wang2017} by different approaches. In Ref.\cite{maruhn2005}, Maruhn et al. have studied the deformation dependence of giant dipole resonances (GDR) and its interplay with Landau fragmentation for some deformed heavy nuclei such as $^{142-150}\text{Nd}$ within TDHF by using two Skyrme forces SkI3\cite{REINHARD1995} and SLy6\cite{CHABANAT1998}. They found that the results for these Nd isotopes are slightly downshift of order 1 MeV compared to experimental data~\cite{carlos1971}. In particular, the weakly deformed isotopes  are  dominated by Landau fragmentation which caused by coupling of the collective strength to 1ph states  that are close in energy. In this work we extended this study to nineteen Nd isotopes by using different Skyrme forces. \\

The goal of the present work is to study the GDR and shape evolution in  even-even Nd isotopes between A=124 and A=160 with TDHF theory~\cite{negele1982} using SKY3D code~\cite{sky3d}, based
on the Skyrme functional. This code uses a fully three dimensional  (3D) Cartesian grid in coordinate space  with no spatial symmetry restrictions and include all time-odd terms. Consequently, it is possible to study both spherical and deformed system within the limitation of mean field theory. Due to the open-shell nature of these nuclei,
pairing and deformation properties must be taken into account
in this study. Firstly, a static calculation  gives some properties of the ground-state of nuclei like $\beta_{2}$, $\gamma$. In dynamic calculation, the ground-state of nuclei is boosted by imposing a dipole excitation  to obtain the GDR spectra and some of their properties (resonance energies, width). \\   

The paper is outlined as follows: in Sec.\ref{sec2}, we give a brief description of  TDHF method and the GDR in deformed nuclei. In Sec.\ref{sec3}, we present  details of the numerical calculations. Our results and discussion are presented in Sec.\ref{sec4}. Finally, Sec.\ref{sec5} gives the summary.

\section{The Time-Dependent Hartree-Fock method to giant resonances}\label{sec2}

\subsection{The TDHF method}
\qquad The TDHF is a self-consistent mean field (SCMF) theory which was proposed by Dirac in 1930 \cite{dirac}. It generalizes the static hartree-Fock(HF) to dynamics problems to treat for example, the vibration of nuclei such as giant resonances  \cite{maruhn2005,stevenson2003,reinhard2007,blocki1979} and Heavy-ion collisions \cite{simenel2018,maruhn2006}. Detailed discussions  of  TDHF theory can be found in several references \cite{engel1975,kerman1976,koonin1977}.
The TDHF equations are obtained from the variation of Dirac action 
\begin{equation}{\label{eq1}}
S \equiv S_{t_0,t_1}[\psi] = \int_{t_0}^{t_1} dt  \bra{\psi(t)} \bigg(i\hbar\frac{d }{dt} - \hat{H} \bigg) \ket{\psi(t)}, 
\end{equation}
where $ \ket{\psi} $ is the Slater determinant, $ t_{0} $ and $ t_{1} $ define the time interval, where the action S is stationary between the fixed endpoints $ t_{0} $ and $ t_{1} $, and $ \hat{H}  $ is the Hamiltonian of the system. The energy of the system is defined as  $ E = \bra{\psi} \hat{H} \ket{\psi} $, and we have 
\begin{equation}{\label{eq2}}
\bra{\psi}\frac{d }{dt}\ket{\psi} = \sum_{i=1}^{N} \bra{\varphi_{i}}\frac{d }{dt}\ket{\varphi_{i}},
\end{equation}
where $ \ket{\varphi_{i}} $ are the occupied single-particle states.
The action S can be expressed as

\begin{eqnarray}{\label{eq3}}
S & = & \int_{t_0}^{t_1} dt \bigg( i\hbar \sum_{i=1}^{N} \bra{\varphi_{i}}\frac{d }{dt}\ket{\varphi_{i}} - E[\varphi_{i}]\bigg) \nonumber \\
& = & \int_{t_0}^{t_1} dt \bigg( i\hbar \sum_{i=1}^{N} \int dx \,\varphi_{i}^{*}(x,t) \frac{d }{dt}\varphi_{i}(x,t) - E[\varphi_{i}] \bigg)
\end{eqnarray}
The variation of the action S
with respect to the wave functions $ \varphi_{i}^* $ reads
\begin{equation}{\label{eq4}}
\frac{\delta S}{\delta \varphi_{i}^*(x,t)} = 0,
\end{equation}
for each $ i = 1....N  $, $ t_{0}\leq t \leq {t_{1}} $ and for all $ x $. More details can be found  for example in Refs.~\cite{kerman1976, simenel2012}. We finally get  the TDHF equation 
\begin{equation}{\label{eq5}}
i\hbar\frac{\partial }{\partial t}\varphi_{i}(t) = \hat{h}[\rho(t)]\varphi_{i}(t) \quad \text{for} \quad 1\leq i \leq \text{N}.
\end{equation}
where $ \hat{h} $ is the single-particle Hartree-Fock Hamiltonian.

The TDHF equations (\ref{eq5}) are solved \textit{iteratively} by a small time step $ \Delta t $  during which we assume that the Hamiltonian remains constant. To conserve the total energy E, it is necessary to apply a symmetric algorithm by time reversal, and therefore to estimate the Hamiltonian at time $ t + \frac{\Delta t}{2} $ to evolve the system between time $ t \;\text{and}\; t+ \Delta t$ \cite{flocard1978,bonche1976}
\begin{equation}{\label{eq6}}
\ket{\varphi(t+\Delta t)} \simeq e^{-i\frac{\Delta t}{\hbar}\hat{h}(t+\frac{\Delta t}{2})}\ket{\varphi(t)}.
\end{equation} 
\subsection{ Giant dipole resonance in deformed nuclei}
\qquad The GDR in deformed axially symetric nuclei splits into  two components with energies $E_{i} \sim R_{i}^{-1} \sim A^{-1/3} $ \cite{speth1981} where R is the nuclear radius,  and even three resonances in the case of asymmetric nuclei. This splitting has been observed experimentally \cite{carlos1971,berman1975,donaldson2018} and treated theoretically by different models \cite{maruhn2005,reinhard2008,yoshida2011}.  For a spherical nucleus, the GDR spectra present one peak i.e the oscillations along the three axes have the same frequency $E_{x} =  E_{y} = E_{z} $. For the axially symmetric prolate nuclei, the GDR spectra present two peaks where the low-energy $E_{z}$ corresponds to the oscillations along the  major axis of symmetry and the high-energy $E_{x}= E_{y}$ corresponds to the oscillations along  transverse minor axes of the nuclear ellipsoid, due to $E \sim R^{-1}$. For an oblate nucleus, it is the opposite situation to the prolate case.

Among the properties of the ground-state of  nuclei,there are the deformation parameters $\beta_{2}$ and $\gamma$ which give an idea on the shape of the nucleus. These deformation parameters are defined as follows \cite{sky3d}
\begin{equation}{\label{eq7}}
\beta = \sqrt{a_{0}^2 + 2a_{2}^2} \qquad , \quad \gamma = atan\bigg(\frac{\sqrt{2}a_{2}}{a_{0}}\bigg),
\end{equation}
where $  a_{m} = \frac{4\pi}{5}\frac{Q_{2m}}{AR^2}   $ with $R = r_{0}A^{1/3}$.

\section{Details of Calculations}\label{sec3}
\qquad In our study of GDR in  isotopes of Nd (Z = 60) from A = 124 to A = 160, we used a  TDHF code Sky3D (v1.1)~\cite{sky3d} . This code solves the static Hartree-Fock(HF) as well as the time-dependent Hartree-Fock (TDHF) equations for Skyrme interactions \cite{SKYRME1958}. This study was performed with four Skyrme forces: SkI3 \cite{REINHARD1995}, SVbas\cite{reinhard2009},  SLy6 \cite{CHABANAT1998}, SLy5\cite{CHABANAT1998}. These Skyrme forces are widely used for the ground state properties (binding energies, radii...) and dynamics (as giant resonances) of nuclei including deformed ones. In particular they provide a reasonable description of the GDR: SLy6 and SkI3\cite{maruhn2005,reinhard2008}, SVbas\cite{reinhard2009}  and SLy5\cite{fracasso2012}.
In Table~\ref{tab1}, we summarize the parameters  of theses  Skyrme forces used in this study.
\begin{table}[!htbp]
	\caption{Parameters ($ t,x $) of the  Skyrme forces used in this study. \label{tab1}}
	% 	\label{1}
	{\begin{tabular}{@{}ccccc@{}} \hline
			Parameters & SkI3 & SVbas &  SLy6 & SLy5 \\
			\hline
			$ t_{0} $ (MeV.fm$^3$) & -1762.880 & -1879.640 &  -2479.500 & -2484.880\\
			$t_{1}$ (MeV.fm$^5$) & 462.180 & 313.749 &  -1762.880 & 483.130\\
			$ t_{2} $ (MeV.fm$^5$)& -227.090 & 112.676  & -448.610 & -549.400\\
			\quad	$ t_{3} $ (MeV.fm$^{3+3\sigma}$)& 8106.200 & 12527.389 &  13673.000 & 13763.000\\
			$ x_{0} $ & 0.308 & 0.258 &  0.825 & 0.778\\
			$ x_{1} $ & -1.172 & -0.381 &  -0.465 & -0.328\\
			$ x_{2} $ & -1.090 & -2.823 &  -1.000 & -1.000\\
			$ x_{3} $ & 1.292 & 0.123 &  1.355 & 1.267\\
			$ \sigma $ & 0.250 & 0.300 &  0.166 & 0.166\\
			W$ _{0} $ (MeV.fm$^5$)& 188.508 & 124.633 &  122.000 & 126.000\\
			\hline
	\end{tabular}}
\end{table}

The calculation is done in two successive steps for a given nucleus:
\begin{itemize}
	\item \textit{Static calculations}: \\
	To carry out a TDHF calculations, an intial state is required. This state is obtained by running  SKY3D code \cite{sky3d} in static mode, which solves the static HF + BCS equations (\ref{eq8})  in a three-dimensional Cartesian
	basis \textit{iteratively} until
	we obtain a convergence, i.e when for example the sum of the single-particle energy fluctuations becomes less than a certain value determined at the beginning of the static calculation. In this work we take as a convergence value 	$ 10^{-5} $ which is sufficient for heavy nuclei \cite{sky3d}.
	\begin{equation}{\label{eq8}}
	\hat{h}\psi_{i}(x)= \epsilon_{i} \psi_{i}(x) \quad \text{for} \quad i=1,....,A,
	\end{equation}
	where $ \hat{h} $ is the single-particle Hamiltonien, and $ \epsilon_{i} $ is the single-particle energy of the state  $ \psi_{i}(x) $ with $ x=(\vec{r},\sigma,\tau) $.\\
	In the input data file of SK3YD we use 24 as the number of grid points in the three cartesian directions ($ n_{x}=n_{y}=n_{z}=24 $) and 1 fm as spacing between grid points ($ d{x}=d{y}=d{z}=1 $ fm). Pairing is treated in the static calculation.  
	\item \textit{Dynamic calculations}:\\
	%	After obtaining the initial states by the static calculations, we apply a dipole boost operator in order  	to start the nucleus in the dipole mode.
	The ground-state wave function obtained by the static calculations is excited by an instantaneous initial dipole boost operator in order to put the nucleus in the dipole mode \cite{maruhn2005,simenel2009,stevenson2008}. 
	\begin{equation}{\label{eq9}}
	\varphi_{\alpha}^{(g.s)}(r) \longrightarrow \varphi_{\alpha}(r,t=0)=\exp(ib\hat{D})\varphi_{\alpha}^{(g.s)}(r),
	\end{equation}
	where $ \varphi_{\alpha}^{(g.s)}(r) $ represents the ground-state of nucleus before the boost, b is the boost amplitude of the studied mode , and $ \hat{D} $ the associated operator. In our case, $ \hat{D} $ represents the isovector dipole operator  defined as
	\begin{eqnarray}{\label{eq10}}
	\hat{D} & = & \frac{NZ}{A} \bigg( \frac{1}{Z}\sum_{p=1}^{Z}\vec{z}_p - \frac{1}{N}\sum_{n=1}^{N}\vec{z}_n \bigg) \nonumber \\
	& = & \frac{NZ}{A}\bigg( \vec{R}_Z - \vec{R}_N \bigg),
	\end{eqnarray}
	where $ \vec{R}_Z $ (resp. $ \vec{R}_N $ ) measures the proton (resp. neutron)  average position on the z axis.
	
	In order to obtain the spectral distribution of the isovector dipole strength, we apply a boost (\ref{eq9}) with a small value of the amplitude b to stay in the linear regime of the excitation. For a long enough time,  the dipole moment $ \hat{D} = \bra{\psi(t)}\hat{D}\ket{\psi(t)} $ is recorded along the dynamical evolution. Finally, the dipole strength $ S_{D}(\omega) $ can be obtained by performing the Fourier transform $ D(\omega) $ of the signal $ \hat{D}(t) $, defined as \cite{ring1980}
	\begin{eqnarray}{\label{eq11}}
	S_{D}(\omega) & = & \sum_{\nu} \delta(E-E_{\nu}) \big| \bra{\nu}\hat{D}\ket{0}\big|^2. 
	\end{eqnarray}
	
	Some filtering is necessary to avoid artifacts in the spectra obtained by catting the signal at a certain final time, in order to the signal vanishes at the end of the simulation time. In practice we use windowing in the time domain by damping the signal $ D(t) $  at the final time with $ cos \big(\frac{\pi t}{2T_{fin}}\big)^n $ \cite{sky3d}.
	\begin{equation}{\label{eq12}}
	D(t) \longrightarrow D_{fil} = D(t). cos \bigg(\frac{\pi t}{2T_{fin}}\bigg)^n,
	\end{equation}
	where n represents the strength of filtering and $ T_{fin} $ is the final time  of the simulation. More details can be founded in Refs. \cite{sky3d, reinhard2006} .\\
	In this work, all dynamic calculations were performed in a cubic space with 24 x 24 x 24 fm$ ^3 $ according to the three directions (x, y, z) and a grid spacing of  1 fm. We chose nt= 5000 as number of time steps to be run, and dt = 0.2 fm/c is the time step, so T$ _ {f}$ = 1000 fm/c is the final time of simulation. Pairing is frozen in the dynamic calculation i.e, the BCS occupation numbers are frozen at their initial values during time evolution.
\end{itemize}
\section{Results and Discussion}\label{sec4}
\subsection{Relation between nuclear shape and  deformation parameters }

\begin{figure}[!htbp]
	\begin{center}
		\includegraphics[scale=0.6]{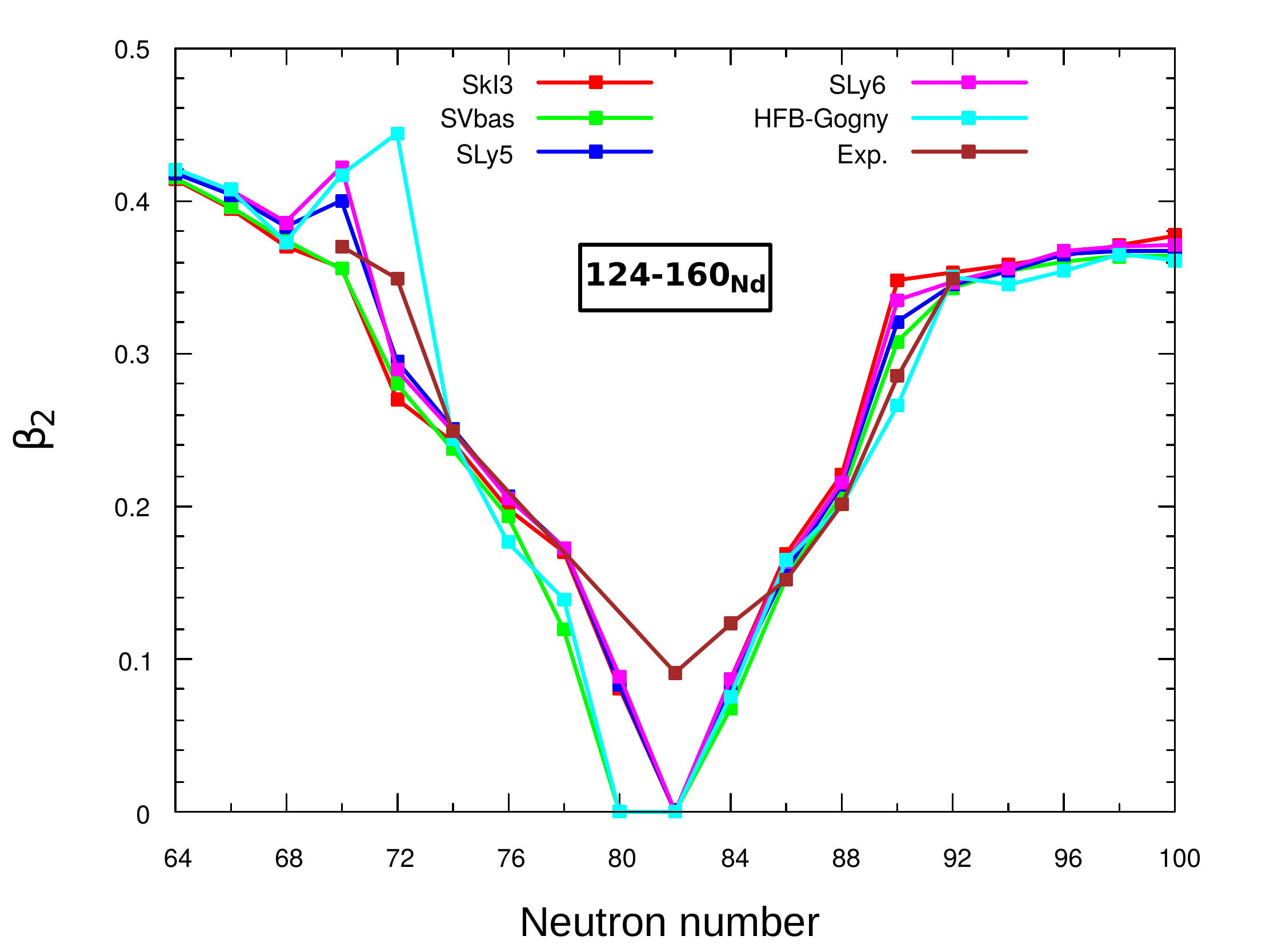}
		\caption{The Quadrupole deformation parameter $ \beta_{2} $ of $ ^{124-160}\text{Nd} $ isotopes as function of their neutron number N. The experimental data are from Ref.  \cite{raman2001}.}
		\label{b2-n}
	\end{center}
\end{figure}

\qquad The chain of  Nd isotopes studied in this work belongs to a region that knows a transition between spherical, where the number of neutrons is close to N=82, and prolate or oblate shape when N increases or decreases \cite{carlos1971,yoshida2011,wang2017}. The deformation parameters  $ \beta_{2} $ and $ \gamma $ give  an estimate of the   nuclei shape \cite{ring1980,takigawa2017}. In Table \ref{tab2}, we summarize the results obtained for the quadrupole deformation parameter $ \beta_{2} $ of $ ^{124-160}\text{Nd} $  isotopes with four Skyrme forces within TDHF method, including the available experimental data from Ref.\cite{raman2001} and the HFB calculations based on the D1S Gogny force~\cite{HFB} for comparison. The variation of $ \beta_{2} $ as a function of neutrons number N is plotted in Fig.~\ref{b2-n}.

%In TABLE. I, the deformation parameters β2  of Nd  isotopes are shown, calculated with four Skyrme forces, including the experimental data from Ref. [42] 
\begin{table}[!htbp]
	\centering
	\caption{The quadrupole deformation parameter $ \beta_{2} $ calculated with SkI3, SVbas,  SLy6, and SLy5 are compared with the experimental data are from Ref.\cite{raman2001}, and data from Ref.\cite{HFB}. \label{tab2} }
	{\begin{tabular}{@{}ccccccc@{}} \hline
			Nuclei & SkI3 & SVbas &  SLy6 & SLy5 & Exp. \cite{raman2001}&HFB$\_$Gogny.\cite{HFB}\\
			\hline
			$ ^{124}\text{Nd} $ & 0.414 & 0.415 &  0.421 & 0.418 & -----&0.421\\
			$ ^{126}\text{Nd} $ & 0.395 & 0.396 &  0.407 & 0.404 & -----&0.407\\
			$ ^{128}\text{Nd} $ & 0.370 & 0.374 &  0.386 & 0.383 & -----&0.373\\
			$ ^{130}\text{Nd} $ & 0.356 & 0.355 &  0.422 & 0.400 & 0.370&0.417\\
			$^{132}\text{Nd}  $ & 0.270 & 0.280 &  0.289 & 0.295 & 0.349&0.444\\
			$ ^{134}\text{Nd} $ & 0.242 & 0.237 &  0.249 & 0.251 & 0.249&0.244\\
			$ ^{136}\text{Nd} $ & 0.198 & 0.193 &  0.205 & 0.207 & -----&0.176\\
			$ ^{138}\text{Nd} $ & 0.170 & 0.119 &  0.173 & 0.173 & -----&0.139\\
			$ ^{140}\text{Nd} $ & 0.081 & 0.000 &  0.089 & 0.083 & -----&0.000\\
			$ ^{142}\text{Nd} $ & 0.000 & 0.000 &  0.000 & 0.000 & 0.091&0.000\\
			$ ^{144}\text{Nd} $ & 0.087 & 0.068 &  0.087 & 0.083 & 0.123&0.076\\
			$ ^{146}\text{Nd} $ & 0.169 & 0.153 &  0.164 & 0.158 & 0.152&0.165\\
			$ ^{148}\text{Nd} $ & 0.221 & 0.208 &  0.216 & 0.214 & 0.201&0.201\\
			$ ^{150}\text{Nd} $ & 0.348 & 0.308 &  0.335 & 0.321 & 0.285&0.266\\
			$ ^{152}\text{Nd} $ & 0.353 & 0.343 &  0.347 & 0.345 & -----&0.350\\
			$ ^{154}\text{Nd} $ & 0.358 & 0.354 &  0.356 & 0.354 & -----&0.345\\
			$ ^{156}\text{Nd} $ & 0.364 & 0.360 &  0.367 & 0.365 & -----&0.354\\
			$ ^{158}\text{Nd} $ & 0.371 & 0.364 &  0.370 & 0.367 & -----&0.365\\
			$ ^{160}\text{Nd} $ & 0.377 & 0.364 &  0.371 & 0.367 & -----&0.361\\
			\hline
	\end{tabular}}
	%\label{table:2}
\end{table}
%%%%%%%%%%%%%%%%%%%%%%%%%%%%%%%%%%%%%%%%%%%%%%%%%%%%%%%%%%%%%%%%
%\begin{comment}
As it can be seen from Fig.\ref{b2-n}, the agreement between the two theoretical calculations: this work and HFB based on the D1S Gogny force\cite{HFB} theory. On the other hand, our calculations are generally in agreement  with experimental data\cite{raman2001}, whose maximum deviation reaches the value $ \simeq 0.090 $ in the case of the spherical nucleus $ ^{142}\text{Nd}$. The discrepancy between theory and experiment is due to that the experimental data are, in fact, a total variance of $ \beta_{2} $ while  theoretical value is an expectation one. Thus theoretical $ \beta_{2} $ values are allowed to be generally a bit smaller \cite{reinhard2019}. We notice also that the minimum of $ \beta_{2} $ corresponds to the magic number of neutrons N = 82 where the nucleus $ ^{142}\text{Nd}$ has a spherical shape ($ \beta_{2} \simeq 0 $) for the two theoretical calculations, while it is slightly deformed with experimental data ($ \beta_{2}\simeq0.091 $).
%\end{comment}
\begin{comment}
\hl{As it can be seen from Fig.\ref{b2-n}, the two theoretical calculations: TDHF (this work) and HFB based on the D1S Gogny force\cite{HFB} produce similar $ \beta_{2} $ values for all Nd isotopes under investigation. Both of these two theoretical calculations predict that $ ^{142}\text{Nd}$ with the magic number of neutrons N = 82 is spherical ($ \beta_{2}\simeq0 $), while it is slightly deformed with experimental data ($ \beta_{2}\simeq0.091 $)\cite{raman2001}.
}
\end{comment}

The results obtained for the deformation parameter $\gamma$ are shown in Table \ref{tab3}, compared with those of HFB  based on the D1S$\_$Gogny  effective nucleon-nucleon interaction~\cite{HFB}.
\begin{table}[!htbp]
	\centering
	\caption{The deformation parameters $ \gamma $,  calculated with SkI3, SVbas,  SLy6, and SLy5 are compared with  data  from Ref.\cite{HFB}.\label{tab3}} 
	{\begin{tabular}{@{}cccccc@{}} \hline
			Nuclei & SkI3 & SVbas &  SLy6 & SLy5 &  HFB$\_$Gogny.\cite{HFB}	\\
			\hline
			$ ^{124}\text{Nd}  $ & $0.0^\circ$ & $0.0^\circ$ &  $0.0^\circ$ & $0.0^\circ$ & $0.0^\circ$\\
			$^{126}\text{Nd}$ & $6.3^\circ$ & $0.0^\circ$ &  $0.0^\circ$ & $0.0^\circ$ & $0.0^\circ$\\
			$ ^{128}\text{Nd}$& $0.0^\circ$ & $0.0^\circ$ &  $0.0^\circ$ & $0.0^\circ$ & $0.0^\circ$\\
			$ ^{130}\text{Nd}  $ & $0.0^\circ$ & $0.0^\circ$ &  $0.0^\circ$ & $0.0^\circ$ & $0.0^\circ$\\
			$^{132}\text{Nd}$ & $18.1^\circ$ & $13.1^\circ$ &  $15.9^\circ$ & $13.2^\circ$ & $0.0^\circ$\\
			$ ^{134}\text{Nd}$& $21.1^\circ$ & $21.7^\circ$ &  $22.1^\circ$ & $22.0^\circ$ & $20.0^\circ$\\
			$ ^{136}\text{Nd}  $ & $21.5^\circ$ & $23.3^\circ$ &  $23.0^\circ$ & $23.5^\circ$ & $21.0^\circ$\\
			$^{138}\text{Nd}$ & $25.0^\circ$ & $0.1^\circ$ &  $25.6^\circ$ & $25.9^\circ$ & $18.0^\circ$\\
			$ ^{140}\text{Nd}$& $0.1^\circ$ & $14.7^\circ$ &  $0.0^\circ$ & $0.0^\circ$ & $0.0^\circ$\\
			$ ^{142}\text{Nd}$& $54.8^\circ$ & $39.5^\circ$ & $29.0^\circ$ & $29.2^\circ$ & $0.0^\circ$\\
			$ ^{144}\text{Nd}$ & $0.2^\circ$ & $0.9^\circ$ &  $0.1^\circ$ & $0.0^\circ$ & $0.0^\circ$\\
			$ ^{146}\text{Nd}$ & $0.4^\circ$ & $0.0^\circ$ &  $0.0^\circ$ & $0.0^\circ$ & $0.0^\circ$\\
			$ ^{148}\text{Nd}$ & $0.0^\circ$ & $0.0^\circ$ &  $0.0^\circ$ & $0.0^\circ$ & $0.0^\circ$\\
			$ ^{150}\text{Nd}$ & $0.0^\circ$ & $0.0^\circ$ &  $0.0^\circ$ & $0.0^\circ$ & $0.0^\circ$\\
			$ ^{152}\text{Nd}  $ & $0.0^\circ$ & $0.0^\circ$ &  $0.0^\circ$ & $0.0^\circ$ & $0.0^\circ$\\
			$^{154}\text{Nd}$ & $0.0^\circ$ & $0.0^\circ$ &  $0.0^\circ$ & $0.0^\circ$ & $0.0^\circ$\\
			$ ^{156}\text{Nd}$& $0.0^\circ$ & $0.0^\circ$ &  $0.0^\circ$ & $0.0^\circ$ & $0.0^\circ$\\
			$ ^{158}\text{Nd}  $ & $0.0^\circ$ & $0.0^\circ$ &  $0.0^\circ$ & $0.0^\circ$ & $0.0^\circ$\\
			$^{160}\text{Nd}$ & $0.0^\circ$ & $0.0^\circ$ &  $0.0^\circ$ & $0.0^\circ$ & $0.0^\circ$\\
			
			\hline
	\end{tabular}}
	%\label{table:2}
\end{table}
We notice that the results obtained are almost the same for the four Skyrme forces and the Gogny interaction. There is a disagreement in the case of  $^{132}\text{Nd}$ nucleus, where the  Skyrme forces  SkI3, SVbas, SLy5 and SLy6 give $ 13.2^\circ\leq \gamma \leq 18.1^\circ $ which shows that this nucleus is weakly  triaxial, whereas  Gogny interaction gives $\gamma = 0.0^\circ$ that  means this nucleus has a prolate shape. For the spherical nucleus  $^{142}\text{Nd}$ where $\beta_{2} \simeq 0$, the triaxiality is undefined that is why $\gamma$ takes different values. \\
According to the results of the couple ($ \beta_{2} $,$\gamma$), we can predict the shape of the nucleus. So for Nd isotopes below N = 82, the  isotopic
chains exhibit a transition from deformed to spherical
shape, and for neutron number higher than N = 82, both the experimental and calculated results
show that the prolate deformation increases gradually and then saturates at a value which closes to $\beta_{2}\simeq$ 0.365.
\subsection{Variation of the dipole moment}

\begin{figure}[!htbp]
	\centering
	\includegraphics[width=1.0\textwidth]{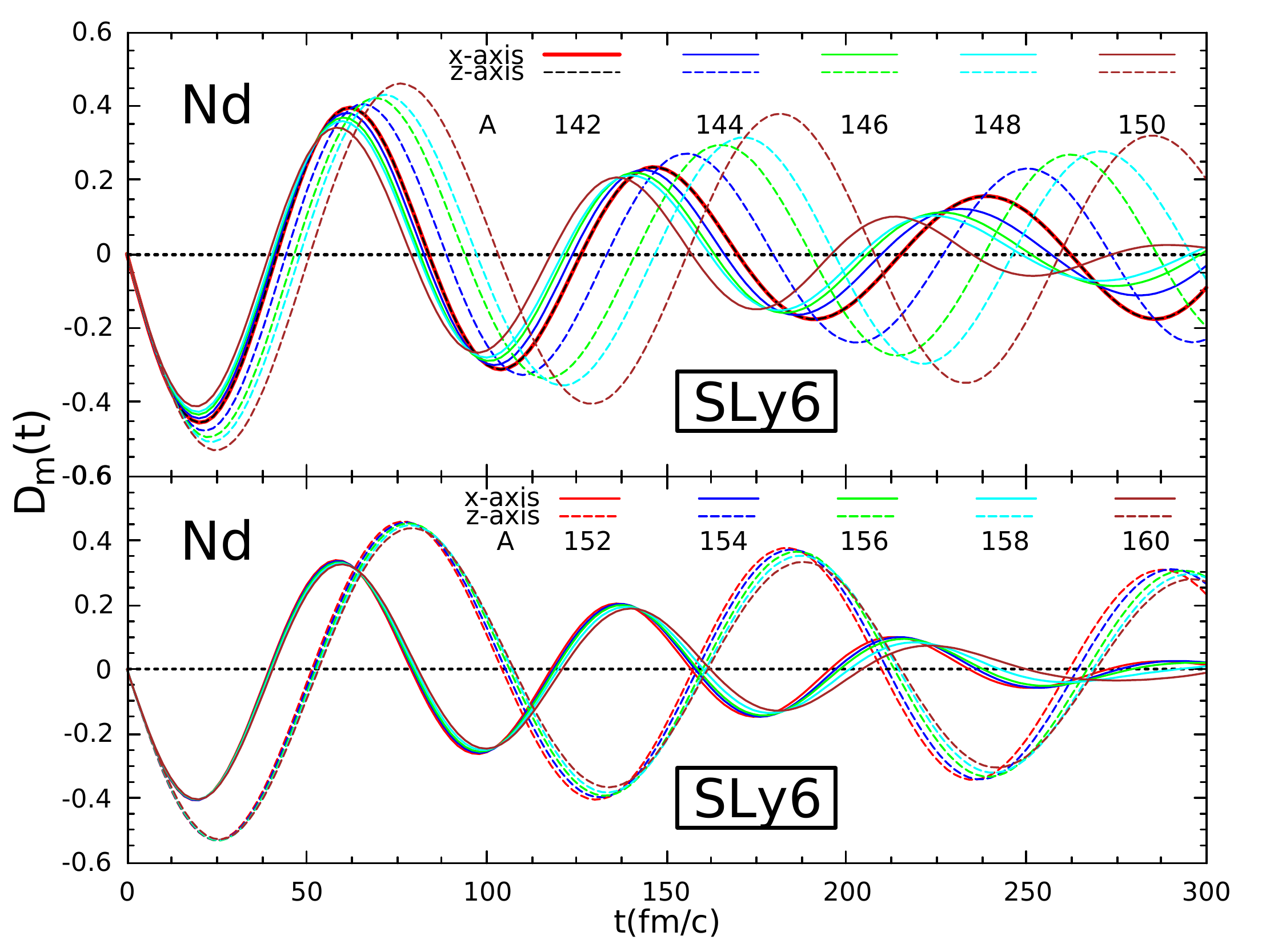} 
	\caption{The dipole moment $ D_{m}(t) $ as function of the simulation time  t(fm/c)  calculated with the Skyrme force SLy6   using the SKY3D code \cite{sky3d} for Nd isotopes from A=142 to A=160.}
	\label{dm-t1}
\end{figure}

\qquad Alongside the static calculation giving some properties of the ground-state of $ ^{124-160}\text{Nd} $ isotopes, the dynamic calculation with the SKY3D code \cite{sky3d} allow to study some dynamic properties of theses nuclei such as the dipole moment $ D_{m}(t) $ which is defined by  Eq.~(\ref{eq10}). Fig.~\ref{dm-t1} shows the time evolution of the dipole moment $ D_{m}(t) $ of $ ^{142-160}\text{Nd} $ isotopes calculated with  Skyrme force SLy6. We note that the collective motion of nucleons in GDR is  done generally along two axes. The oscillation frequency $\omega_{i}$ is related to the nuclear radius $ R_{i} $ by $\omega_{i} \propto R_{i}^{-1}$ where i$\in$\{x,y,z\}. For $ ^{124-130}\text{Nd}$ and $ ^{144-160}\text{Nd}$ isotopes, the oscillation frequencies alone the symmetry z-axis (the dashed line) are lower than that  along the two other axes x and y (solid line) perpendicular to z-axis, which shows that these isotopes are deformed in prolate shape because $\omega_{z}< \omega_{x}=\omega_{y}$\cite{Masur2006}. That is why deformed nuclei  have usually two splitting peaks in their GDR spectrum\cite{carlos1971,berman1975}. For the nucleus $ ^{142}\text{Nd}$, the oscillation frequencies along the three axes are equal ( $\omega_{x}=\omega_{y}=\omega_{z}$), which show  that it is spherical shape. We note also that the time evolution of dipole moment $ D_{m}(t) $ is almost the same for the others  Skyrme forces (SLy5, SkI3, SVbas).\\

%%%%%%%%%%%%%%%%%%%%%%%%%%%%%%%%%%%%%%%%%%%%%%%
\begin{comment}
\begin{figure}[!htbp]
	\includegraphics[scale=0.3]{figs/figure2a.pdf} \hfill
	\includegraphics[scale=0.3]{figs/figure2b.pdf}
	\includegraphics[scale=0.3]{figs/figure2c.pdf} \hfill
	\includegraphics[scale=0.3]{figs/figure2d.pdf}
	\caption{The dipole moment $ D_{m}(t) $ as function of the simulation time  t(fm/c)  calculated with the Skyrme forces SkI3, SVbas, SLy5 and SLy6   using the SKY3D code \cite{sky3d} for Nd isotopes from A=124 to A=130.}
	\label{dm-t1}
\end{figure}
\end{comment}
%%%%%%%%%%%%%%%%%%%%%%%%%%%%%%%%%%%%%%%%%%%%%%%%%%%%%%%%%%%%%%%%%%%%%

\begin{figure}[!htbp]
	\includegraphics[width=1.0\textwidth]{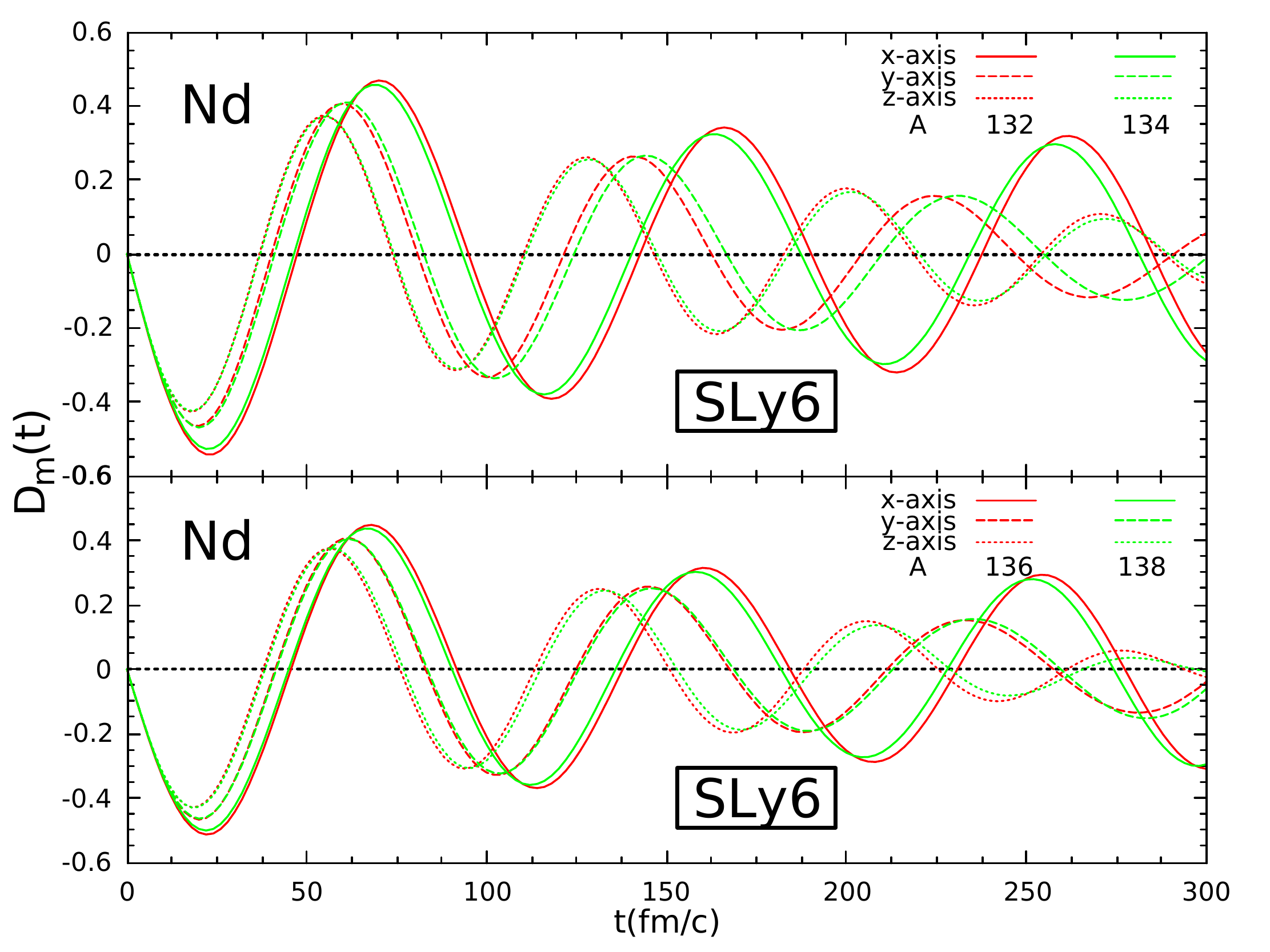}
	\caption{Same as in Fig.~2 for isotopes Nd from A=132 to A=138.}
	\label{dm-t2}
\end{figure}

Fig.~\ref{dm-t2} displays the time evolution  of D(t) for    $ ^{132-138}\text{Nd} $ isotopes calculated with the  Skyrme force SLy6. We notice that the oscillation frequencies $\omega_{i}$ along the three axes  are different from each other $\omega_{x}\neq \omega_{y}\neq \omega_{z}$, which shows that these  isotopes are weakly triaxial where $ 13.0^\circ\leq \gamma \leq 26^\circ $. We point out that $ ^{132-138}\text{Nd} $ is weakly triaxial with our calculation, but other works\cite{xiang2018} found that it  has  prolate shape. Thus we tend to think that shape coexistence  exist in this isotope.
%%%%%%%%%%%%%%%%%%%%%%%%%%%%%%%%%%%%%%%%%%%%%%%%%%%%%%%%%%%%%%%%
%%%%%%%%%%%%%%%%%%%%%%%%%%%%%%%%%%%%%%%%%%%%%%%%%%%%%%%%%%%%%%%%%ù

%%%%%%%%%%%%%%%%%%%%%%%%%%%%%%%%%%%%%%%%%%%%%%%%%%%%%%%%%%%%%%%%%
\begin{comment}
\begin{figure}[!htbp]
	\includegraphics[scale=0.6]{figs/figure5.pdf} 
	\caption{Same as in Fig.~2 for $ ^{140}\text{Nd} $ in case SLy6 and SVbas forces.}
	\label{dm-t4}
\end{figure}
\end{comment}
%%%%%%%%%%%%%%%%%%%%%%%%%%%%%%%%%%%%%%%%%%%%%%%%%%%%%%%%%%%%%%%%%
\subsection{GDR Spectrum}
\begin{figure}[!htbp]
	\begin{center}
		\begin{minipage}[t]{0.46\textwidth}
			\includegraphics[scale=0.5]{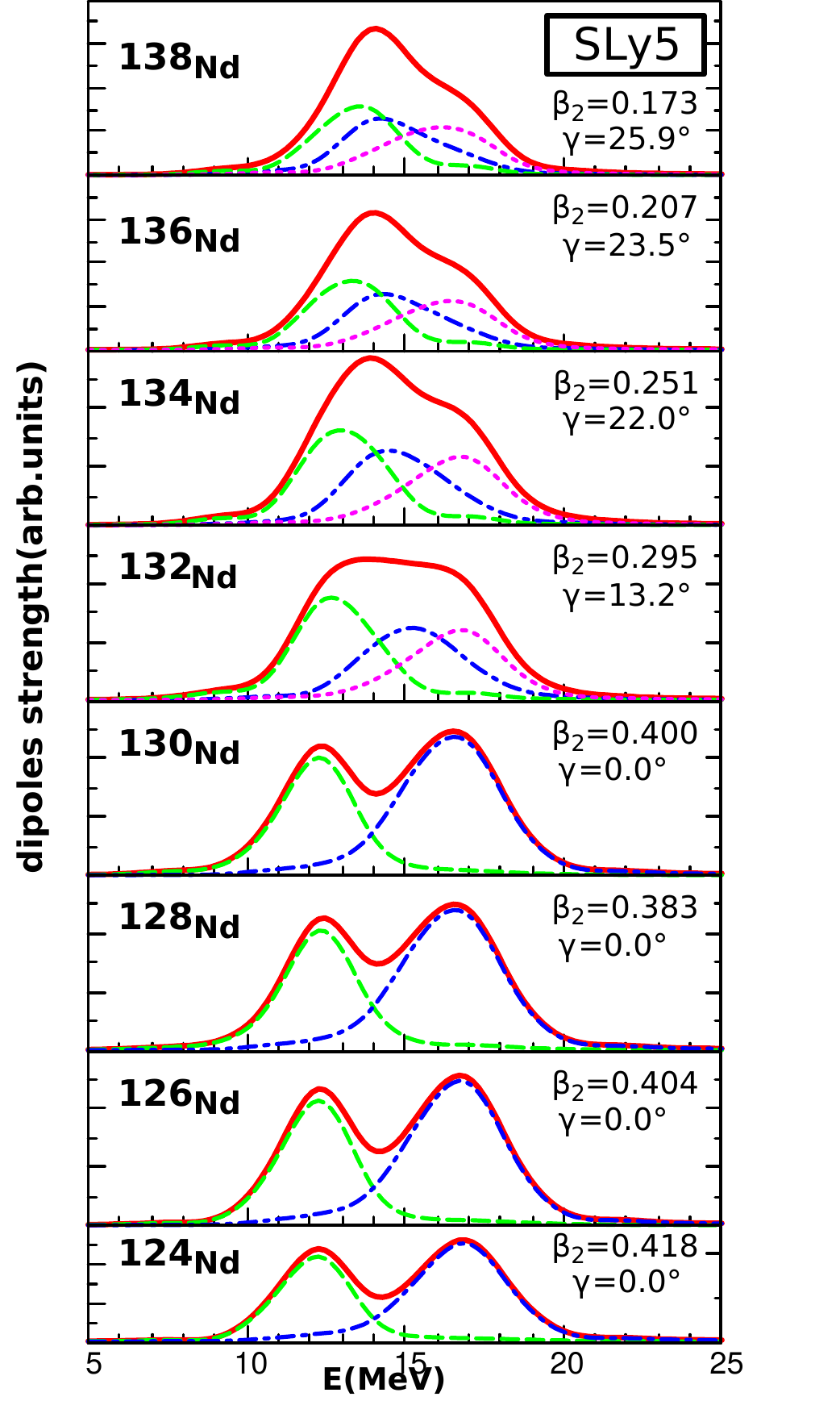}
		\end{minipage}
		\vspace{0.5cm}
		\hspace{0.5cm}
		\begin{minipage}[t]{0.46\textwidth}
			\includegraphics[scale=0.5]{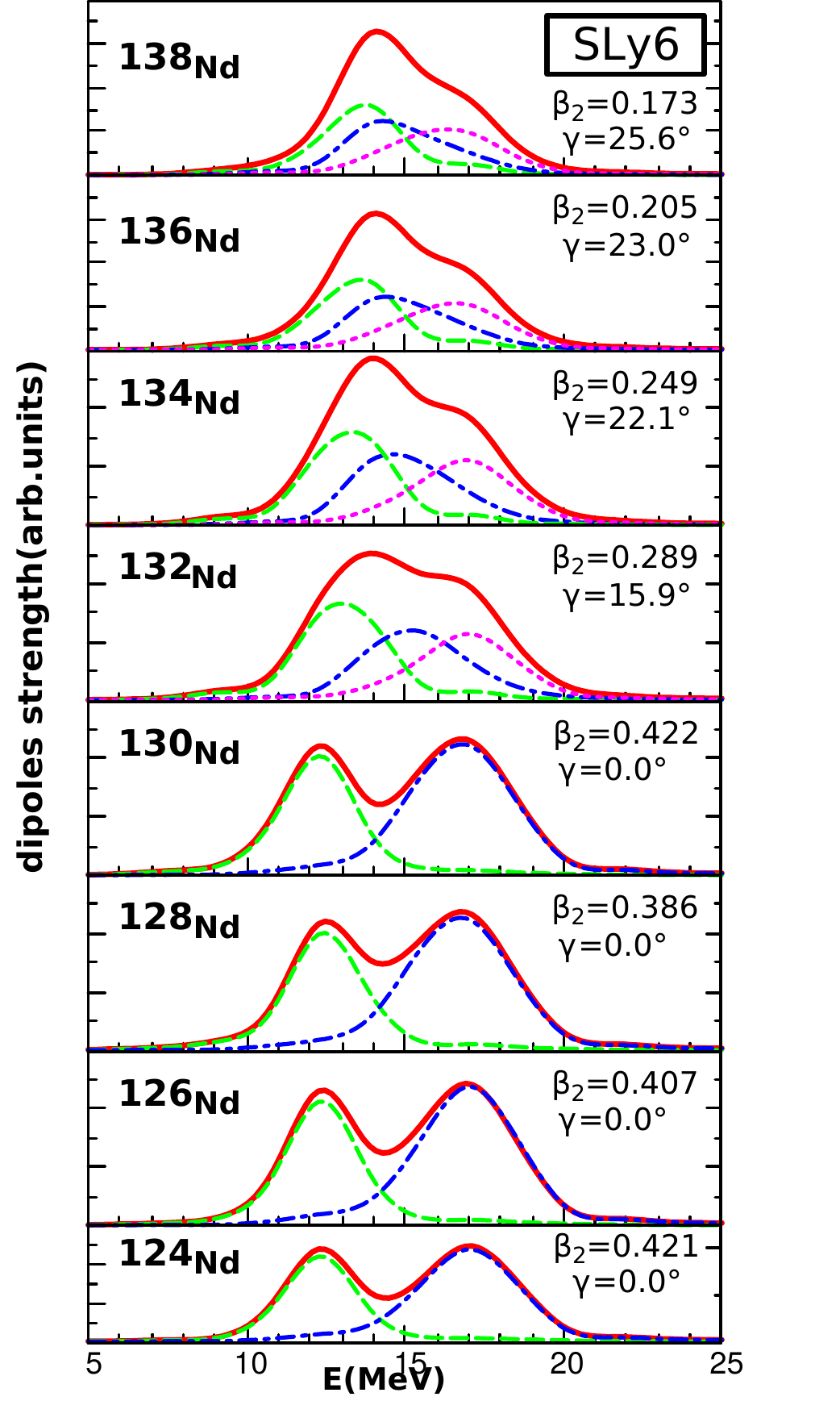}
		\end{minipage}
		\begin{minipage}[t]{0.46\textwidth}
			\includegraphics[scale=0.5]{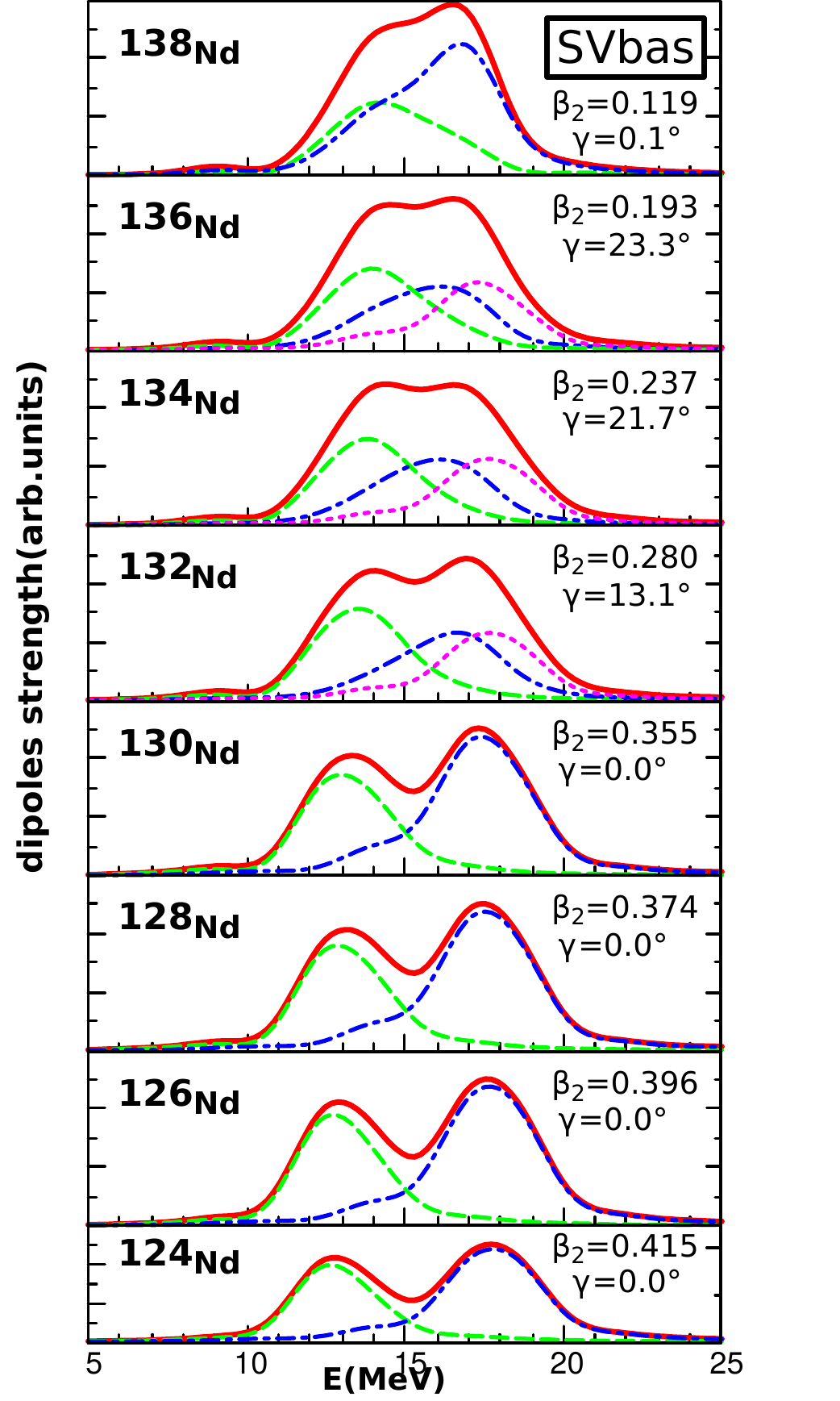}
		\end{minipage}
		\hspace{0.5cm}
		\begin{minipage}[t]{0.46\textwidth}
			\includegraphics[scale=0.5]{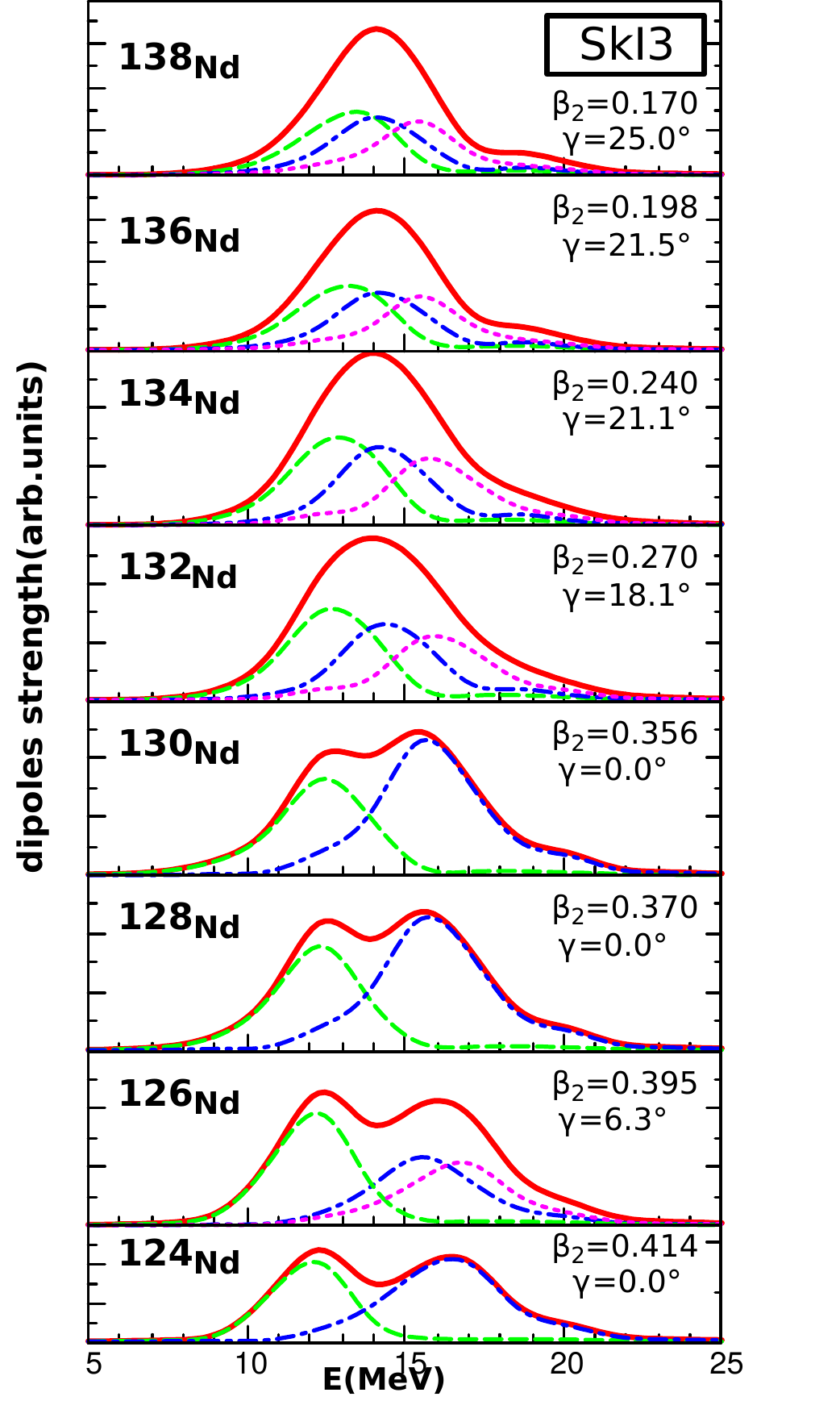}
		\end{minipage}
		\caption{ GDR spectra in the chain of $ ^{124-138}\text{Nd} $ calculated with  SLy5, SLy6, SVbas and SkI3 . The solid(red), dashed(green) and dotted-dashed(magenta) lines denote the dipole strengths: total, along the long axis and the short axis(multiplied by 2 except $ ^{132-138}\text{Nd} $  ) respectively. The dotted (brown) line denote the strength along the third middle axis in the case of the triaxial nuclei $ ^{132-138}\text{Nd} $. The calculated strength total is compared with the experimental data \cite{carlos1971} depicted by blue dots.} 
		\label{gr24-38}
	\end{center}	
\end{figure}

\begin{figure}[!htbp]
	\begin{center}
		\begin{minipage}[t]{0.46\textwidth}
			\includegraphics[scale=0.5]{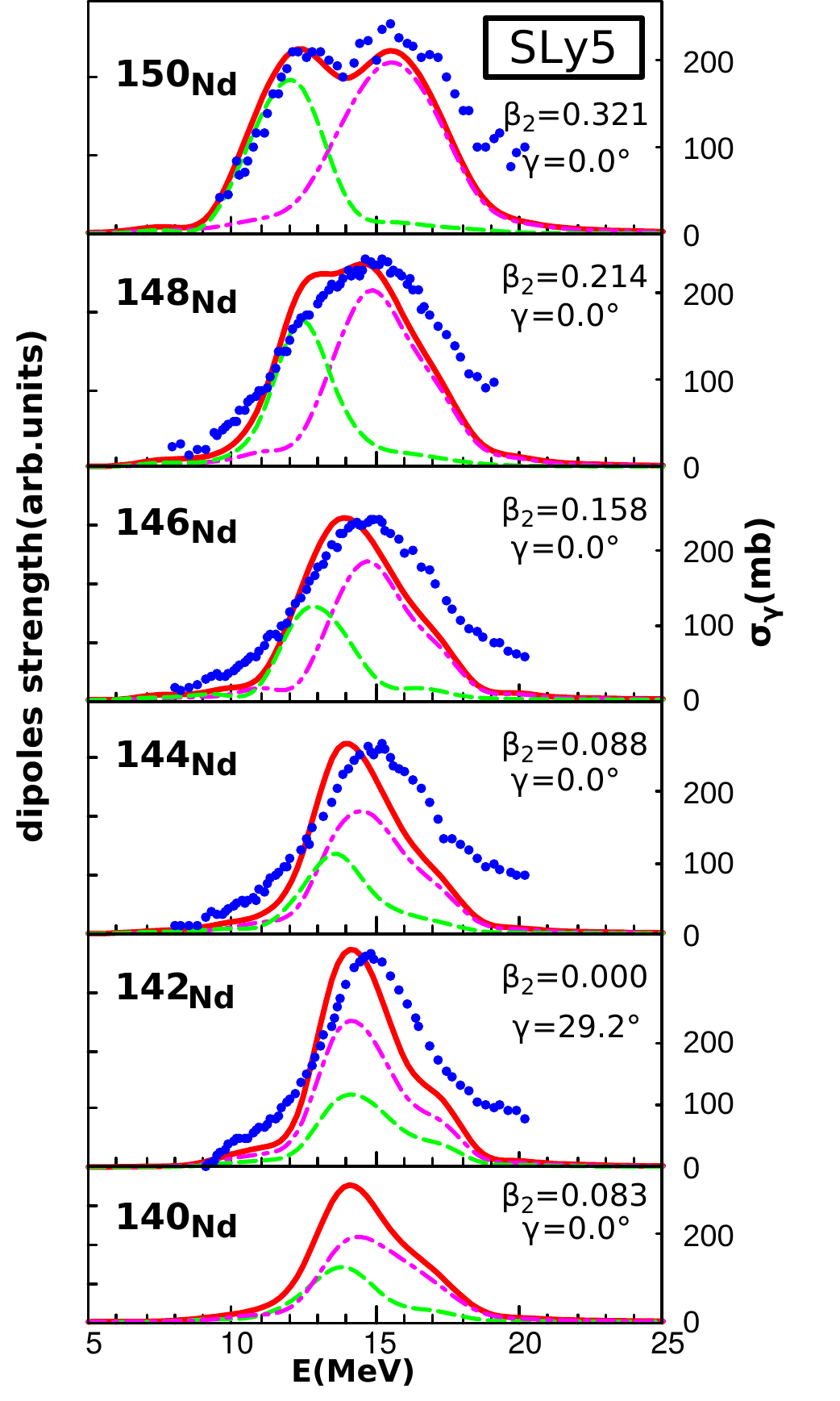}
		\end{minipage}
		\vspace{0.5cm}
		\hspace{0.5cm}
		\begin{minipage}[t]{0.46\textwidth}
			\includegraphics[scale=0.5]{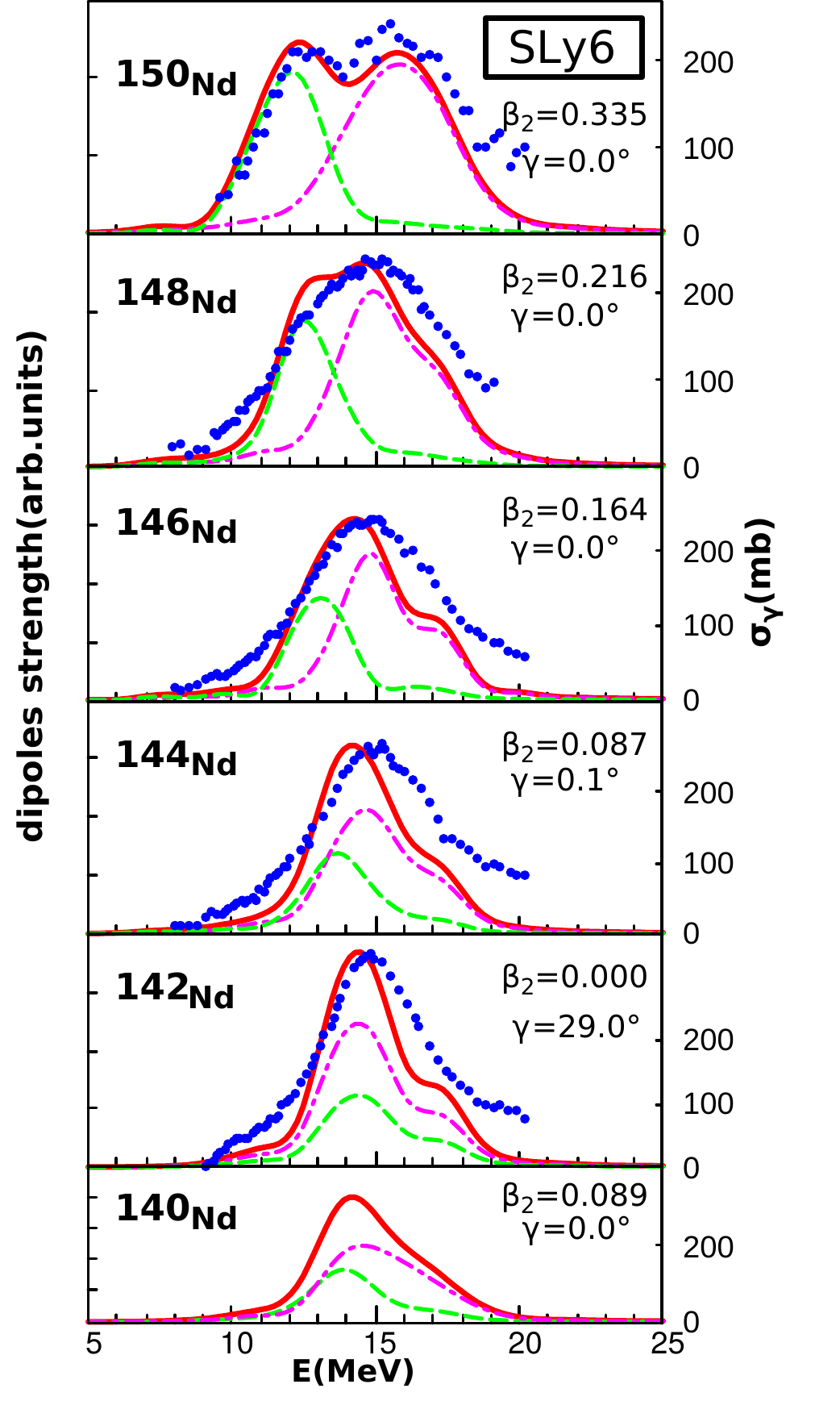}
		\end{minipage}
		\begin{minipage}[t]{0.46\textwidth}
			\includegraphics[scale=0.5]{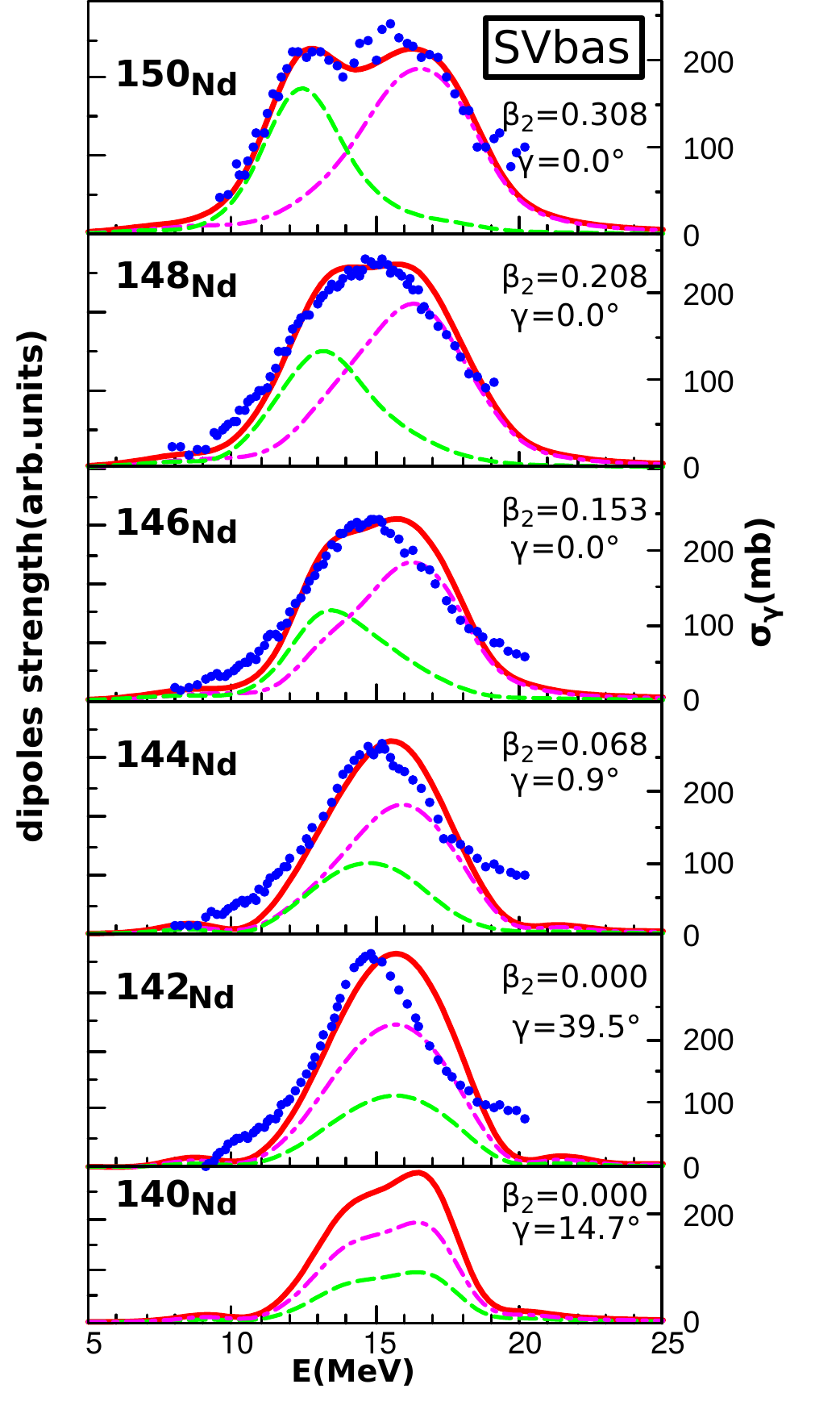}
		\end{minipage}
		\hspace{0.5cm}
		\begin{minipage}[t]{0.46\textwidth}
			\includegraphics[scale=0.5]{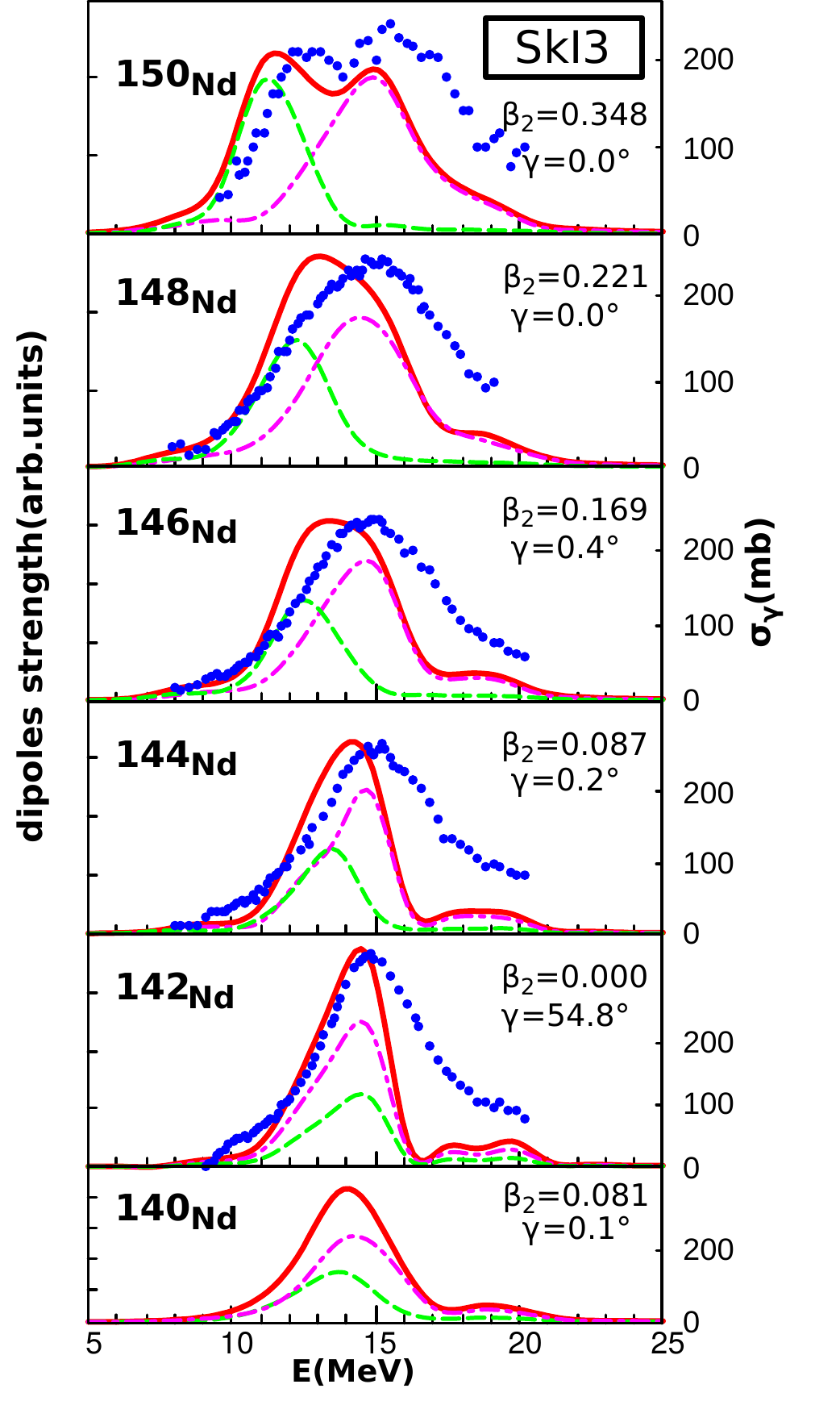}
		\end{minipage}
		\caption{ GDR spectra in the chain of $ ^{140-150}\text{Nd} $ calculated with  SLy5, SLy6, SVbas and SkI3 . The solid(red), dashed(green) and dotted-dashed(magenta) lines denote the dipole strengths: total, along the long axis and the short axis (multiplied by 2) respectively. The calculated strength total is compared with the experimental data \cite{carlos1971} depicted by blue dots.} 
		\label{gr40-50}
	\end{center}	
\end{figure}

\begin{figure}[!htbp]
	\begin{center}
		\begin{minipage}[t]{0.46\textwidth}
			\includegraphics[scale=0.5]{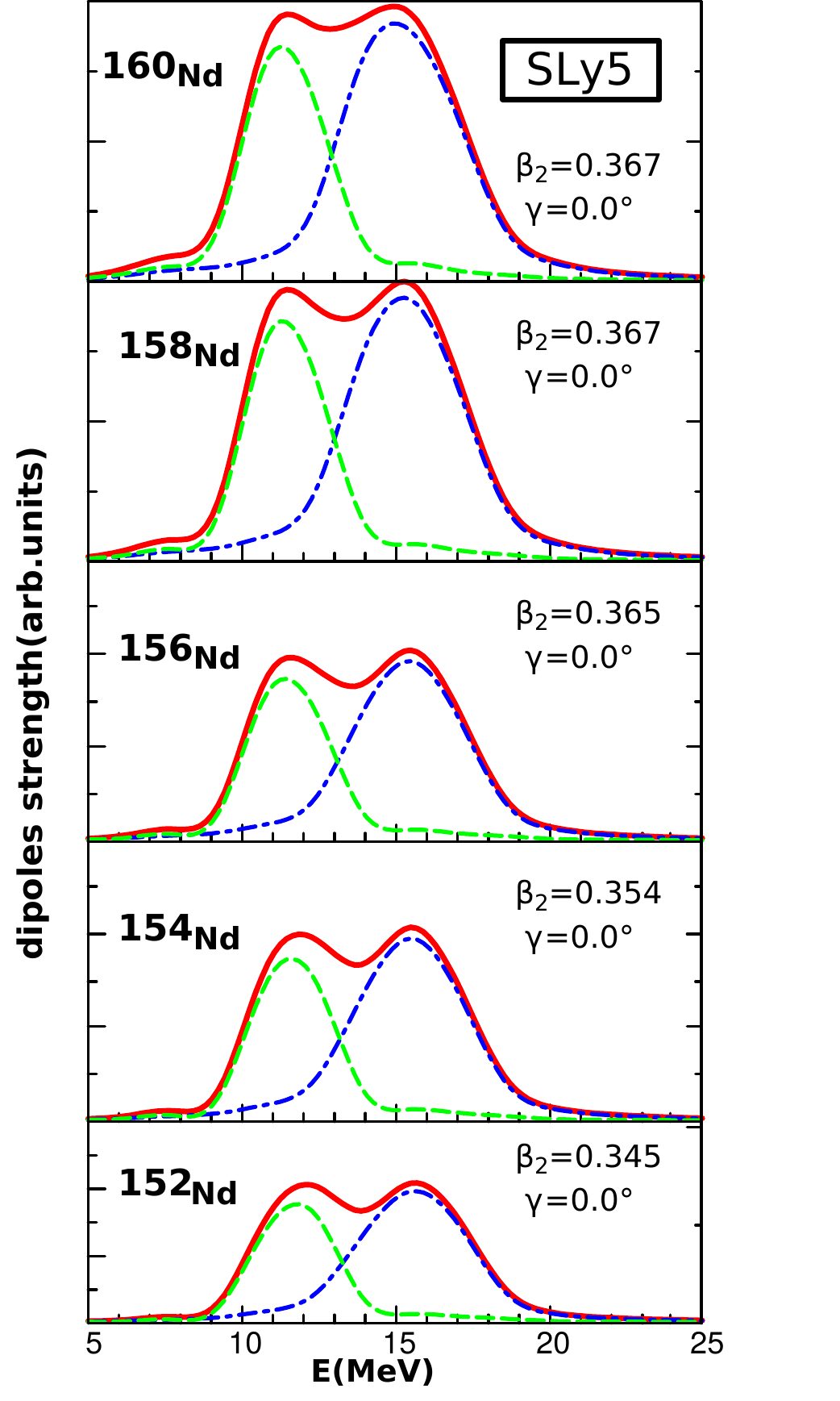}
		\end{minipage}
		\vspace{0.5cm}
		\hspace{0.5cm}
		\begin{minipage}[t]{0.46\textwidth}
			\includegraphics[scale=0.5]{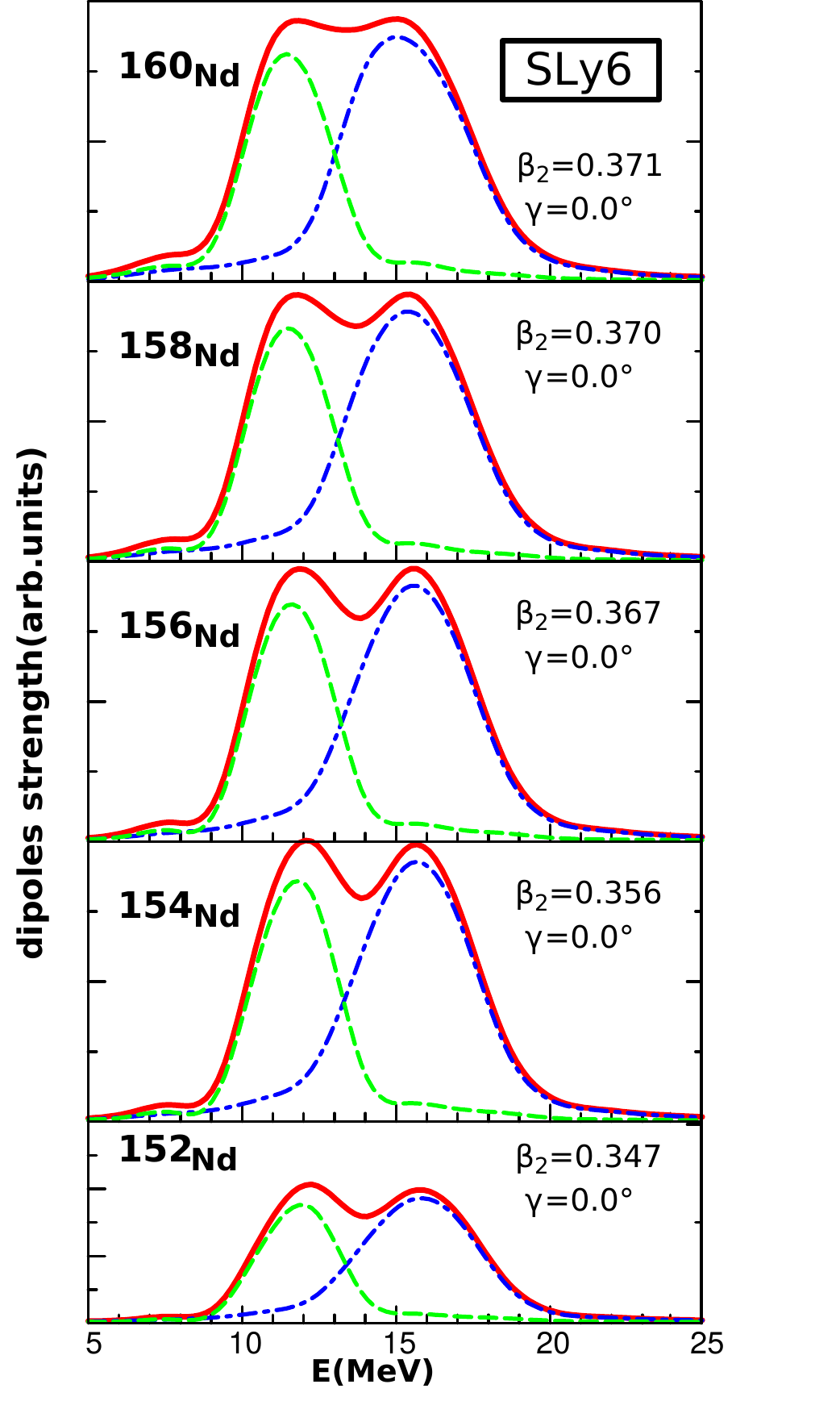}
		\end{minipage}
		\begin{minipage}[t]{0.46\textwidth}
			\includegraphics[scale=0.5]{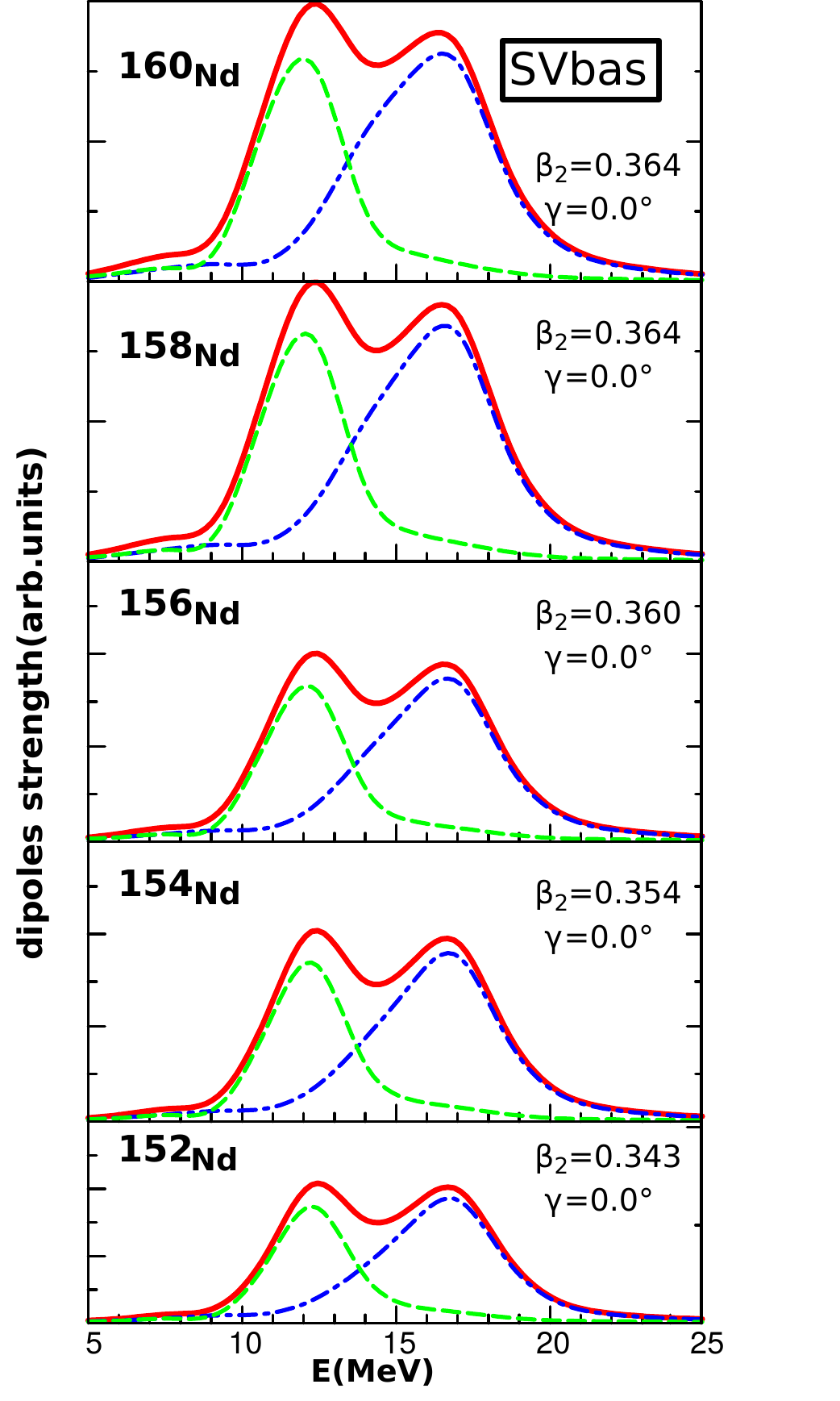}
		\end{minipage}
		\hspace{0.5cm}
		\begin{minipage}[t]{0.46\textwidth}
			\includegraphics[scale=0.5]{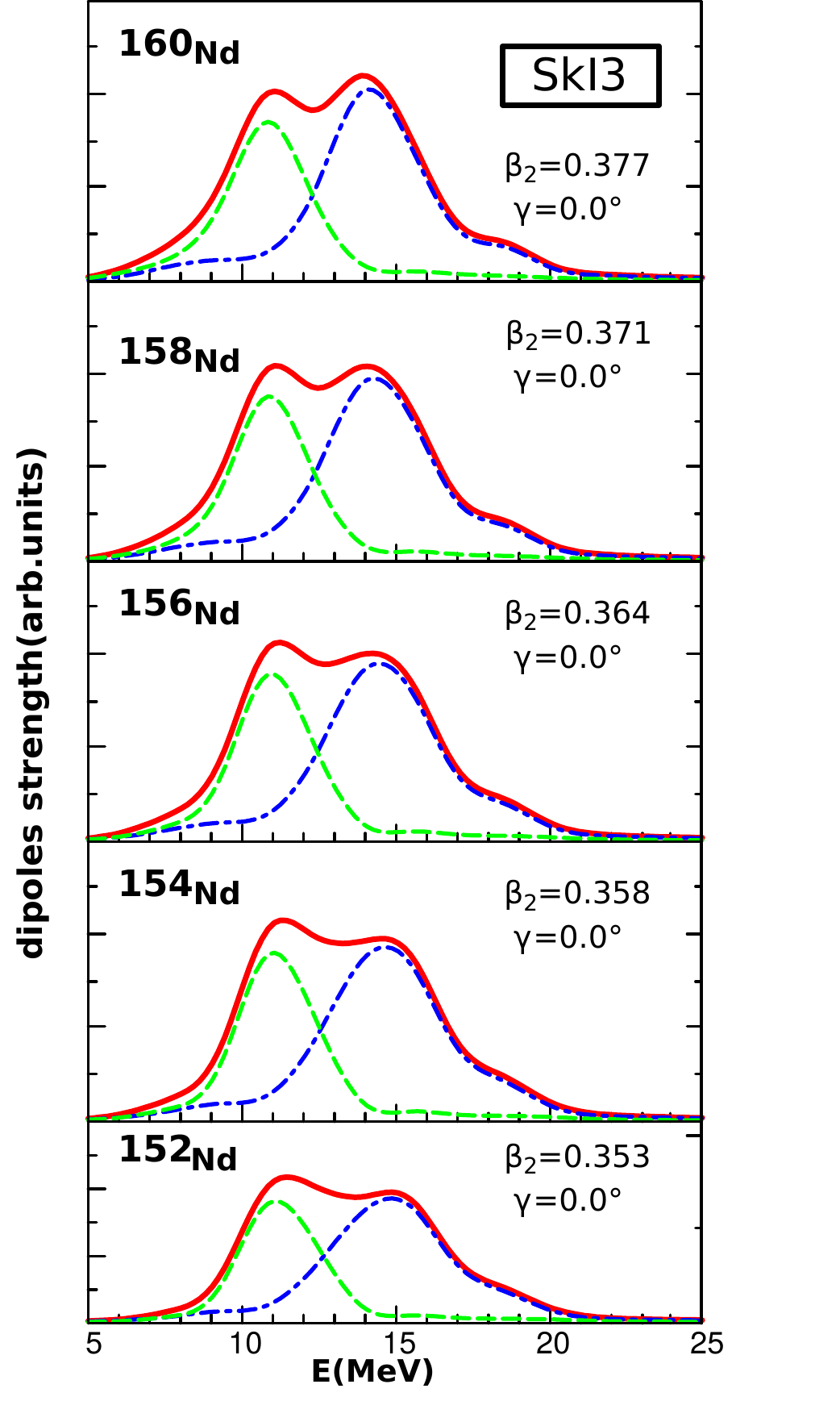}
		\end{minipage}
		\caption{ GDR spectra in the chain of $ ^{152-160}\text{Nd} $ calculated with  SLy5, SLy6, SVbas and SkI3 . The solid(red), dashed(green) and dotted-dashed(magenta) lines denote the dipole strengths: total, along the long axis and the short axis(multiplied by 2) respectively. The calculated strength total is compared with the experimental data \cite{carlos1971} depicted by blue dots.} 
		\label{gr52-60}
	\end{center}	
\end{figure}

\begin{figure}[!htbp]
	\centering
	%\hypertarget{b2-n}{}
	\includegraphics[scale=0.72]{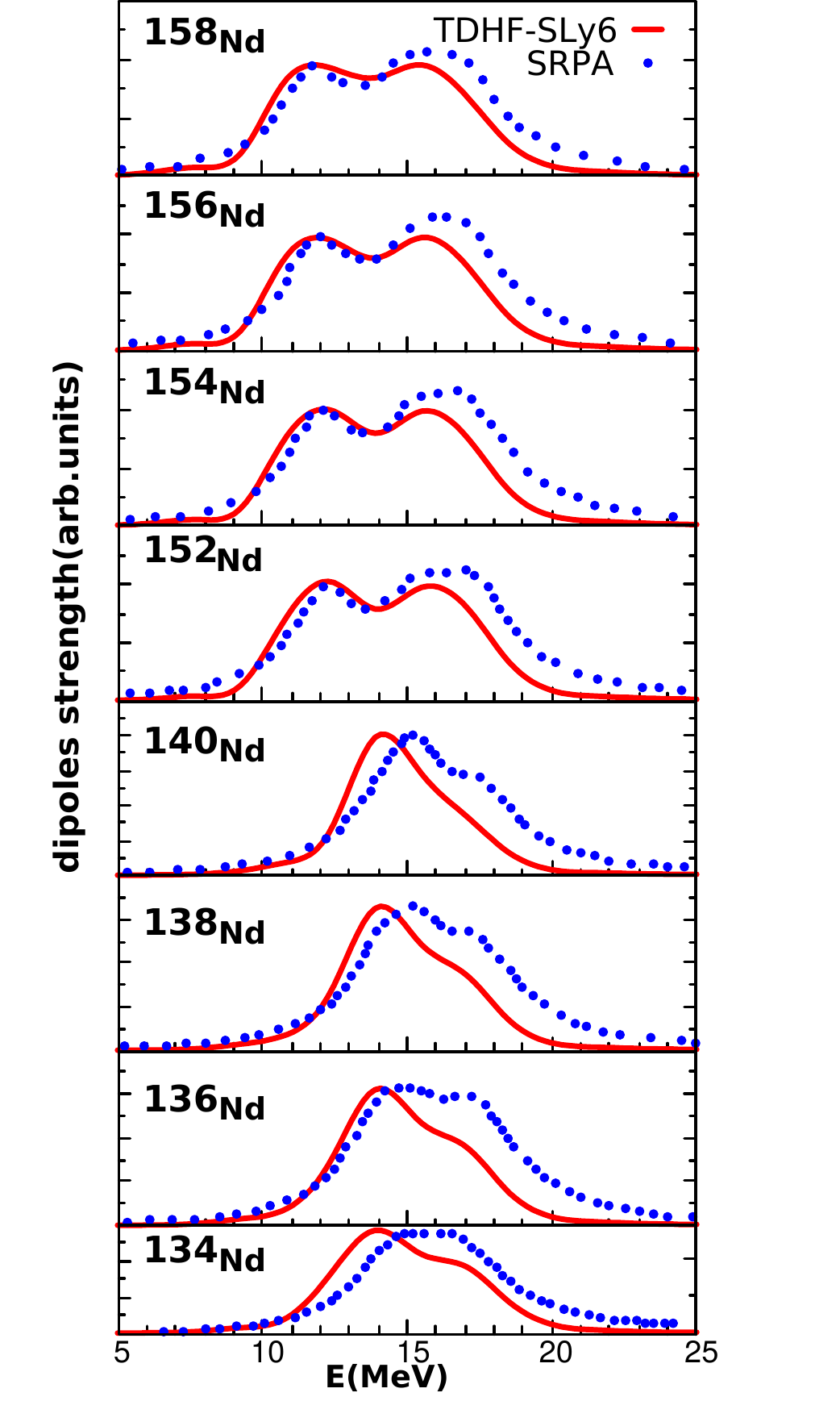} \hfill
	\includegraphics[scale=0.72]{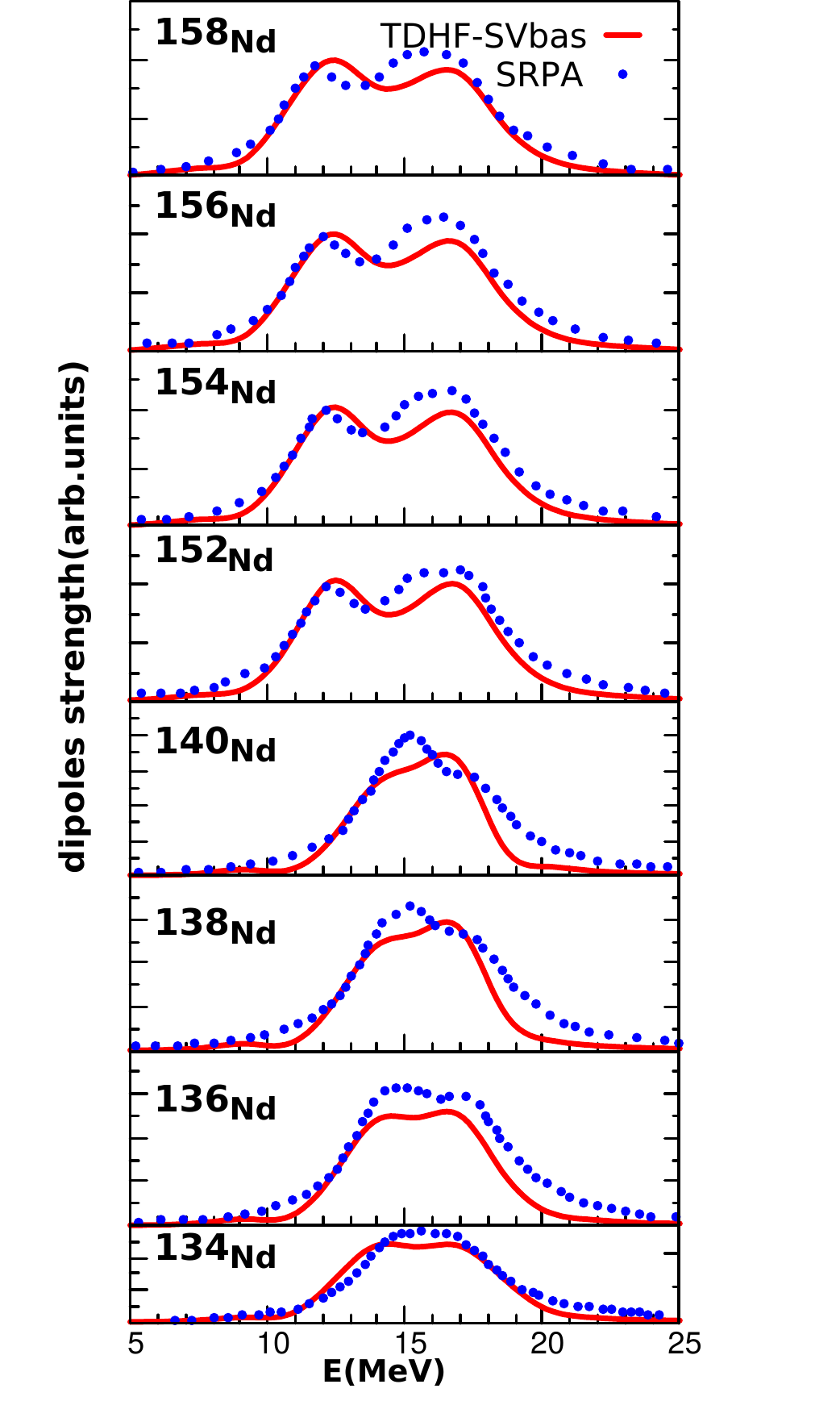}
	\caption{The GDR spectra in $ ^{134-140}\text{Nd} $ and $ ^{152-158}\text{Nd} $ calculated with the Skyrme forces SLy6 and SVbas (red line) are compared with SRPA theory \cite{nesterenko2008} (blue dot).}
	\label{gr34-58}
\end{figure}

\qquad In order to obtain the energy spectrum of the GDR, we calculated the Fourier transform of the isovector signal D(t). The spectral strength  S(E)  (\ref{eq11}) is  simply the imaginary part of the Fourier transform of D(t).\\
Figs.~\ref{gr24-38} -~\ref{gr52-60}  show the GDR spectra  in Nd isotopes computed with the four Skyrme forces, compared with the available experimental data \cite{carlos1971}. It should be noted that the experimental GDR spectra for Nd isotopes from A=124 to A=140  and from A=152 to A=160 are not yet available, so we compared our results concerning theses isotopes with those of SRPA theory~\cite{nesterenko2008} shown in Fig.~\ref{gr34-58}. From Fig.\ref{gr24-38}-~\ref{gr52-60}, we can note  that the deformation decreases gradually when the number of neutrons increases from N=64 ($ ^{124}\text{Nd} $) to number magic N=82 ($ ^{142}\text{Nd} $), and increases again until   N=100 ($ ^{160}\text{Nd} $), which corresponds to a shape transition region of the Nd isotopes  between deformed and spherical shape for N=82. We notice also that the resonance width $\Gamma$ is related to  deformation parameter $\beta_{2}$ as  indicated on the right at the top  of the panel in Figs.~\ref{gr24-38},~\ref{gr40-50},~\ref{gr52-60}.\\

For all isotopes studied in this work except  $ ^{132-138}\text{Nd} $ and $ ^{142}\text{Nd} $, we notice that the oscillations along the major axis correspond to the lower centroid energy E$_{m}^{1}$ indicated by a dashed line in green, and along the minor axis correspond to the high energy E$_{m}^{2}$ indicated by a dotted-dashed line in blue, which shows that these nuclei are more or less deformed according to the values of the deformation parameters ($ \beta_{2} $, $\gamma$). For the nucleus $ ^{142}\text{Nd} $ which has a magic number N=82, the oscillations along the three axes correspond to the same resonance energy E$ _{m} $, which means that $ ^{142}\text{Nd} $ has a spherical shape.\\
For the isotopes $ ^{132}\text{Nd} $, $^{134}\text{Nd} $, $^{136}\text{Nd}$ and $ ^{138}\text{Nd} $ the oscillations along the three axes correspond to  different energies  $ E_{x}\neq E_{y}\neq E_{z}$, so they have a degree of triaxiality. We point out that $ ^{138}\text{Nd} $ has prolate shape in case of Skyrme force SVbas where $\gamma=0.1$, and is triaxial in case of the others Skyrme force (SLy6, SLy5, SkI3) where $\gamma\simeq 25$. Also for $ ^{140}\text{Nd} $ shown in Fig.~\ref{gr40-50}, SVbas gives a spherical shape ($\beta_{2}$=0), while the others Skyrme forces predict weakly deformed ($\beta_{2}$$\simeq$0.081). The region with masse $\simeq$ 140 is known by a variety of coexisting shape.

Fig.~\ref{gr40-50} compares the results of the GDR spectra for different Skyrme forces with the available experimental data\cite{carlos1971}. It is seen that all four Skyrme forces provide in general acceptable agreement with the experimental data with  a slight down-shift of the order of 0.5 MeV  for SLy5, SLy6 and SkI3 in the case of the weakly deformed isotopes  $^{142-148}\text{Nd} $, and slight up-shift ($\sim$ 0.5 MeV) for SVbas force. For spherical $ ^{142}\text{Nd}  $ and weakly deformed $ ^{144}\text{Nd} $ and $ ^{146}\text{Nd} $, the GDR spectrum is fragmented where appears a small  shoulder due to the landau fragmentation. This fragmentation depends on the choice of the Skyrme force. For example, SVbas gives a weak fragmentation among the four forces of Skyrme. For $ ^{148}\text{Nd} $, the small shoulder disappears in GDR spectrum only SkI3. The agreement is better for deformed isotope $ ^{150}\text{Nd} $, where all Skyrme force produce the deformation splitting. In general, the force SVbas shows a further performance in respect to the other forces. We can see also, that  Sly6 and Sly5  have good results with a slight advantage for Sly6.\\

In a previous work \cite{nesterenko2008}, such a problem has been studied with other different forces of ours, including SLy6. So, in Fig.~\ref{gr34-58}, we compare  our obtained results with SLy6 with those of that work \cite{nesterenko2008}. From this figure, we can see an overall agreement between both calculations particularly for more deformed nuclei. We explain this mismatch of about 1 MeV: the results of two theoretical calculations (TDHF,SRPA) depend on their numerical realisations (size of configuration space, completeness of residual interaction, ...). SRPA was done in a limited space of 1ph configuration while TDHF includes
effectively a much larger computational space. Larger is better and thus the TDHF results is more reliable.
 In Fig.~\ref{gr34-58}, we compare also SRPA with SVBas results. Their results are very close to each other because SVbas produce a slightly higher peak position.\\

\begin{figure}[!htbp]
	\centering
	%\hypertarget{b2-n}{}
	\includegraphics[scale=0.7]{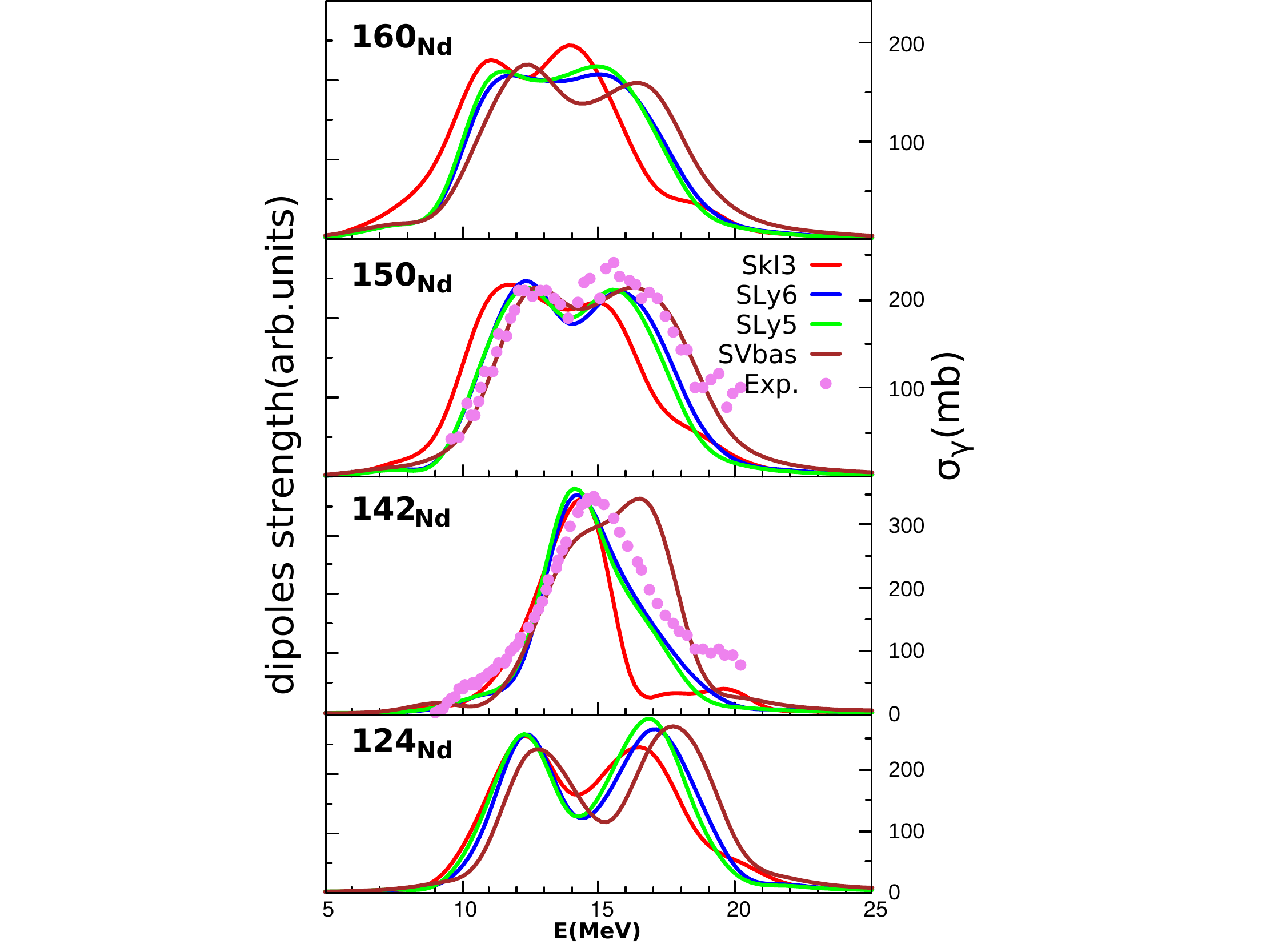}
	\caption{Comparison of calculated GDR spectra with Skyrme forces SkI3,  SLy6, SLy5 and SVbas in case $^{124}\text{Nd} $, $ ^{142}\text{Nd} $, $ ^{150}\text{Nd} $ and $ ^{160}\text{Nd} $. the experimental data \cite{carlos1971}are  depicted by purple dots.}
	\label{gdr-com}
\end{figure}

Fig.~\ref{gdr-com} shows a comparison of the strength GDR for the isotopes $^{124}\text{Nd} $, $ ^{142}\text{Nd} $, $ ^{150}\text{Nd} $ and $ ^{160}\text{Nd} $ with the four Skyrme forces. We can see a small shift of the average peak position between these forces around $\pm$ 1 MeV. That  means there is a dependence of the GDR spectra on Skyrme  forces~\cite{reinhard2007c}. This dependence is related to some basic features  and nuclear properties of the Skyrme functional as listed in Table \ref{tab4} .
High sum rule enhancement factor  $\kappa = (m_{1}^*/m)^{-1} -1$  (i.e low isovector effective mass $ m_{1}^*/m $) leads to a shift of the GDR strength towards the higher energy region as indicated in Ref.~\cite{reinhard2009} in the case of GDR in $ ^{208}\text{Pb} $ and in Ref. \cite{oishi2016} in the case of GDR in $ ^{174}\text{Yb} $. For example, the  large collective shift in SVbas can be related to a very high enhancement factor $\kappa$=0.4 compared to other Skyrme forces. For the deformed nucleus $ ^{150}\text{Nd} $, we see that all these forces reproduce well the experimental data  except for SkI3. The giant dipole resonance for SkI3 is lower than  other forces because it has larger symmetry energy ($\sim$ 35MeV) while 30-32 MeV is the standard value. 
%%%%%%%%%%%%%%%%%%%%%%%%% gdr124-138 %%%%%%%%%%%%%%%%%%%%%%
%%%%%%%%%%%%%%%%%%%%%%%%%%%%%%%%%%%%%%%%%%%%%%%%%%%%%%%%%%%%%%%%%%%%%
%%%%%%%%%%%%%%%%%%%%%%% gr140-150%%%%%%%%%%%%%%%%%%%%%%%%%%%%%%%%%%
%%%%%%%%%%%%%%%%%%%%%%%%%%%%%%%%%%%%%%%%%%%%%%%%%%%%%%%%%%%%%%%%%%%%
%%%%%%%%%%%%%%%%%%%%%%%%% gr152-160%%%%%%%%%%%%%%%%%%%%%%%%%%%%%%%%%
%%%%%%%%%%%%%%%%%%%%%%%%%%%%%%%%%%%%%%%%%%%%%%%%%
%%%%%%%%%%%%%%%%%%%%%%%%%%%%%%%%%%%%%%%%%%%%%%%%%%%%%%%%
%%%%%%%%%%%%%%%%%%%%%%%%%%%%%%%%%%%%%%%%%%%%%%%%%%%%%%%%%%%%%%%%%%%%%%%%%%%%%%%%%%%%
%%%%%%%%%%%%%%%%%%%%%%%%%%%%%% isovector mass for forces %%%%%%%%%%%%%%%%%%%%%%%%%
\begin{table}[!htbp]
	\centering
	\caption {The isovector effective mass $m_{1}^*/m$, sum rule enhancement  factor $\kappa = (m_{1}^*/m)^{-1} -1$, and symmetry energy $ a_{sym} $ for the Skyrme forces in this work  \cite{reinhard2009},\cite{CHABANAT1998}.\label{tab4}} 
	{\begin{tabular}{@{}ccccccccc@{}} \hline
			Forces && $m_{1}^*/m$ &&& $\kappa$&&& $ a_{sym}(MeV) $  \\
			\hline
			SLy6  && 0.80&&&  0.25&&& 31.96\\
			SLy5   && 0.80&&&  0.25&&&32.03  \\
			SkI3  && 0.80 &&& 0.246&&&34.80  \\
			SVbas  && 0.715 &&& 0.4&&&30.00  \\
			\hline
	\end{tabular}}
	%\label{table:2}
\end{table}
%%%%%%%%%%%%%%%%%%%%%%%%%%%%%%%%%%%%%%%%%%%%%%%%%%%%%%%%%%%%%%%%%%%%%%%%%%%%%%%%%

\begin{figure}[!htb]
	\centering
	\begin{tabular}{@{}cccc@{}}
		\includegraphics[width=.25\textwidth]{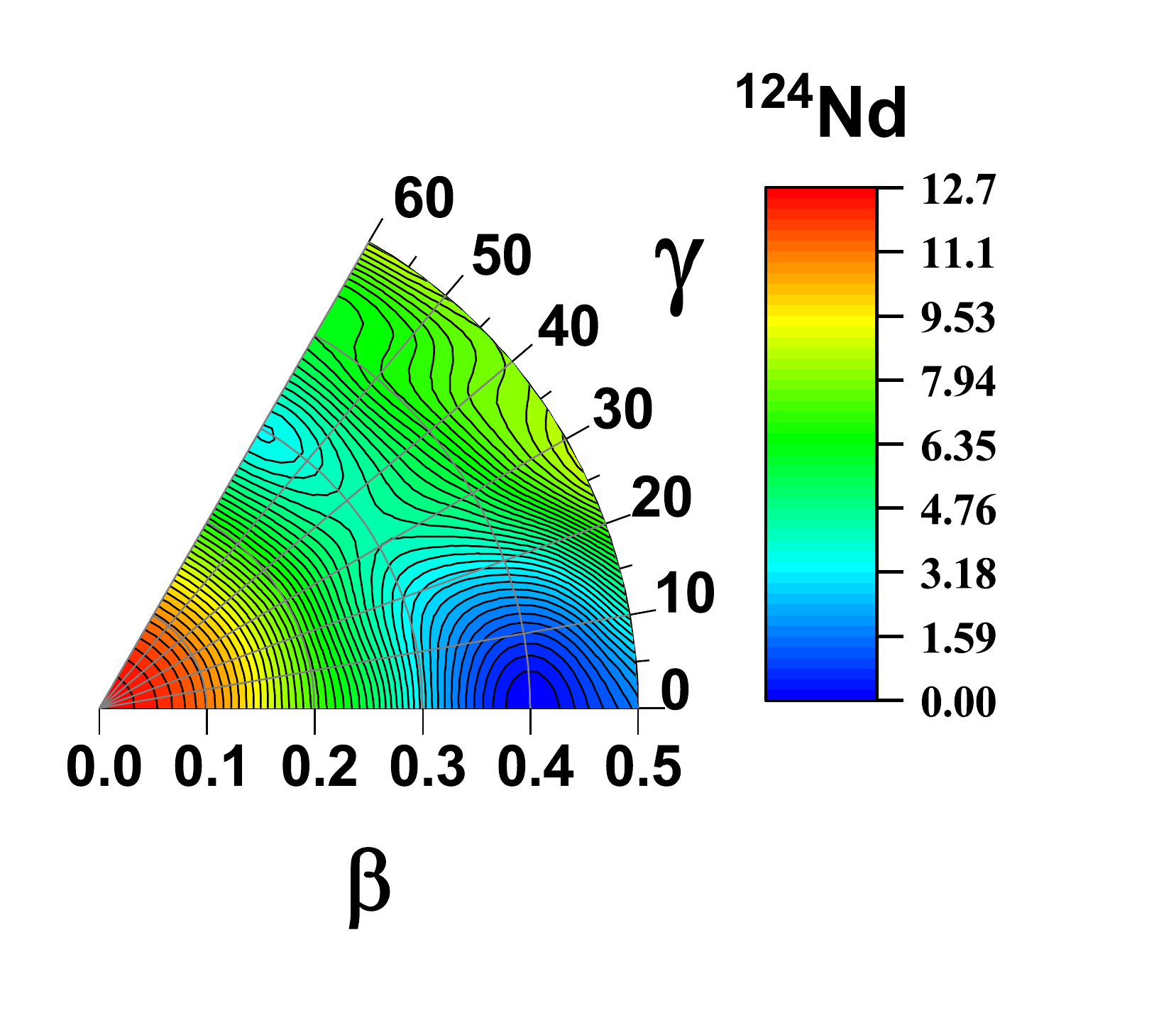} &
		\includegraphics[width=.25\textwidth]{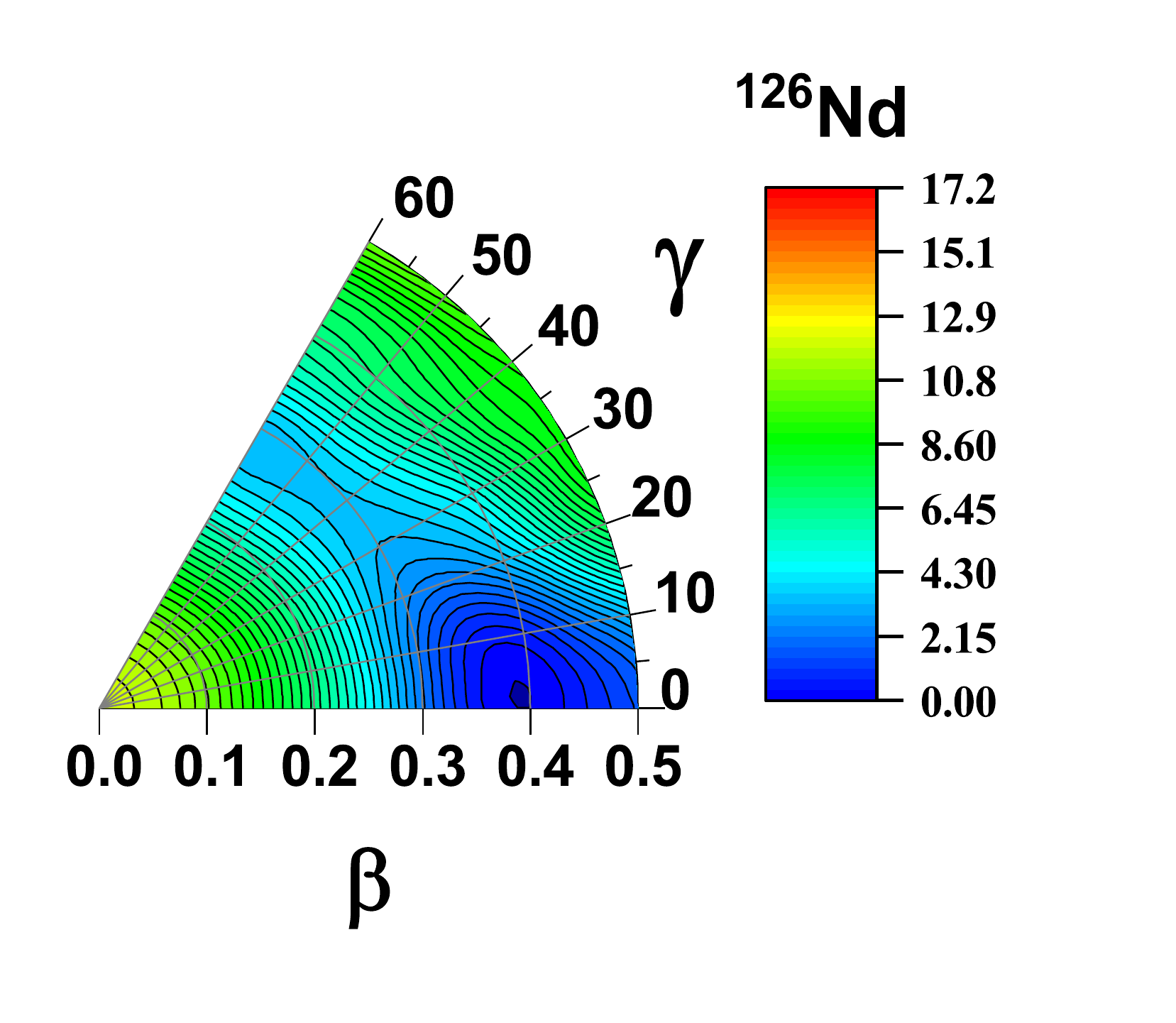} &
		\includegraphics[width=.25\textwidth]{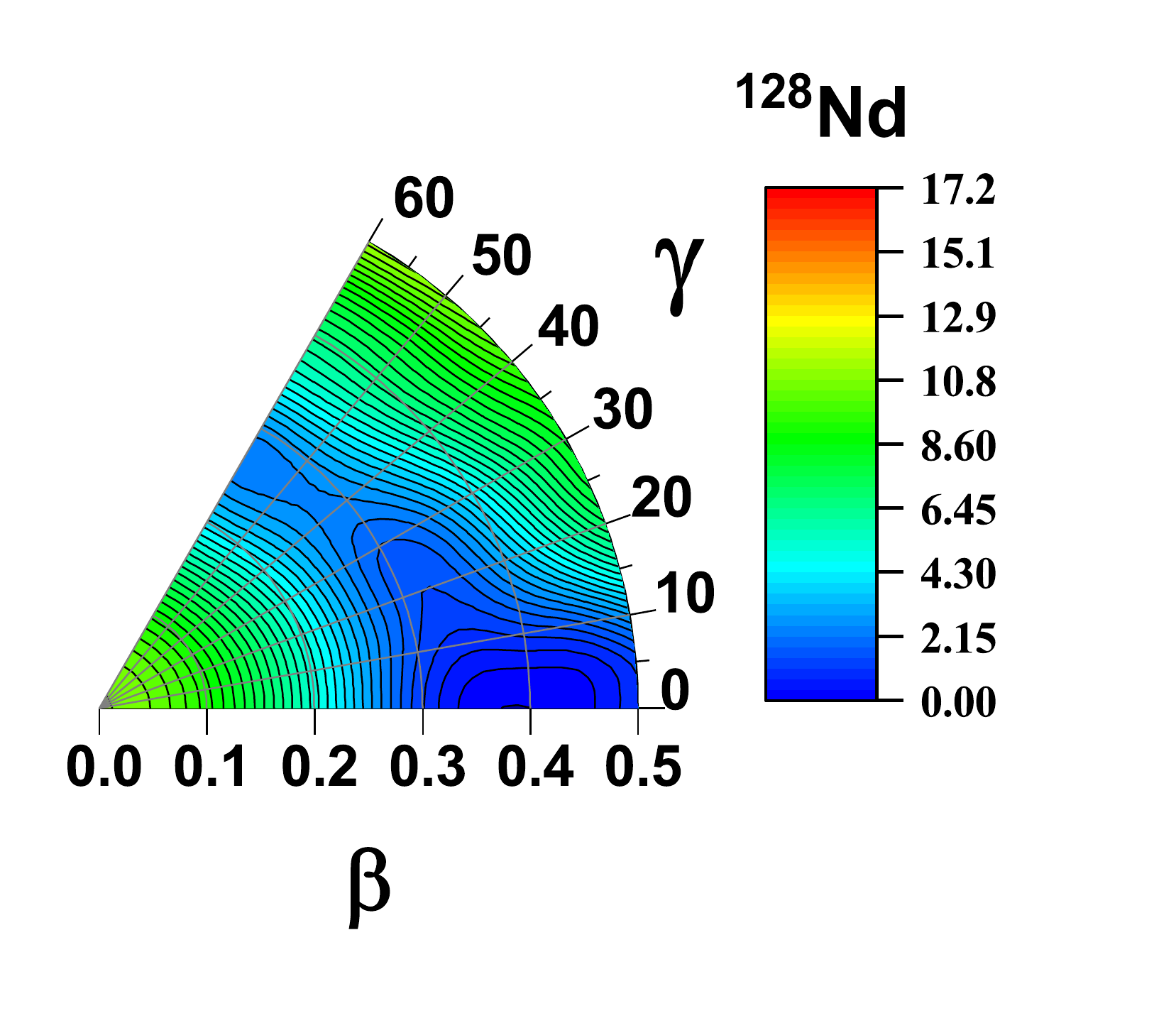} &
		\includegraphics[width=.25\textwidth]{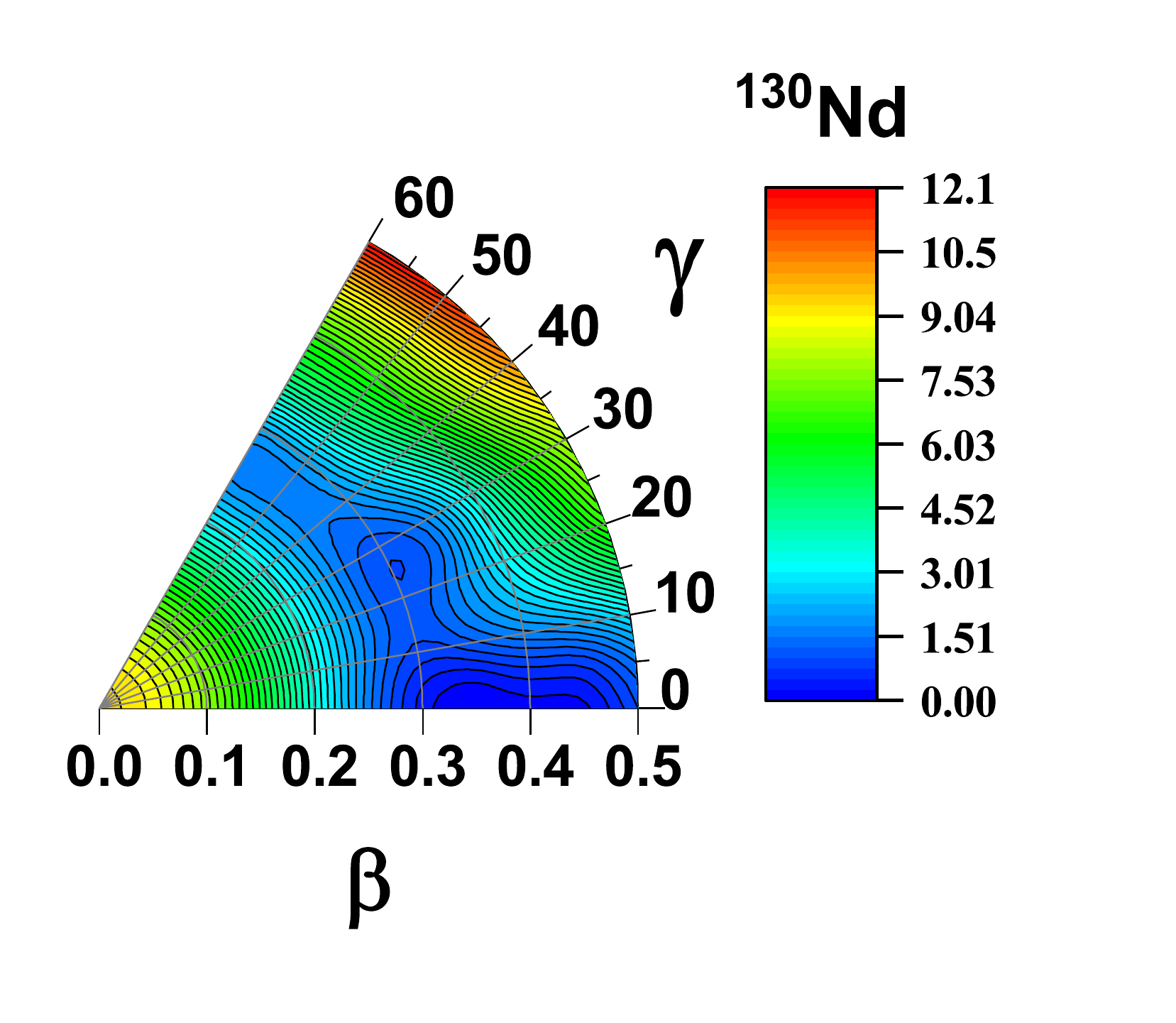}   \\
		\includegraphics[width=.25\textwidth]{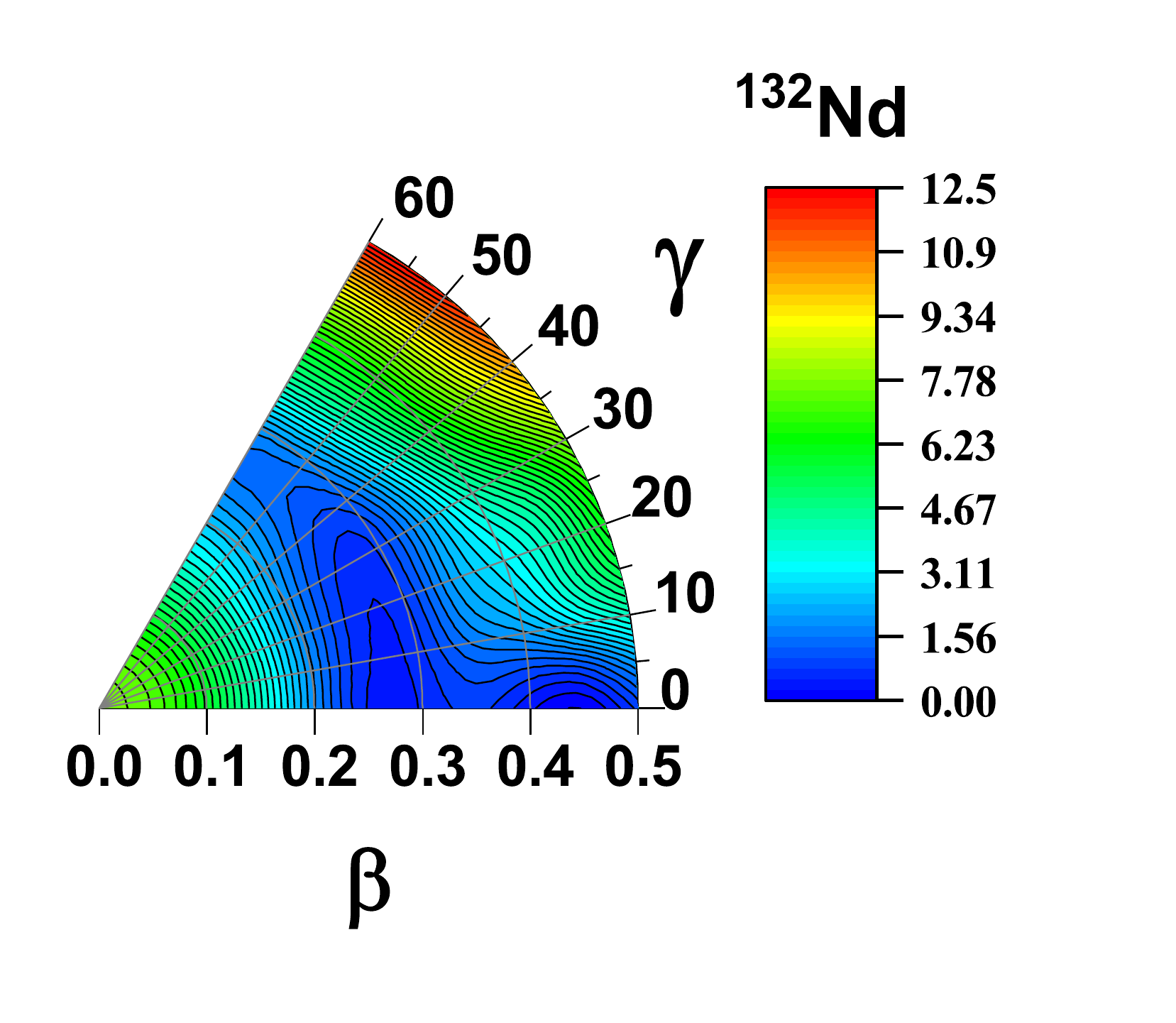} &
		\includegraphics[width=.25\textwidth]{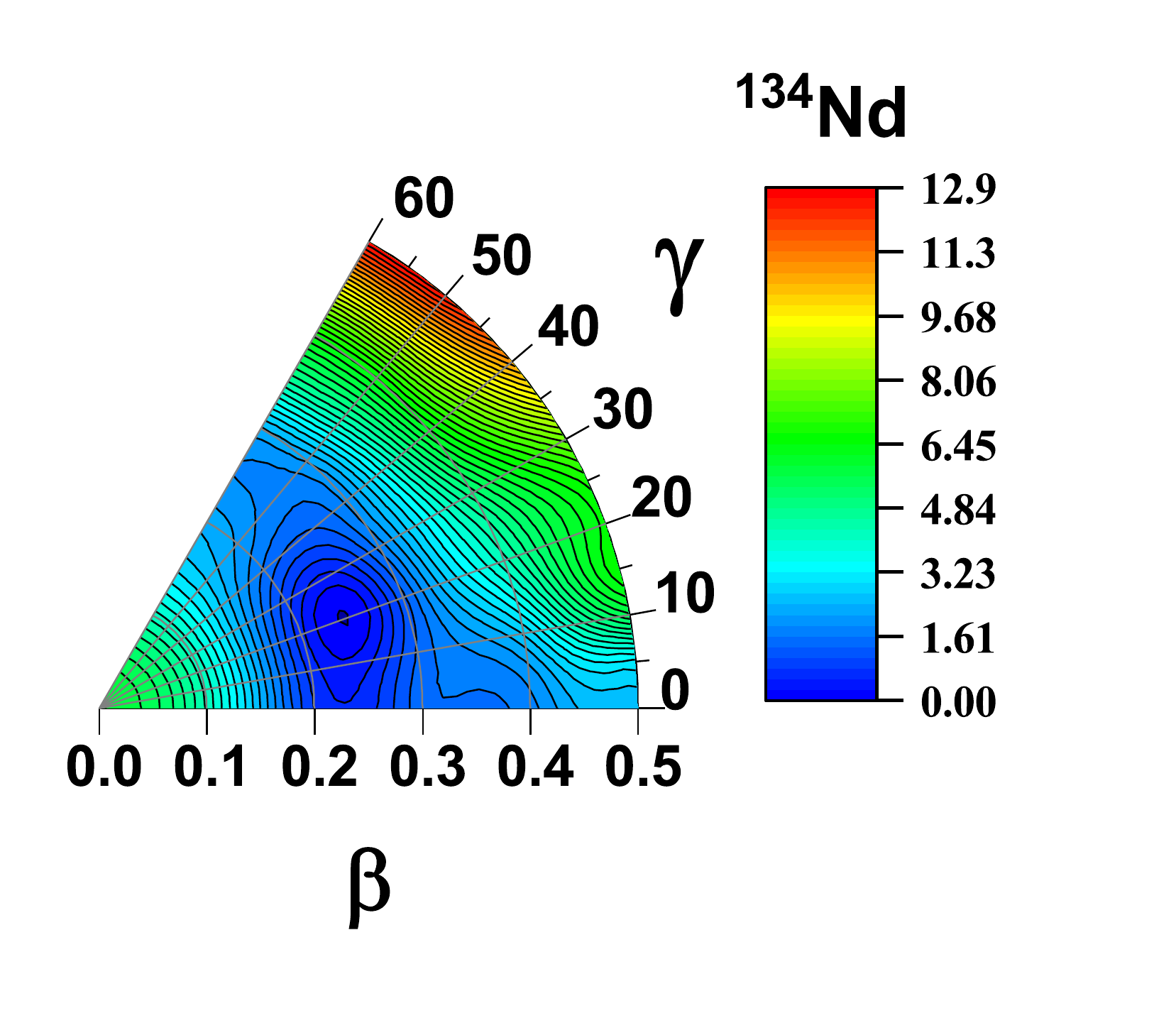} &
		\includegraphics[width=.25\textwidth]{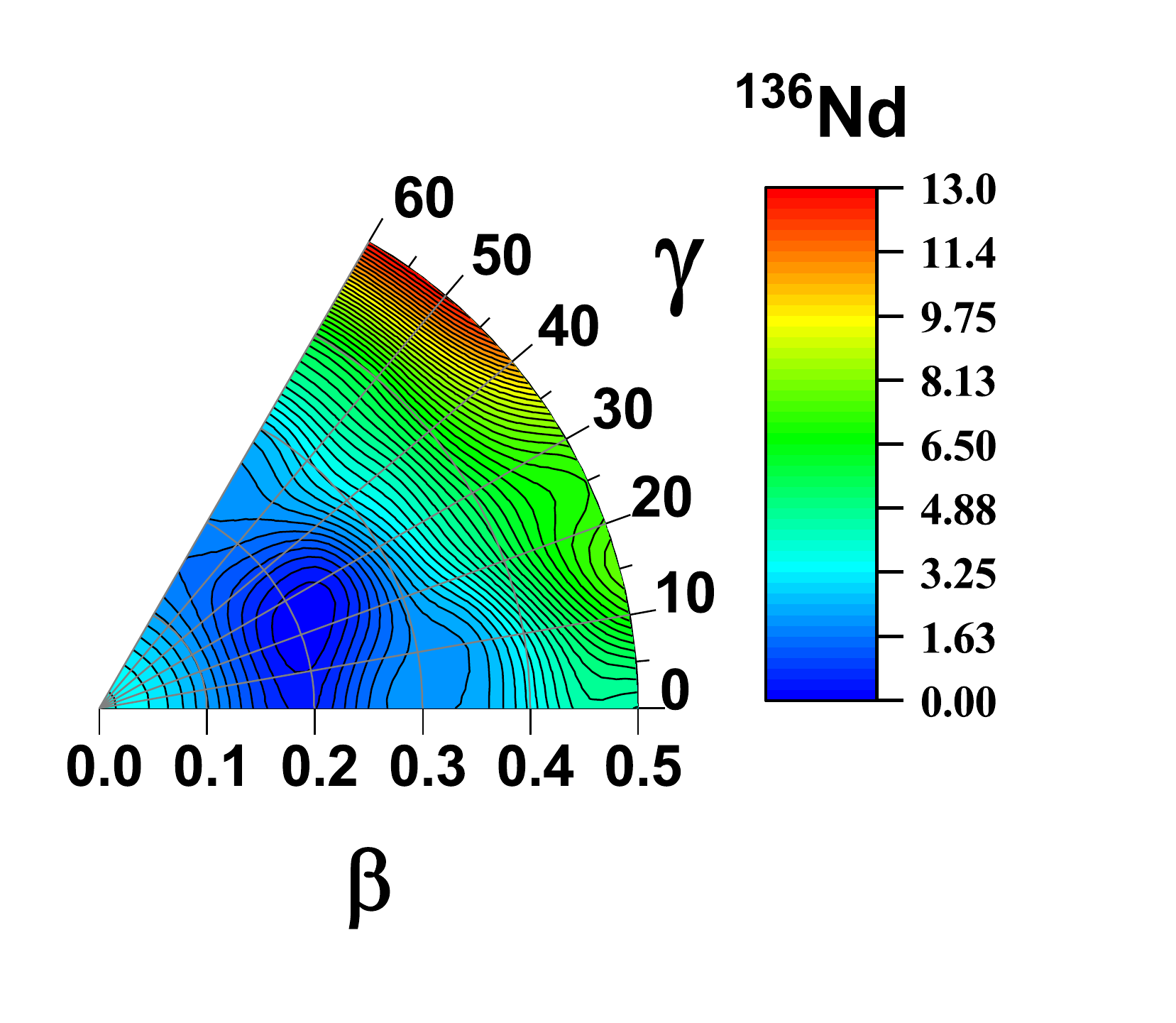} &
		\includegraphics[width=.25\textwidth]{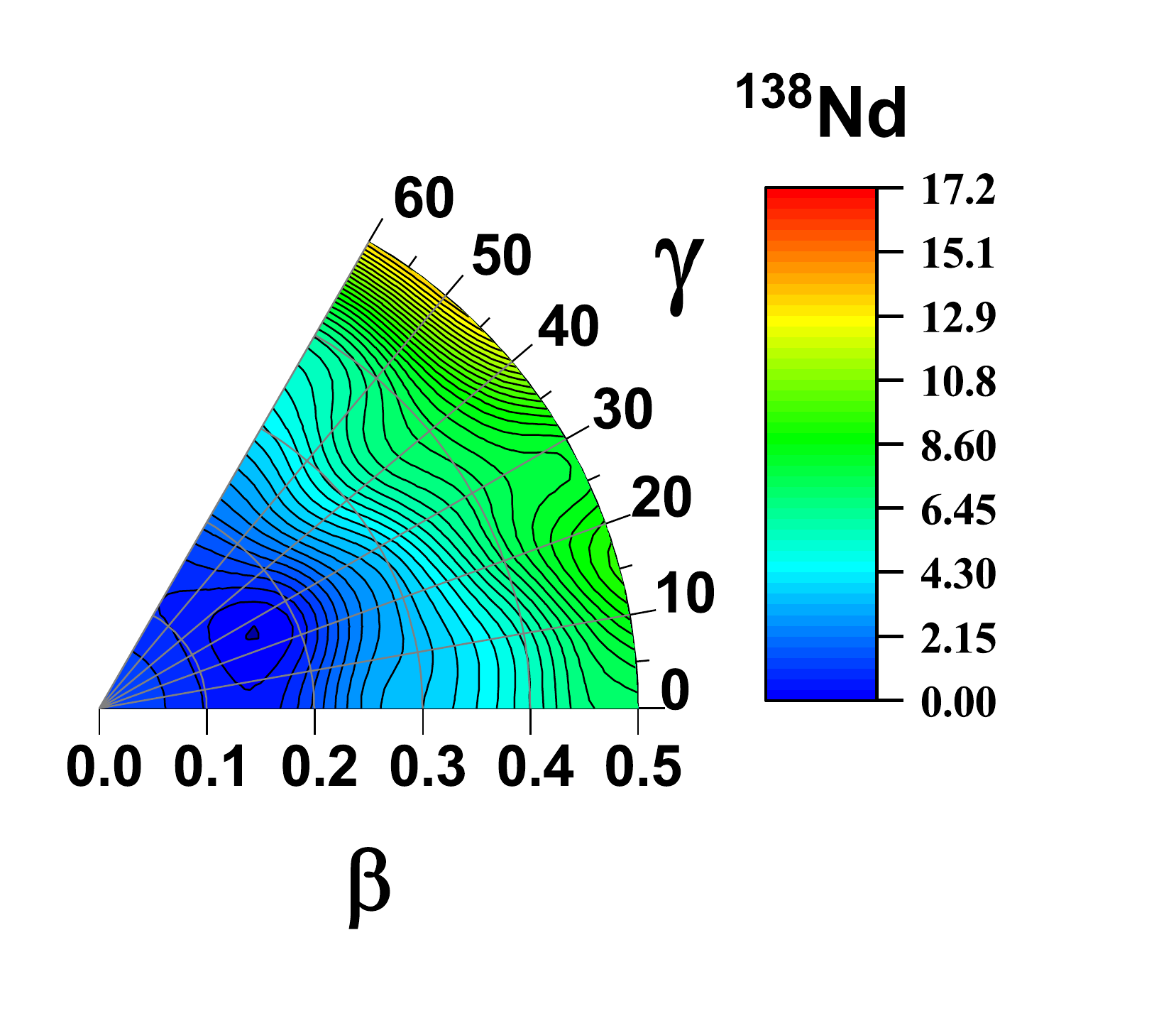}   \\
		\includegraphics[width=.25\textwidth]{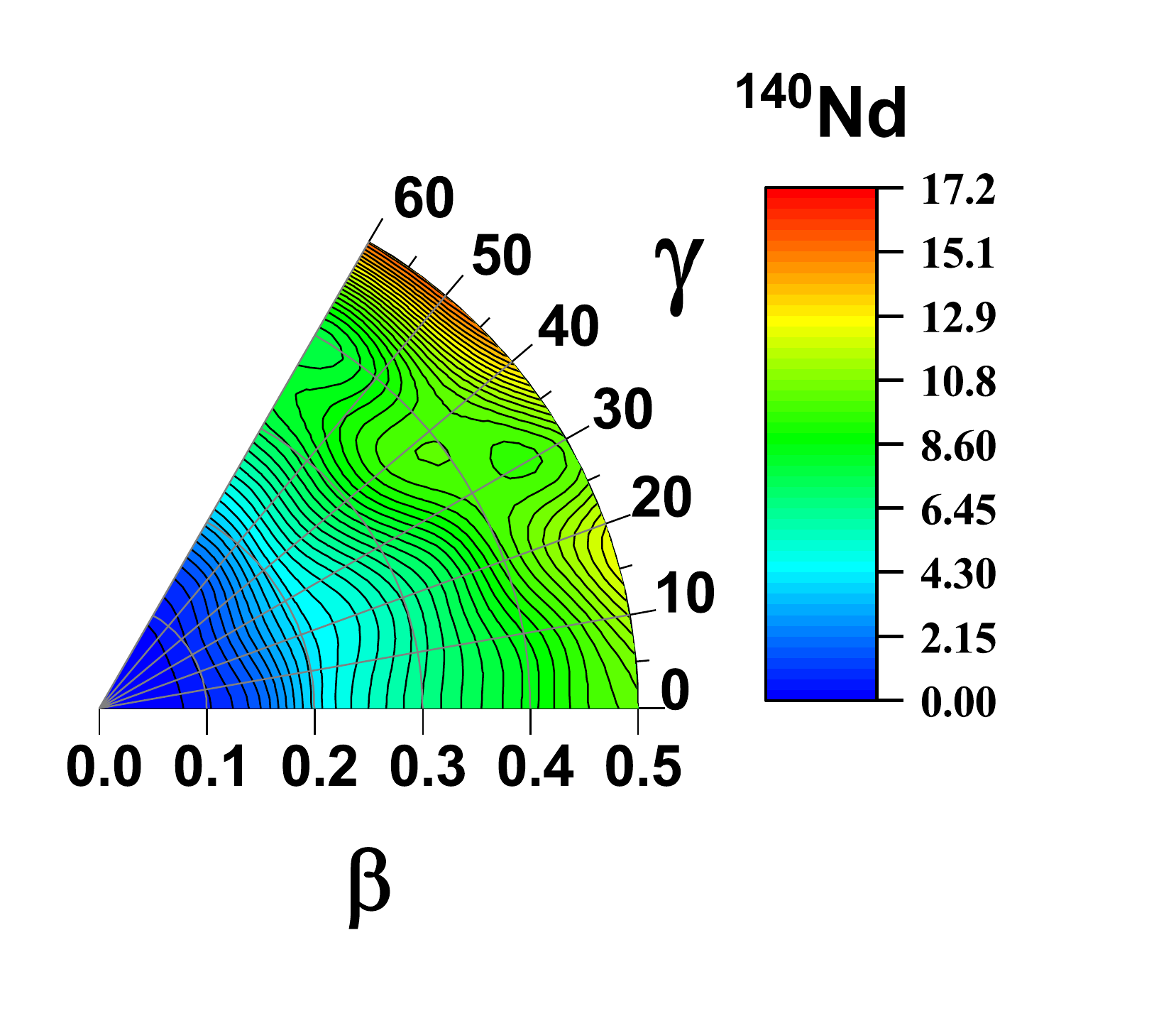} &
		\includegraphics[width=.25\textwidth]{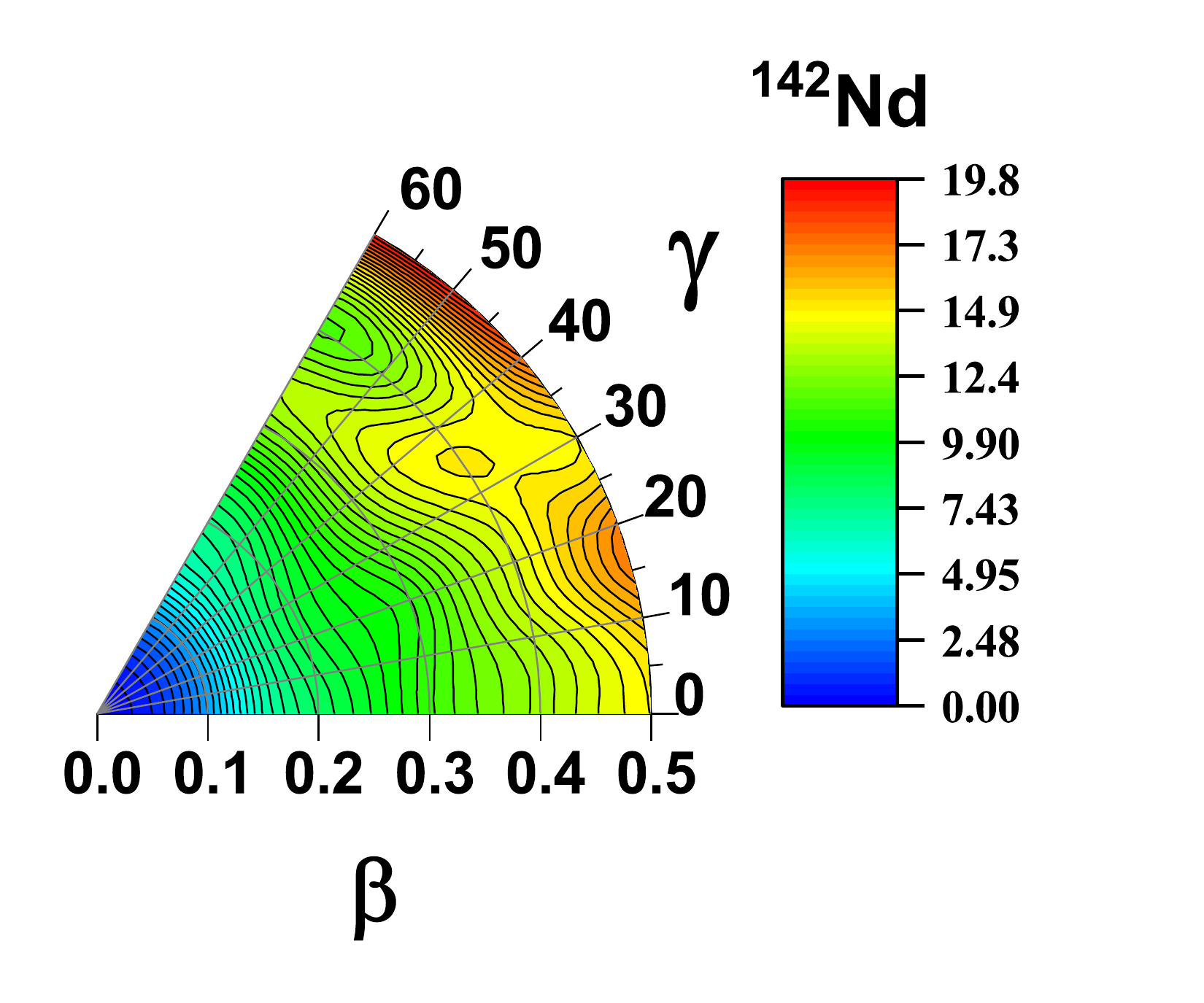} &
		\includegraphics[width=.25\textwidth]{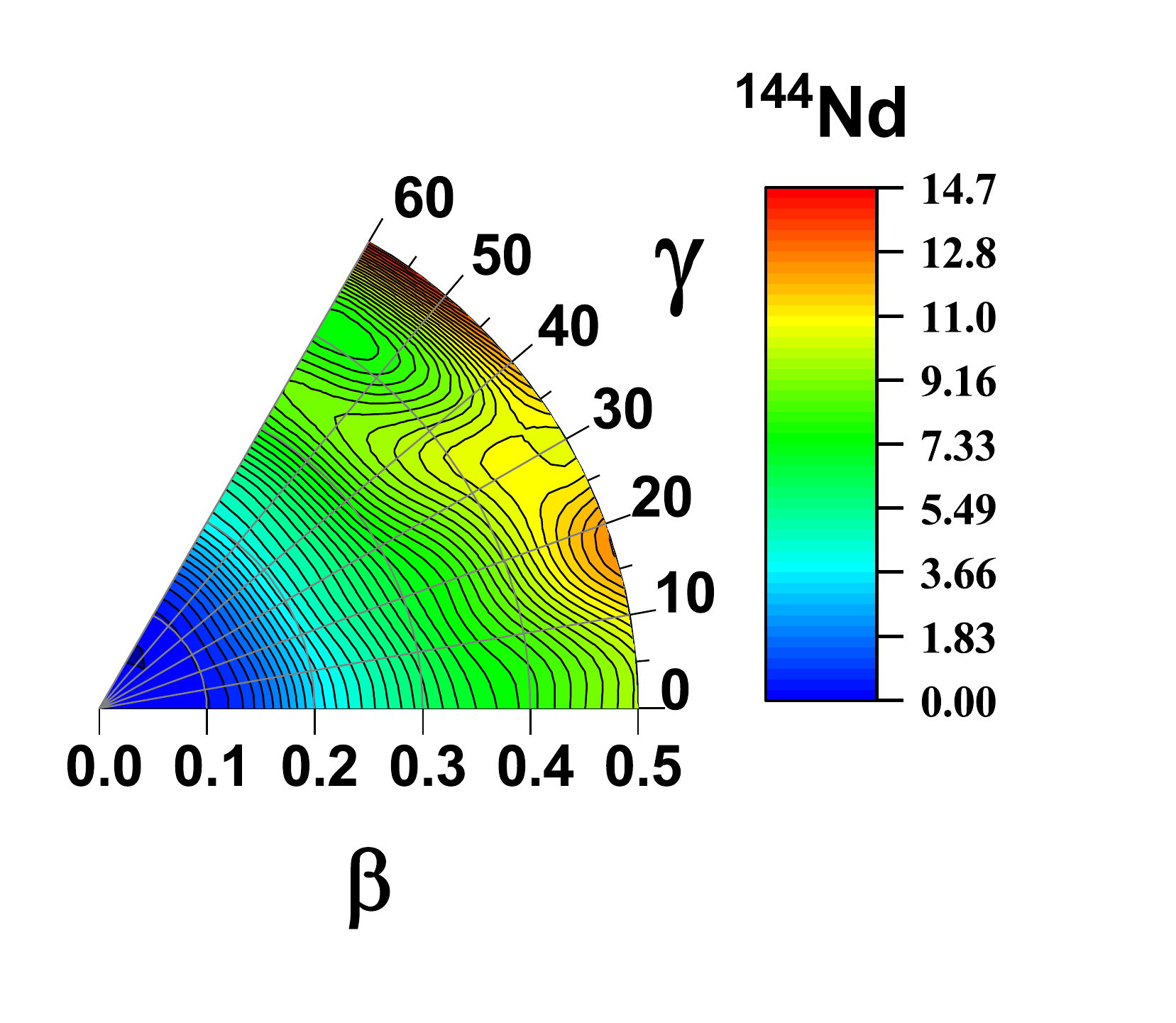} &
		\includegraphics[width=.25\textwidth]{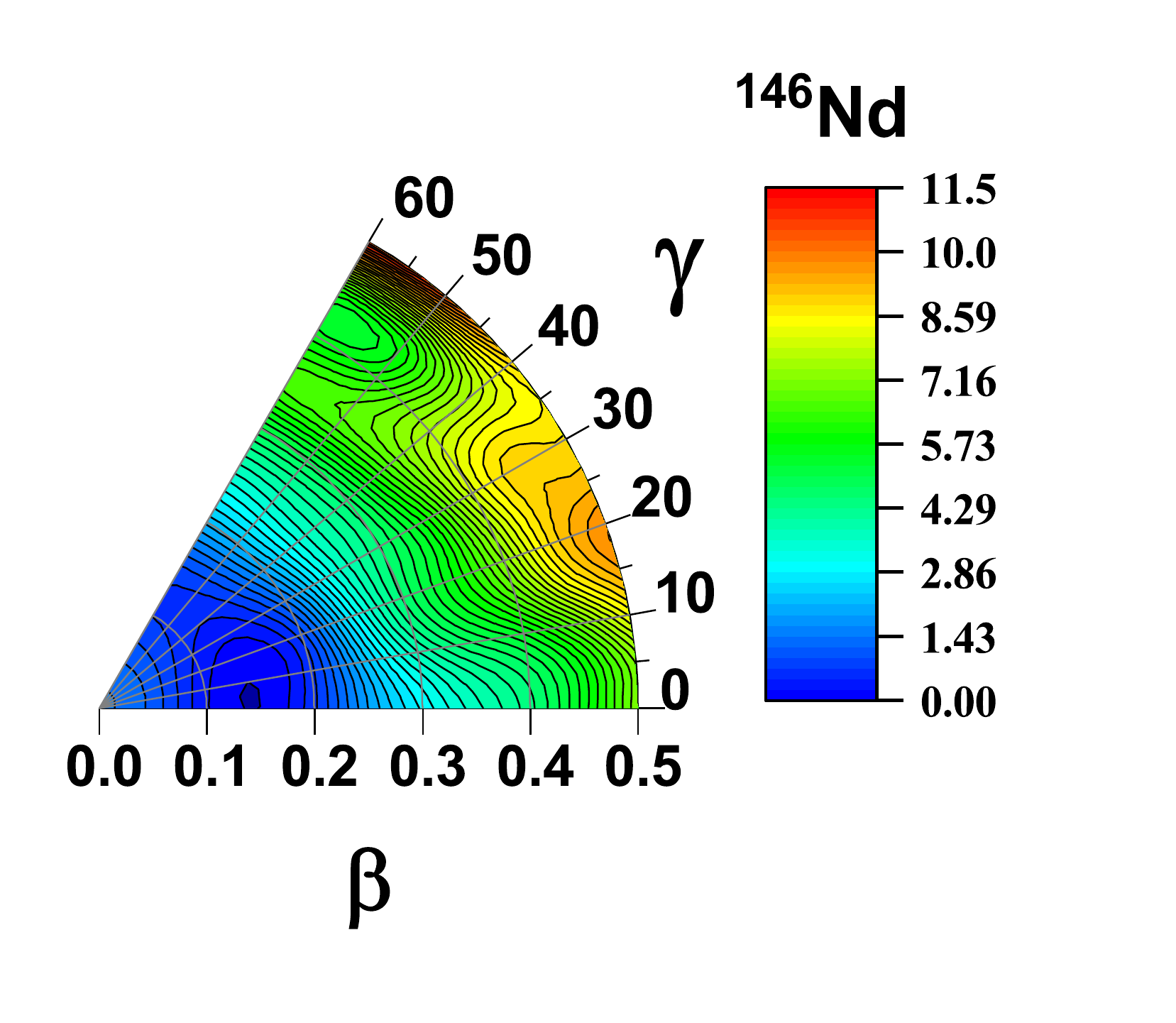}   \\
		\includegraphics[width=.25\textwidth]{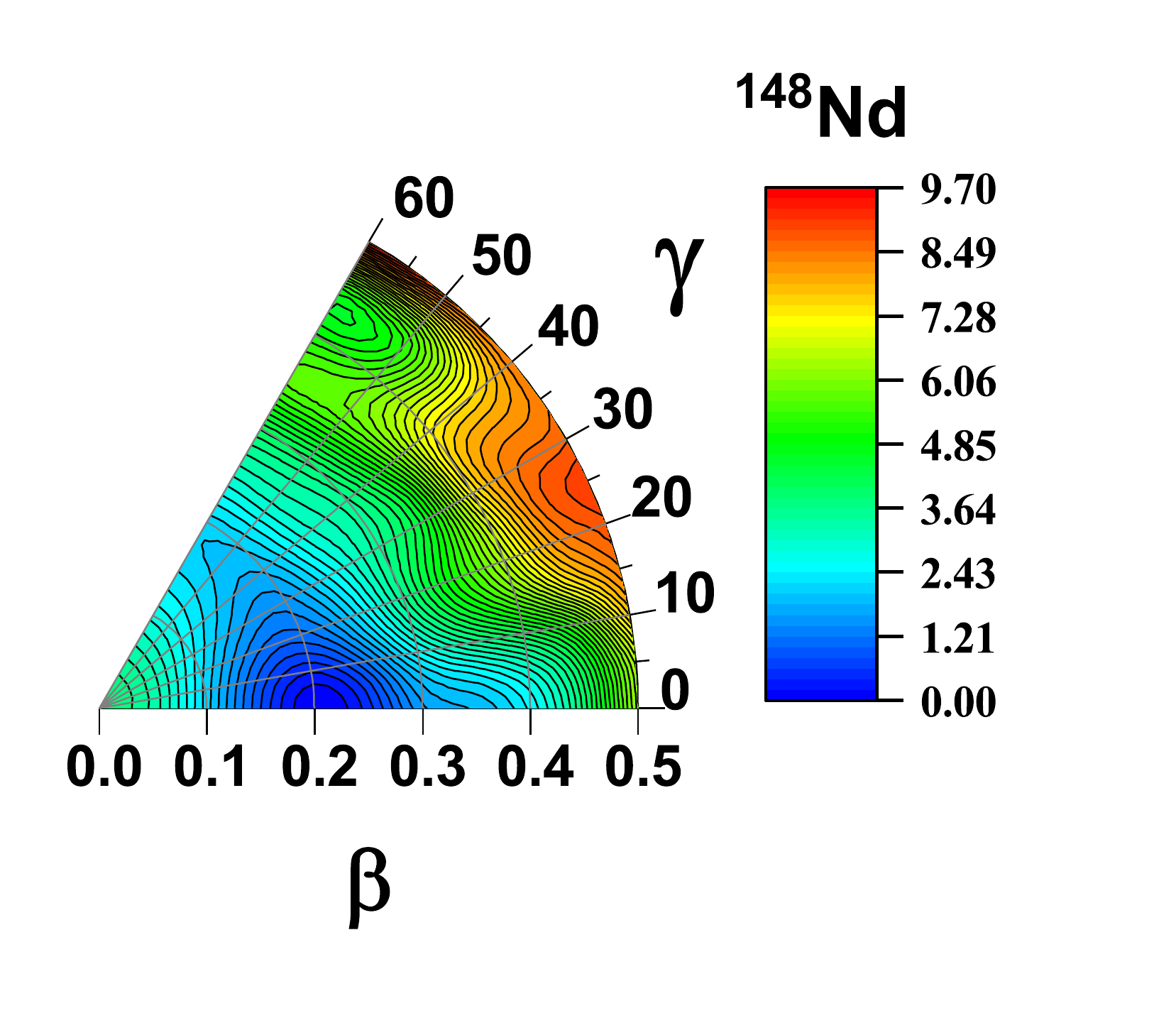} &
		\includegraphics[width=.25\textwidth]{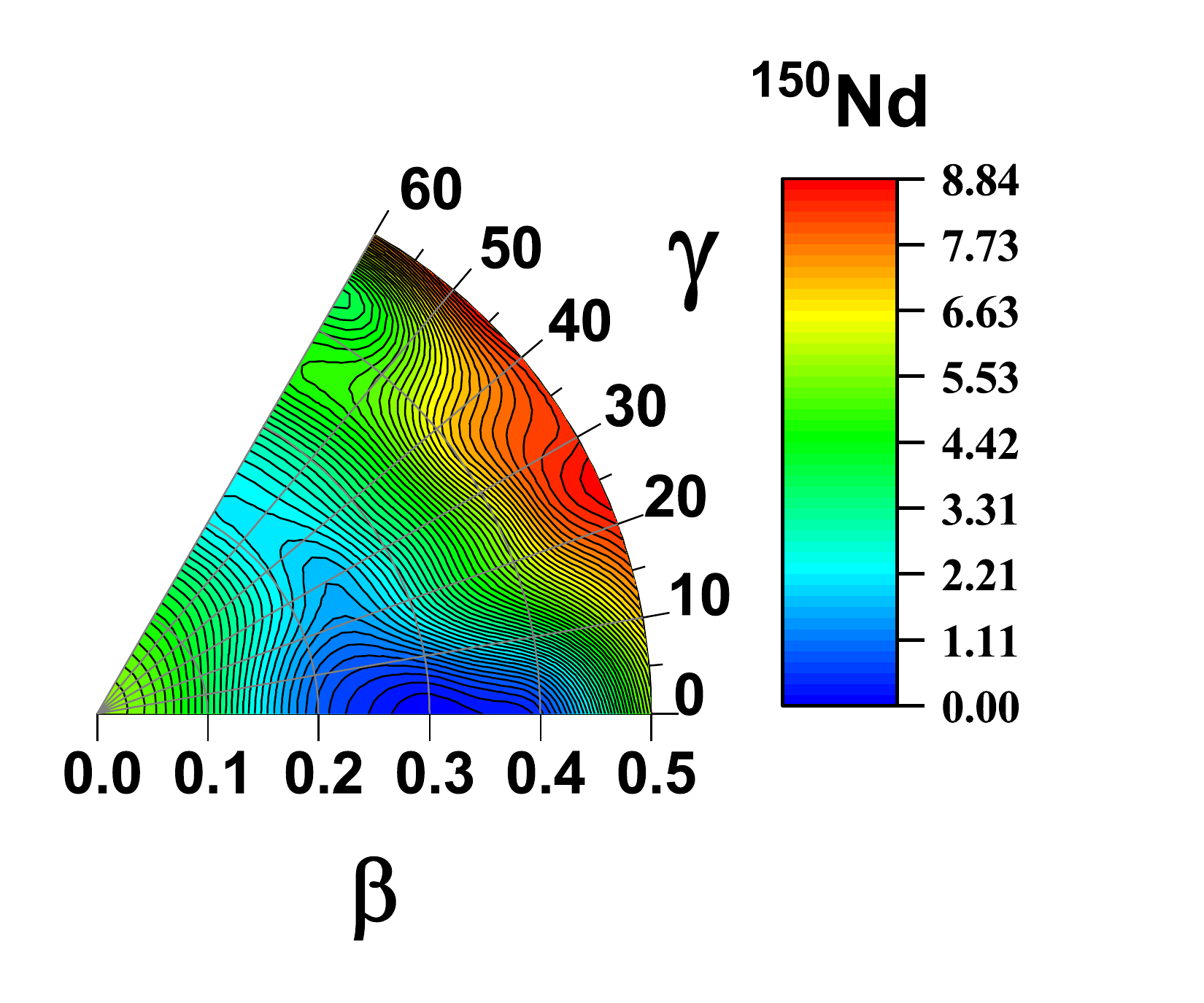} &
		\includegraphics[width=.25\textwidth]{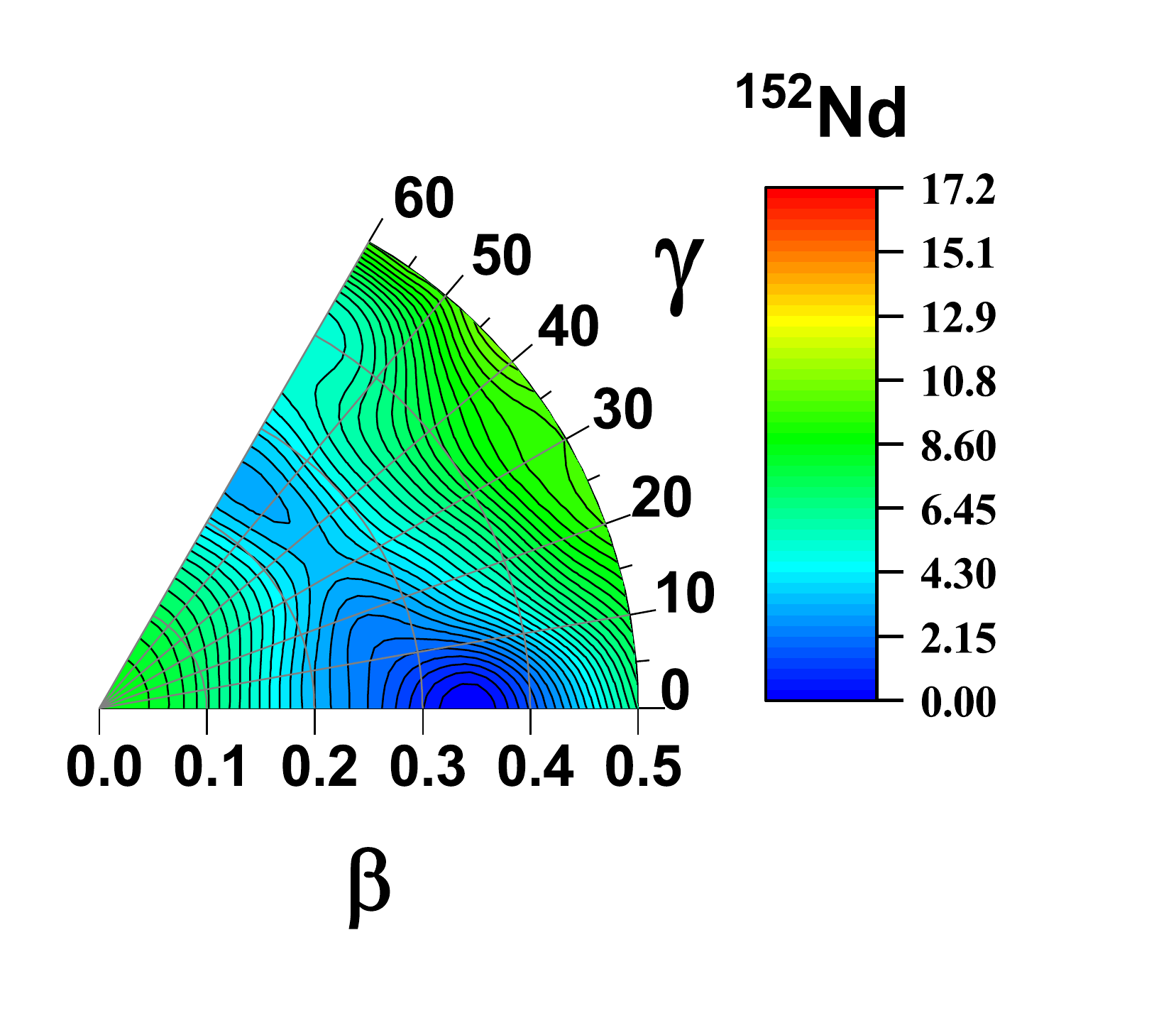} &
		\includegraphics[width=.25\textwidth]{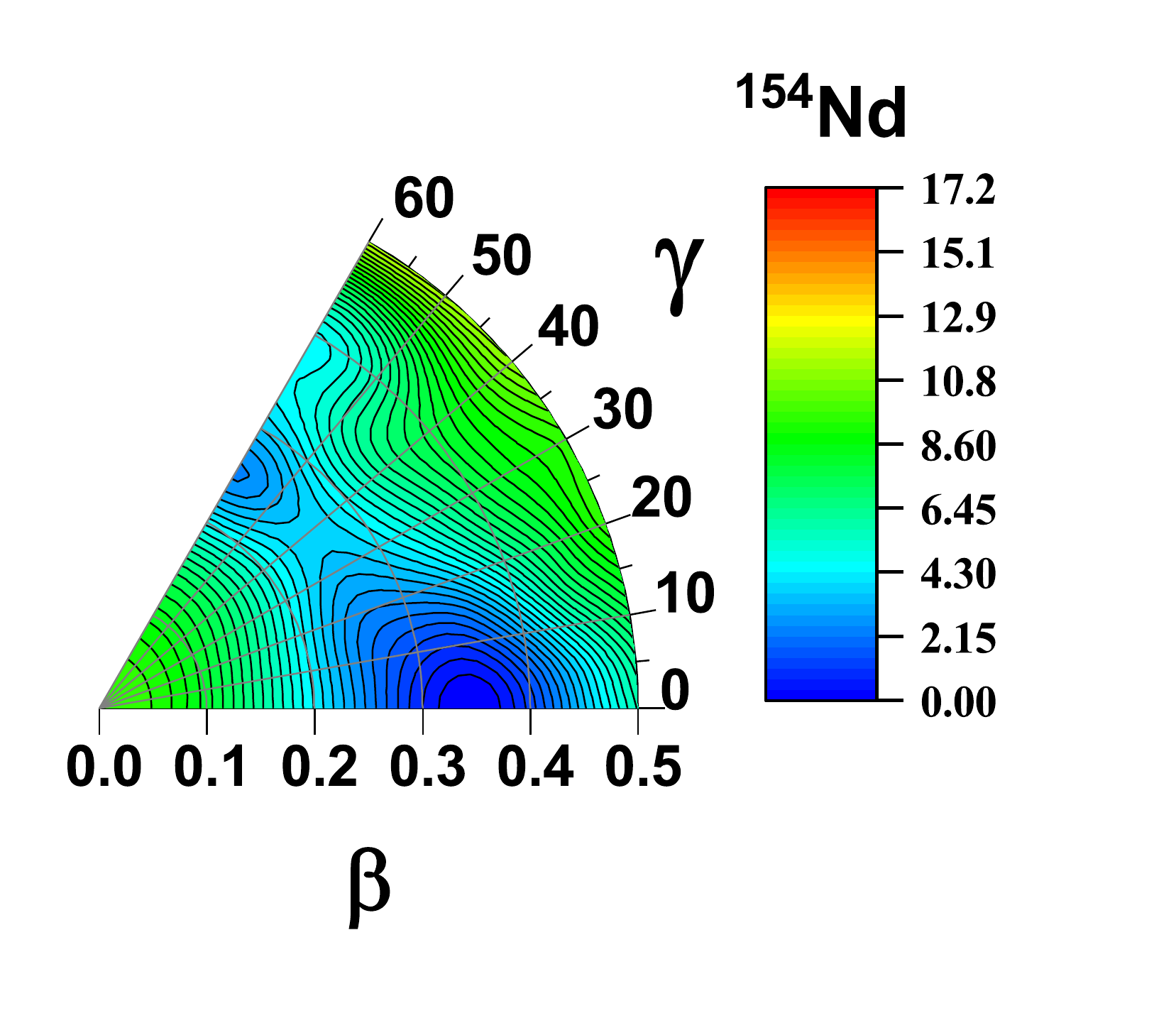}   \\
		\includegraphics[width=.25\textwidth]{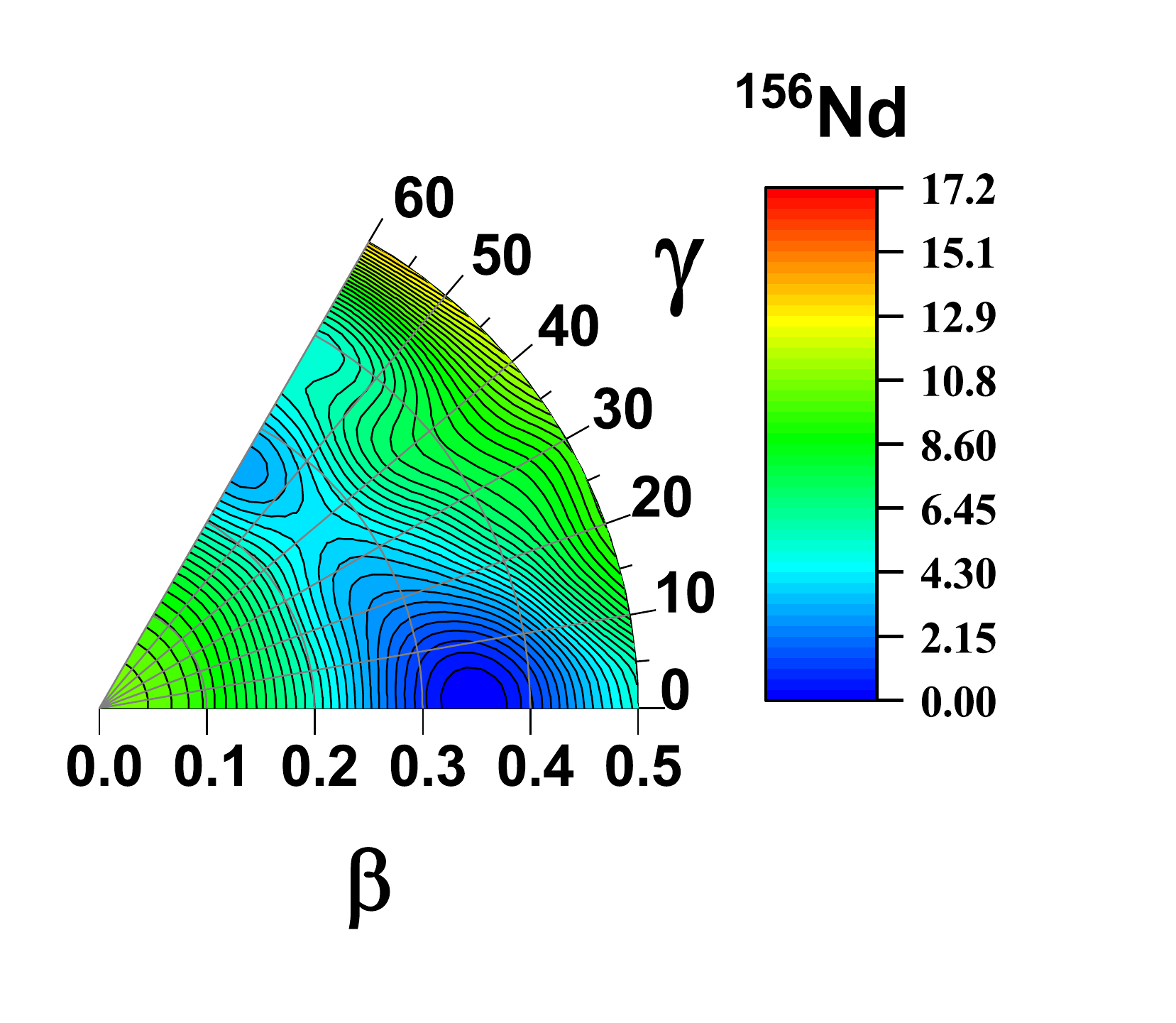} &
		\includegraphics[width=.25\textwidth]{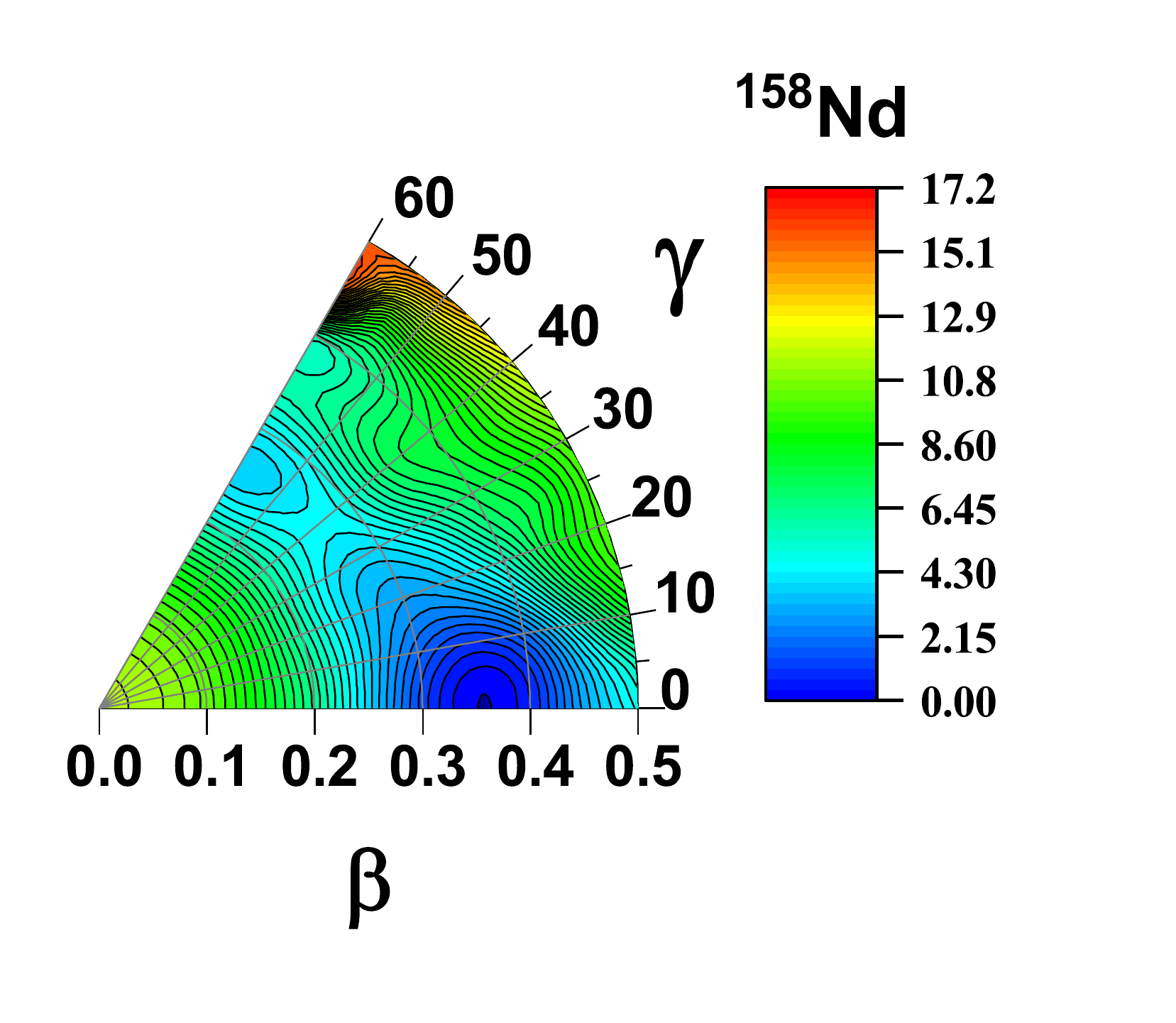} &
		\includegraphics[width=.25\textwidth]{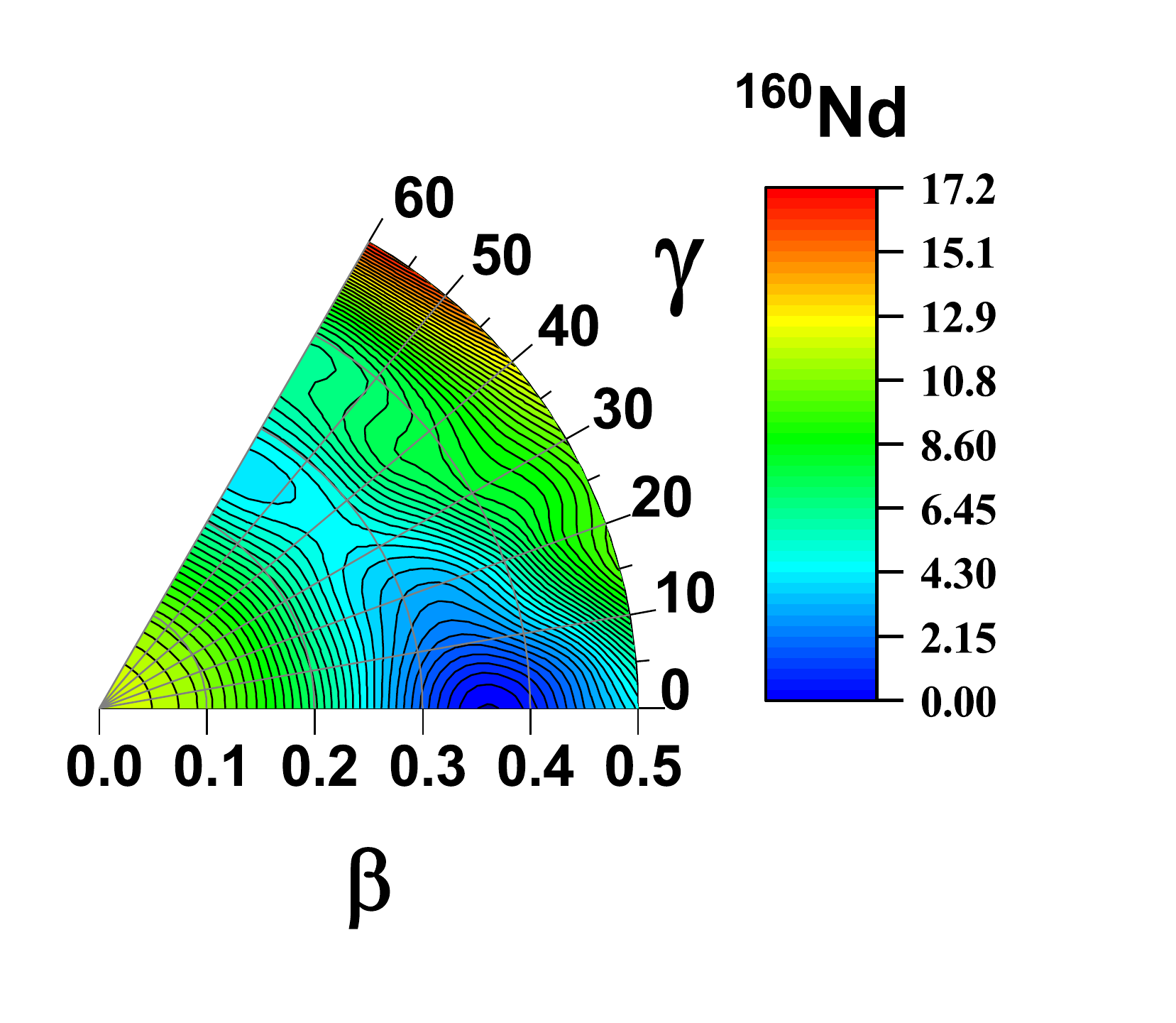} &
	\end{tabular}
	\caption{(Color online) The total energy curves for $^{124-160}$Nd obtained within CDFT framework with DD-ME2 sets as a function of the axial quadrupole deformation parameter, $\beta_2$.}
	\label{PES}
\end{figure}

In order to compare our results obtained concerning the shape of Nd isotopes from GDR spectra  with other theory, we perform a calculation of the total energy curves for $^{124-160}$Nd with the covariant density functional theory (CDFT) \cite{nik2008,nik2014,nik2005,roca2011,nik2002} by using the very successful density-dependent meson-exchange relativistic energy functional DD-ME2~\cite{nik2005}. In Fig.~\ref{PES} we display  triaxial contour plots of $^{124-160}$Nd isotopes in the $\beta \gamma$-plane. To study the dependency on $\gamma$, a systematic constrained triaxial calculation has been done for mapping the quadrupole deformation space defined by $\beta_2$ and $\gamma$. The obtained energies are normalized with respect to the binding energy of the global minimum. 
%%%%%%%%%%%%%%%%%%%%%%%%%%%%%%%%% curve surface energy %%%%%%%%%%%%%%%%%%%

%%%%%%%%%%%%%%%%%%%%%%%%%%%%%%%%%%%%%%%%%%%%%%%%%%%%%%%%%%%%%%%%%%%%%%%%%%%%
From Fig.~\ref{PES}, we can notice that, at the beginning of the chain,
Nd isotopes have an axial prolate shape for $^{124-130}$Nd. Then, the location of the ground state minimum moves to triaxial  shape at $^{132-138}$Nd. $^{140-144}$Nd have a spherical shape, while the isotope $^{146}$Nd appears as more deformed with a prolate shape. The ground state remains prolate up to the end of the isotopic chain (N=100).
These  results are in  agreement with ours  shown in Figs.~\ref{dm-t1} - \ref{gr52-60} except for $^{132}$Nd. The calculation with TDHF theory gives a weakly triaxial shape for $^{132}$Nd as shown in Fig.~\ref{dm-t2}, while a prolate shape is obtained with CDFT theory as shown in Fig.~\ref{PES}.
\subsection{Relation between deformation splitting $\Delta E$ and quadrupole deformation $\beta_{2}$}

\begin{figure}[!htbp]
	\centering
	%\hypertarget{b2-n}{}
	\includegraphics[scale=0.7]{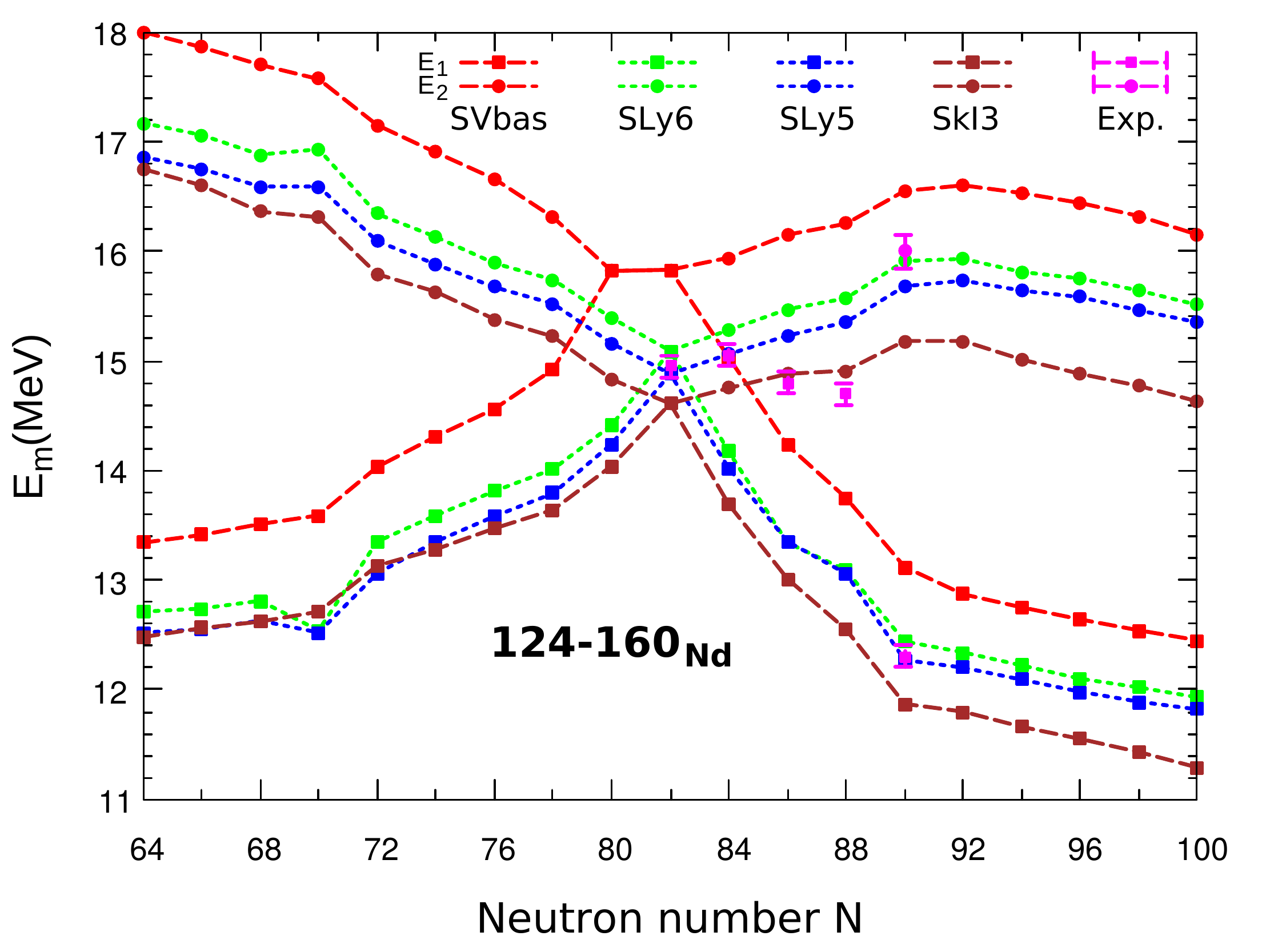}
	\caption{The peak position E$_{m}$ of GDRs along major axis (circle symbol) and minor axis (square symbol) in  Nd isotopes.}
	\label{e1e2n}
\end{figure}

\qquad Among the properties of  giant resonance, thers is the resonance energy centroid defined by the formula
\begin{equation}{\label{eq13}}
E_{m} = \frac{\int_{0}^{+\infty} S(E)E dE}{\int_{0}^{+\infty} S(E) dE},
\end{equation}
where S(E) (\ref{eq11}) is the strength function of GR. The total GDR spectrum generally presents two peaks for  deformed nuclei, one corresponds to a resonance energy  E$_{m}^{1}$ along the major axis and the other corresponds to a resonance energy E$_{m}^{2}$ along the minor axis. In Table \ref{tab5} %and \ref{tab6}%
, we summarize the TDHF calculation with the four Skyrme forces concerning  E$_{m}^{1}$ and  E$_{m}^{2}$ compared with the available experimental data from Ref.\cite{carlos1971}. In Fig.~\ref{e1e2n}, we plot E$_{m}$ as function as number of neutrons. E$_{m}^{1}$ and E$_{m}^{2}$ denote the peak position of GDR along the minor axis and the major axis respectively. It is seen  that the resonance energy E$_{m}^{1}$ along the minor axis decreases until N=142 , and then trends to increase when the neutron number  N increases. On the contrary, the resonance energy E$_{m}^{2}$ along the major axis increases when N increases until N=142, and then gradually decreases. We note also, that the values of E$_{m}^{1}$ and  E$_{m}^{2}$ calculated by these Skyrme forces are generally in agreement with the available experimental data with a small shift $\pm$ 0.5 MeV. The agreement is better for the Skyrme force SLy6, which was shown to be optimal for the description of the IVGDR in medium-heavy deformed nuclei \cite{reinhard2008}.\\

\begin{figure}[!htbp]
	\centering
	%\hypertarget{b2-n}{}
	\includegraphics[scale=0.7]{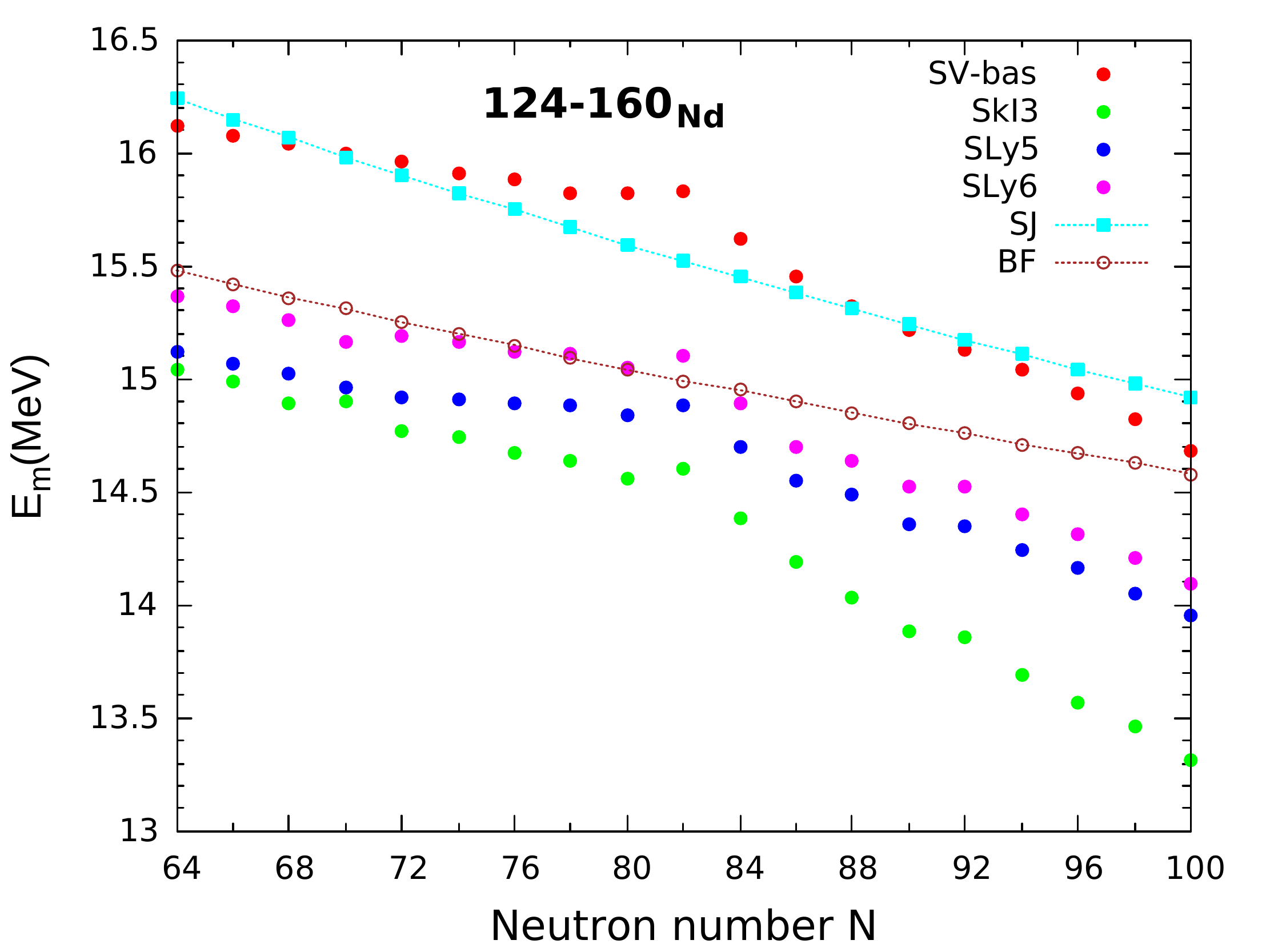}
	\caption{The resonance energies E$_{m}$ in  Nd isotopes compared with the estimates SJ((\ref{eq14})) and BF(\ref{eq15}).}
	\label{Em}
\end{figure}

In Fig.~\ref{Em}, the resonance energy E$_{m}$ calculated with the four Skyrme forces is compared with the collective models Steinwedel-Jensen(SJ) \cite{steinwedel1950} 
\begin{equation}{\label{eq14}}
E_{SJ}= 81.A^{-1/3} MeV
\end{equation}
and Berman-Fultz (BF) \cite{berman1975}
\begin{equation}{\label{eq15}}
E_{BF}= (31.2A^{-1/3} + 20.6A^{-1/6} )MeV
\end{equation}
The latter treats dipole resonance as a combination of  Steinwedel-Jensen($ E_{SJ}\sim A^{-1/3} $)~\cite{steinwedel1950} and  Goldhaber-Teller($ E_{GT}\sim A^{-1/6}$)~\cite{goldhaber1948} models. It is seen that the calculated energies with SLy6 agree quite well with the BF estimate~\cite{berman1975} until the N=82 shell closure, while those calculated with SVbas are close to the SJ  estimate~\cite{steinwedel1950}.  But all Skyrme forces reproduce the overall trend of decreasing energy with increasing mass number A.
\begin{table}[!htbp]
	\centering
	\caption {The resonance energy centroids E$_{m}^{1}$ and E$_{m}^{2}$  correspond respectively to oscillation along the major axis  and the minor one. The experimental data are from ref. \cite{carlos1971}. \label{tab5}} 
	%	{\begin{tabular}{ccccccccccc}
 
			{\begin{tabular}{ccccccccccc}
			\cline{2-11}
			& \multicolumn{2}{c}{SkI3} & \multicolumn{2}{c}{SVBas}  &    \multicolumn{2}{c}{SLy5} &  \multicolumn{2}{c}{SLy6}& \multicolumn{2}{c}{Exp.\cite{carlos1971}} 	\\
			\hline Nuclei &    E$_{m}^{1}$ & E$_{m}^{2}$ &  E$_{m}^{1}$ & E$_{m}^{2}$ & E$_{m}^{1}$  & E$_{m}^{2}$ &  E$_{m}^{1}$ & E$_{m}^{2}$ &  E$_{m}^{1}$ &  E$_{m}^{2}$	\\
			\hline	$^{\textbf{124}}\textbf{Nd}$ & 12.48 & 16.75 &  13.34 & 18.00 &  12.52 & 16.86 &12.71  & 17.17 & ---  &  ---	\\
			\hline	$^{\textbf{126}}\textbf{Nd}$ & 12.56 &16.60 &  13.41 &  17.86 &  12.55 & 16.75 & 12.74 & 17.06  & --- &  ---	\\
			\hline	$^{\textbf{128}}\textbf{Nd}$& 12.62 & 16.36 &  13.51 & 17.70 &  13.63 & 16.59  & 12.81 & 16.88 & ---  &  ---	\\
			\hline	$^{\textbf{130}}\textbf{Nd}$ & 12.71 & 16.31 &  13.59 & 17.57 &  12.52 & 16.59 &12.54  & 16.93 & ---  &  ---	\\
			\hline	$^{\textbf{132}}\textbf{Nd}$ & 13.13 &15.79 &  14.03 &  17.15 &  13.06 & 16.09 & 13.35 & 16.34  & --- &  ---	\\
			\hline	$^{\textbf{134}}\textbf{Nd}$& 13.28 & 15.63 &  14.31 & 16.90 &  13.35 & 15.88  & 13.59 & 16.13 & ---  &  ---	\\
			\hline	$^{\textbf{136}}\textbf{Nd}$ & 13.47 & 15.38 &  14.56 & 16.66 &  13.58 & 15.67 &12.54  & 16.93 & ---  &  ---	\\
			\hline	$^{\textbf{138}}\textbf{Nd}$ & 13.64 &15.22 &  14.92 &  16.31 &  13.79 & 15.52 & 13.35 & 16.34  & --- &  ---	\\
			\hline	$^{\textbf{140}}\textbf{Nd}$& 14.04 & 14.83 &  15.82 & 15.82 &  14.24 & 15.16  & 13.59 & 16.13 & ---  &  ---	\\
			\hline	$^{\textbf{142}}\textbf{Nd}$& 14.61 & 14.61 &  15.83 & 15.83  &   14.88 & 14.88  & 15.09 & 15.09  & 14.95$\pm$ 0.10 & --- 	\\
			\hline	$^{\textbf{144}}\textbf{Nd}$ & 13.69 & 14.75 &  15.03 & 15.94 &   14.02 & 15.06  & 14.17 & 15.28 & 15.05$\pm$ 0.10  &  ---	\\
			\hline	$^{\textbf{146}}\textbf{Nd}$ & 13.01 & 14.88 &  14.23 & 16.15  &  13.35 & 15.23  & 13.34 & 15.47  & 14.80$\pm$ 0.10 &  ---	\\
			\hline	$^{\textbf{148}}\textbf{Nd}$ & 12.55 & 14.91 &  13.75 & 16.25   &13.06  &15.35  & 13.09 & 15.57  & 14.70$\pm$ 0.15 &  ---	\\
			\hline	$^{\textbf{150}}\textbf{Nd}$ & 11.87 & 15.18 &  13.12 & 16.55   & 12.27 &15.68  & 12.44 & 15.91  & 12.30$\pm$ 0.15 & 16.00$\pm$ 0.15 \\
			\hline	$^{\textbf{152}}\textbf{Nd}$ & 11.80 & 15.18 &  12.87 & 16.60 &  12.20 & 15.73 &12.34  & 15.93 & ---  &  ---	\\
			\hline	$^{\textbf{154}}\textbf{Nd}$ & 11.66 &15.01 &  12.75 &  16.53 &  12.09 & 15.64 & 12.22 & 15.81  & --- &  ---	\\
			\hline	$^{\textbf{156}}\textbf{Nd}$& 11.55 & 14.88 &  12.64 & 16.44 &  11.98 & 15.58  & 12.10 & 15.75 & ---  &  ---	\\
			\hline	$^{\textbf{158}}\textbf{Nd}$ & 11.43 & 14.77 &  12.54 & 16.32 &  11.89 & 15.46 &12.02  & 15.64 & ---  &  ---	\\
			\hline	$^{\textbf{160}}\textbf{Nd}$ & 11.29 &14.63 &  12.45 &  16.15 &  11.82 & 15.35 & 11.93 & 15.51  & --- &  ---	\\
			\hline	
	\end{tabular}}
	%\label{table:2}
\end{table}
%%%%%%%%%%%%%%%%%%%%%%%%%%%%%%%%%%%%%%%%%%%%%%%%%%%%%%%%%%%%%%%%%%%%%%%%%%%%%%%%
%%%%%%%%%%%%%%%%%%%%%%%%%%%%%%%%%%%%%%%%%%%%%%%%%%%%%%%%%%%%%%%%%%%%%%%%%%%%%%%%
%%%%%%%%%%%%%%%%%%%%%%%%%%%%%%%%%%%%%%%%%%%%%%%%%%%%%%%%%%%

\begin{figure}[!htbp]
	\centering
	%\hypertarget{b2-n}{}
	\includegraphics[scale=0.5]{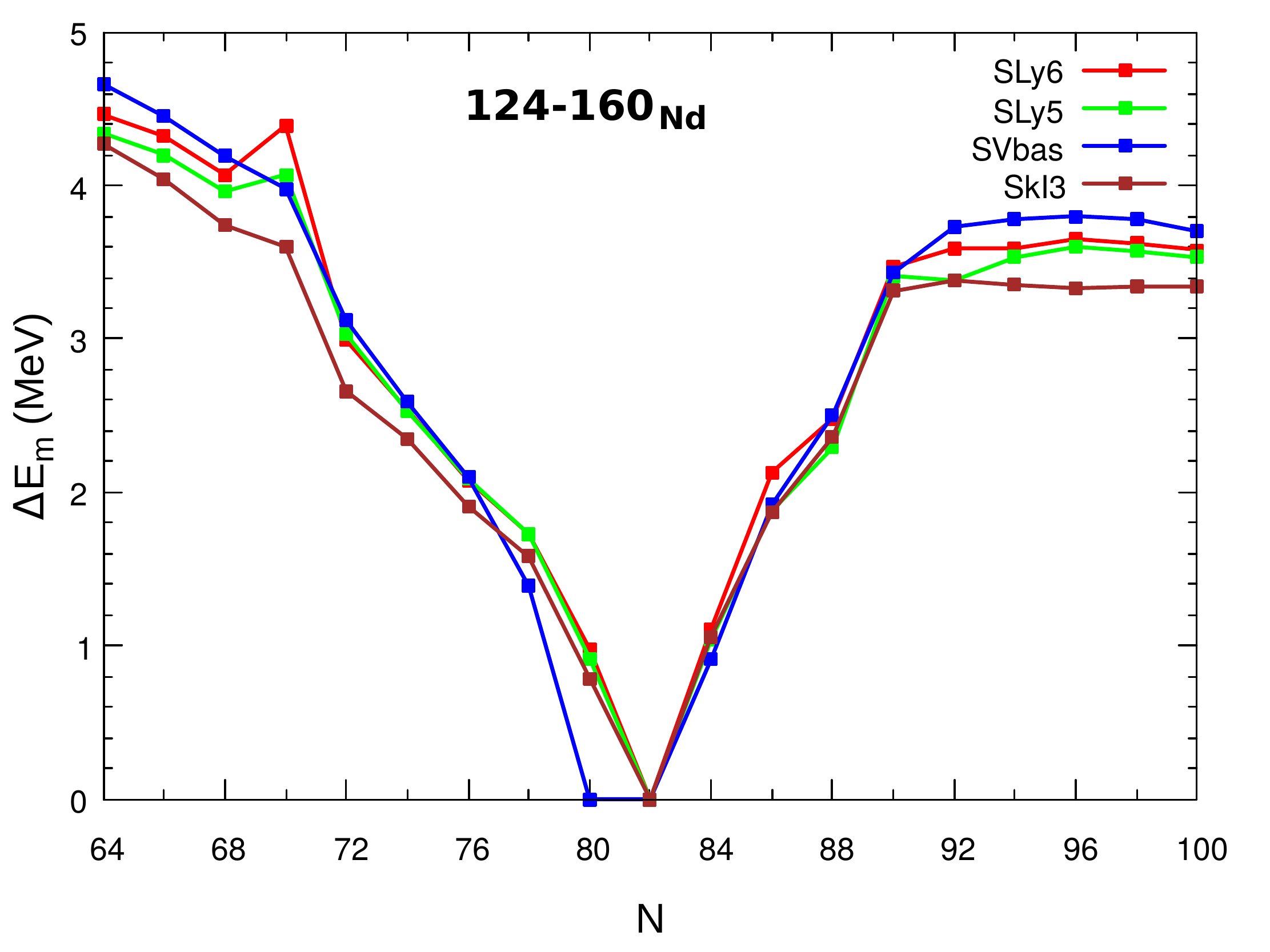}
	\caption{The GDR splitting as a function of the number of neutrons N in the case  of $ ^{124-160}\text{Nd} $ isotopes calculated with SLy6, SLy5, SVbas and SkI3.}
	\label{dE-N}
\end{figure}

The deformation splitting $\Delta$E  is the difference between the resonance energy centroids along two axis $\Delta$E = E$_{m}^{2}$ - E$_{m}^{1}$. From  Fig.~\ref{dE-N}, it can be easily seen for all Skyrme forces that the GDR splitting $\Delta$E decreases firstly when N increases from N=64 until N=82 (magic Number ) which corresponds to spherical nucleus $ ^{142}\text{Nd} $ and then increases again until N=100 which correspond to well deformed nucleus $ ^{160}\text{Nd} $. It confirms that the deformation structure of the nuclei is responsible to the division peak in the deformed nuclei.

\begin{figure}[!htbp]
	\begin{center}
		\begin{minipage}[t]{0.4\textwidth}
			\includegraphics[scale=0.3]{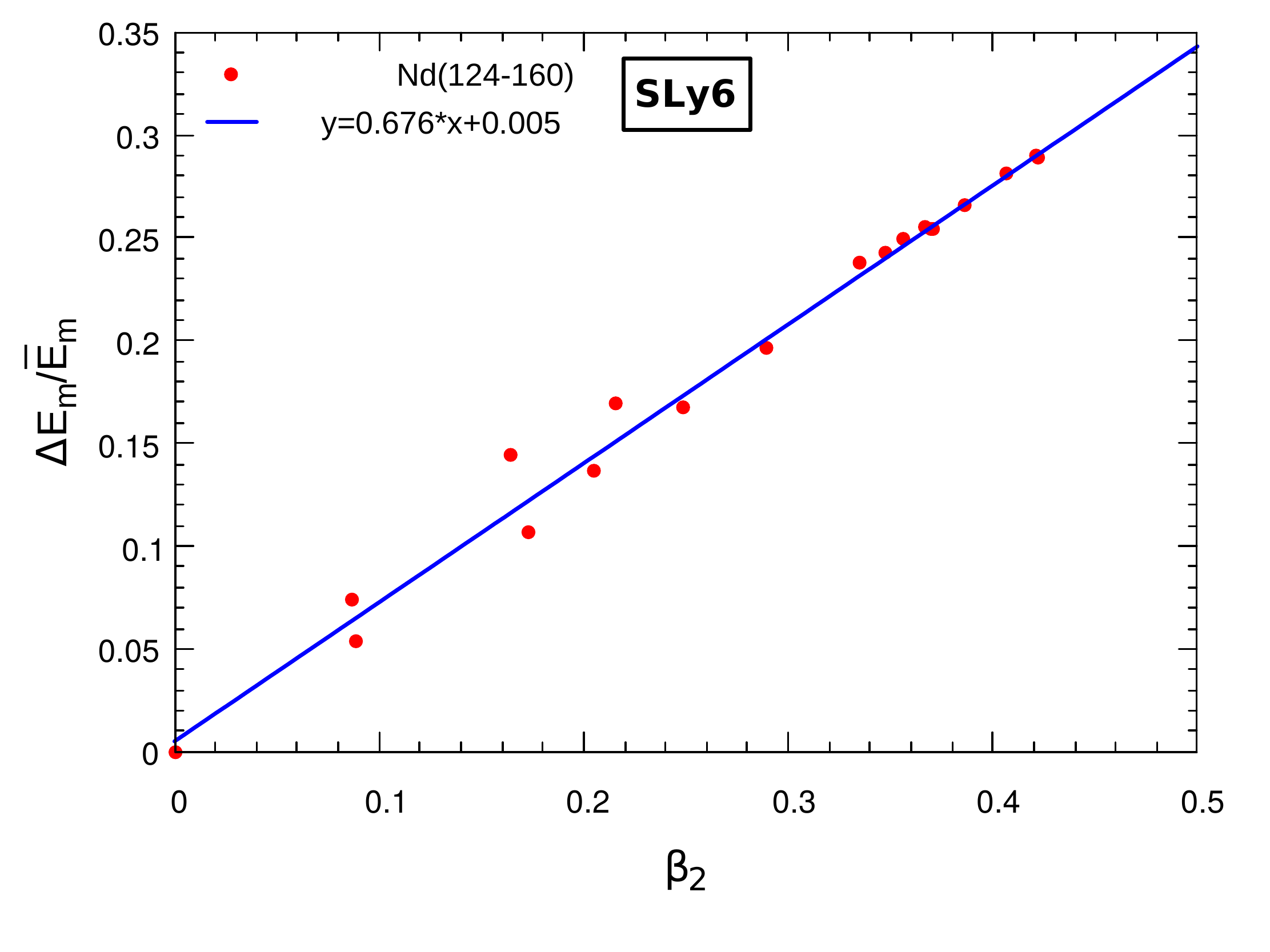}
		\end{minipage}
		\vspace{0.5cm}
		\hspace{2cm}
		\begin{minipage}[t]{0.4\textwidth}
			\includegraphics[scale=0.3]{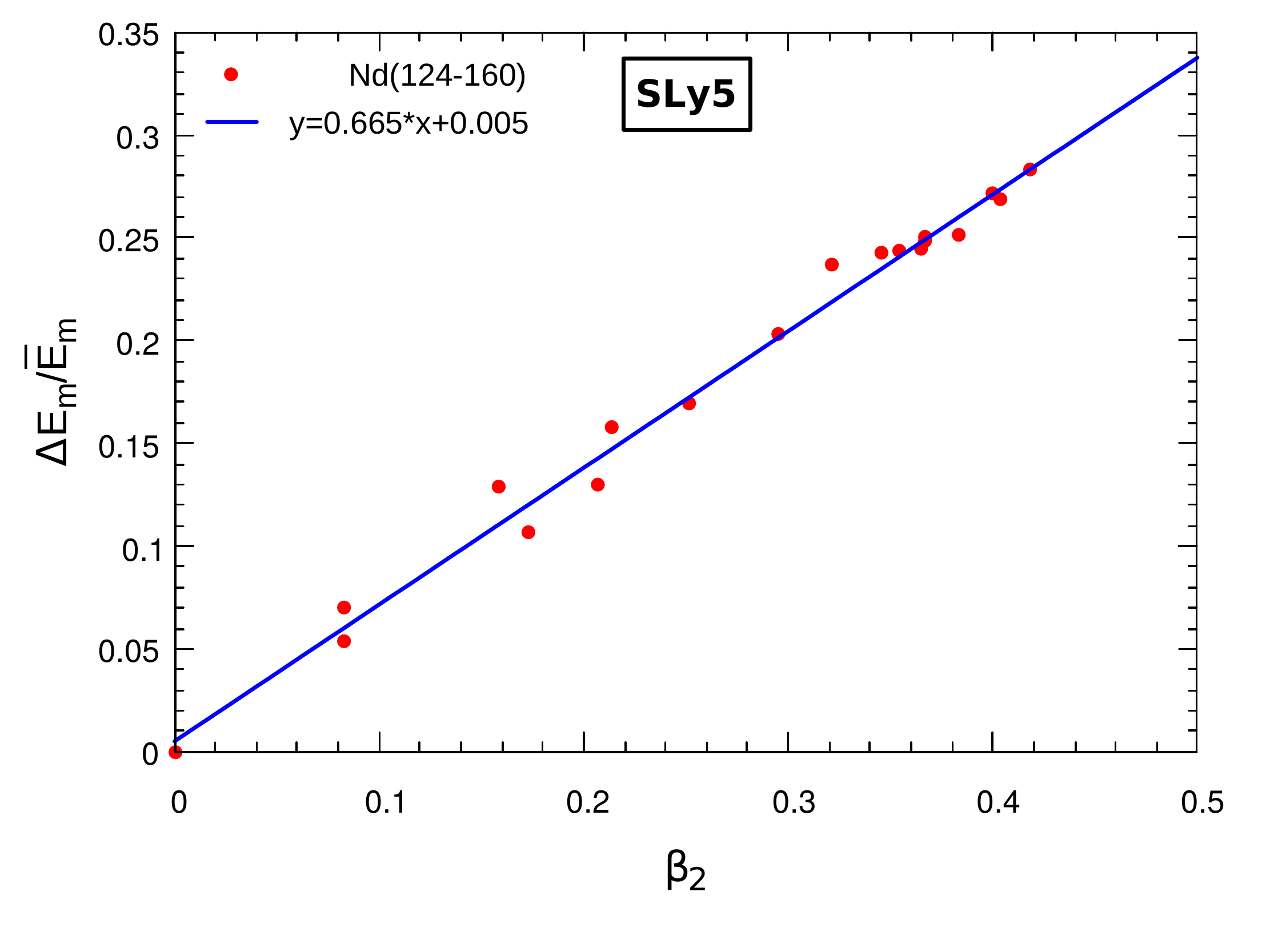}
		\end{minipage}
		\begin{minipage}[t]{0.4\textwidth}
			\includegraphics[scale=0.3]{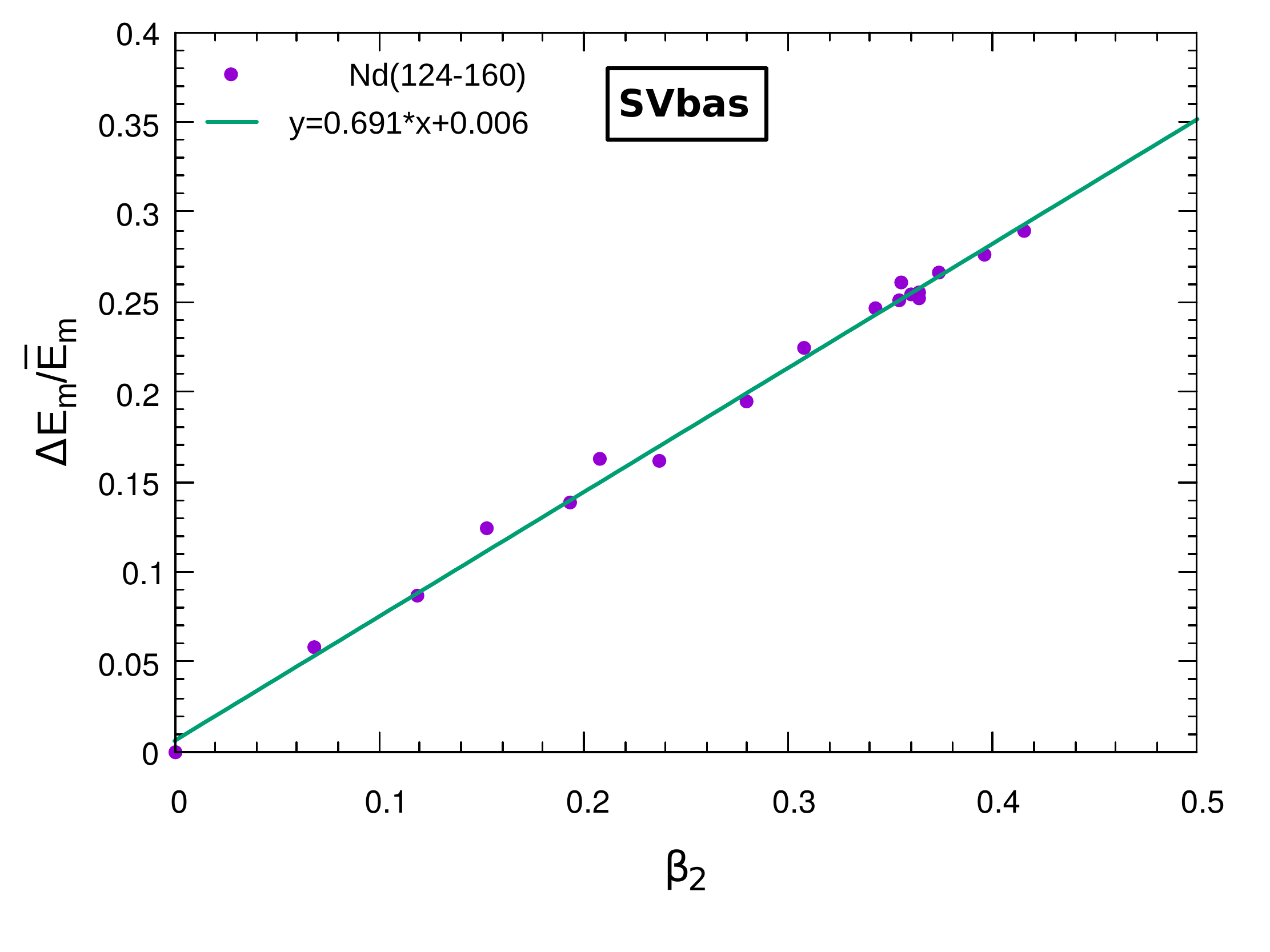}
		\end{minipage}
		\vspace{0.5cm}
		\hspace{2cm}
		\begin{minipage}[t]{0.4\textwidth}
			\includegraphics[scale=0.3]{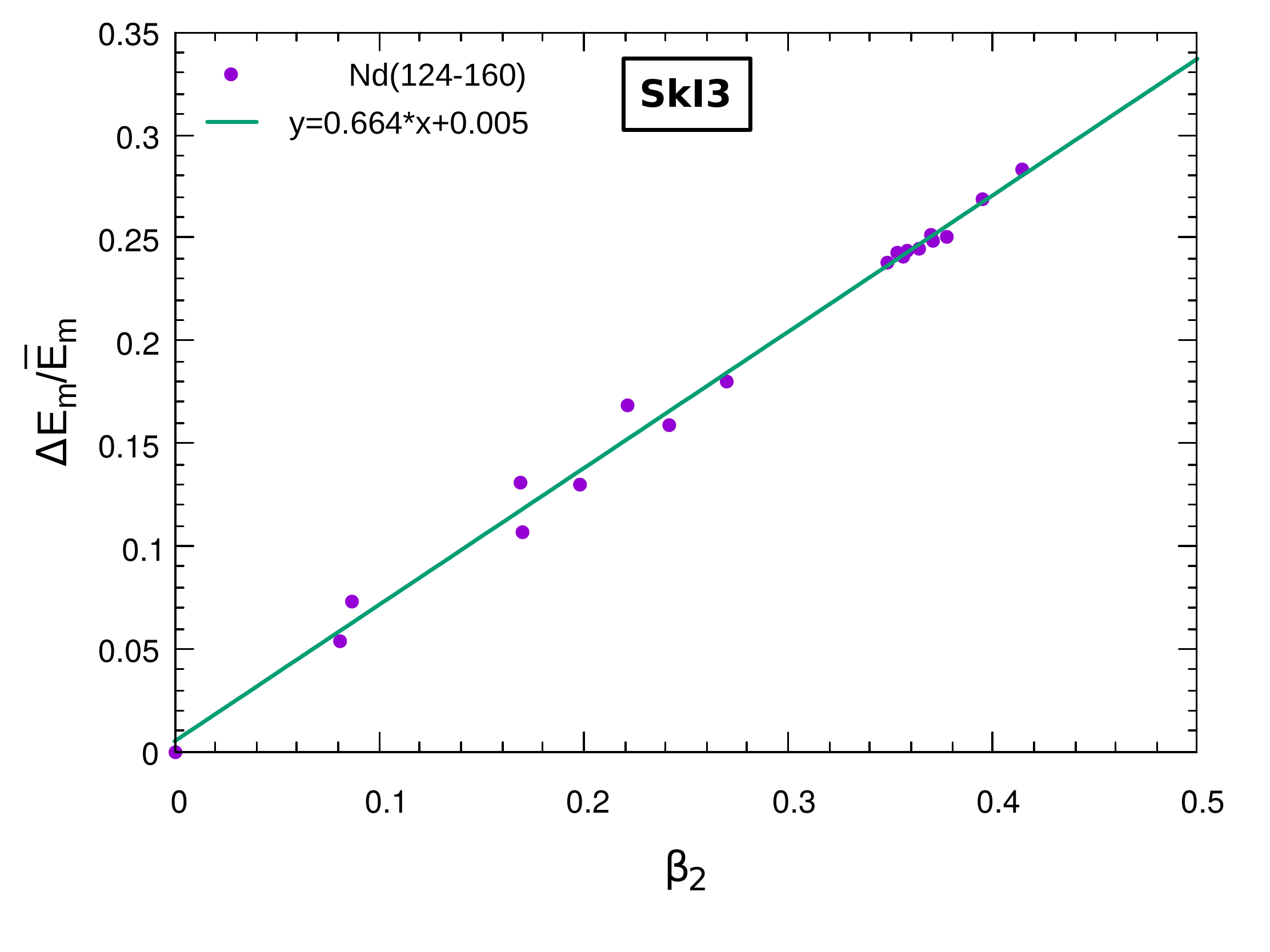}
		\end{minipage}
		\caption{The correlation between the deformation parameter $\beta_{2}$ and the ratio $\Delta E/ \bar{E}$. circles denote the data in the  Nd isotopes and lines are the fitting results. } 
		\label{de-b2}
	\end{center}	
\end{figure}

Since the GDR splitting is caused by the deformation, it is possible to relate the nuclear deformation with the distance $\Delta$E between two peaks of GDR spectra. In Fig.~\ref{de-b2} is plotted the ratio $\Delta E/\bar{E}_{m}$  for $ ^{124-160}\text{Nd}$ isotopes as a function of the quadrupole deformation parameter $\beta_{2}$. For the four Skyrme forces, we see that there is an approximate linear relation between $\Delta E/\bar{E}_{m}$ and $\beta_{2}$ of the general form 
\begin{equation}{\label{eq16}}
\Delta E/ \bar{E}_{m}  = a.\beta_{2} + b, 
\end{equation}
where a and b are  parameters depend on the Skyrme force, but they have very close values. For example, in the  case of SVbas: a $\simeq$ 0.691 and b $\simeq$ 0.006 and in the case of SLy6:  a $\simeq$ 0.676 and b  $\simeq$ 0.005. The relation (\ref{eq16})  was already studied in Refs. \cite{okamoto1958,ring2009,wang2017}. 

%%%%%%%%%%%%%%%%%%%%%%%%%%%%%%%%%% beta2 & N %%%%%%%%%%%%%%%%%%%%%%%%%%%%
%%%%%%%%%%%%%%%%%%%%%%%%%%%%%%%%%%%%%%%%%%%%%%%%%%%%%%%%%%%%%%%%%%%%%%%%%%%
%%%%%%%%%%%%%%%%%%%%%%%%%%% dE/E & beta %%%%%%%%%%%%%%%%%%%%%%%%%%%%%%%%
%%%%%%%%%%%%%%%%%%%%%%%%%%%%%%%%%%%%%%%%%%%%%%%%%%%%%%%%%%%%%%%%%%%%%%%%%%%
\section{Conclusion}\label{sec5}
\qquad We have studied the GDR  in $ ^{124-160}\text{Nd} $ isotopes with  TDHF method based on  Skyrme functional, using  SKY3D code~\cite{sky3d}. This study covers 19 isotopes of Nd with four Skyrme forces SkI3, SVbas, SLy5 and SLy6. Deformation parameters ($\beta, \gamma$), resonance energy centroids (E$_{m}$, E$_{m}^{1}$, E$_{m}^{2}$) have been calculated for  $ ^{124-160}\text{Nd} $ isotopes. The even isotopes of Nd from A=124 to A=130 and from A=144 to A=160  are of prolate shape with $\gamma \simeq 0 ^\circ$. The nuclei $^{132}\text{Nd} $, $ ^{134}\text{Nd} $, $^{136}\text{Nd} $ and $ ^{138}\text{Nd} $ are weakly triaxial. The nucleus $^{142}\text{Nd} $ is of a spherical shape with $\beta \simeq 0 $.
The dipole moment D$_{m}$(t) of GDR calculation showed that oscillation frequency along the major axis is lower than that along the minor axis of the deformed axially nuclei as $ ^{150}\text{Nd} $, and is the same  along the three axes  in spherical nuclei as $ ^{142}\text{Nd} $. 

In addition, the  GDR strength calculated is compared with the experimental data. The results showed that  TDHF method can reproduce the shape of the GDR spectra with a small shift depending on the used Skyrme force. The Skyrme forces SLy6, SLy5 and SVbas reproduce well the experimental data compared to SkI3, with a slight advantage for SLy6. 

Resonance energies and GDR splitting $\Delta$E are well described. We have found a correlation between the ratio $\Delta E/ \bar{E}$ and the deformation parameter $\beta_{2}$. For the isotopic chain of Nd, $\Delta E/ \bar{E}_{m} \simeq 0.670\beta_{2} + 0.005$. It confirms that the splitting of GDR spectra is proportional to the deformation of the nucleus. 
\section*{Acknowledgement}  Discussion with  Prof. Paul Stevenson  from university of serrey, Prof. Paul-Gerhard Reinhard and Prof. Sait Umar from Vanderbilt University is acknowledged.
\section*{References}
\bibliographystyle{iopart-num}
\bibliography{nd}

\end{document}